\documentclass[fleqn,usenatbib]{mnras}

\usepackage{newtxtext,newtxmath}

\usepackage[T1]{fontenc}

\DeclareRobustCommand{\VAN}[3]{#2}
\let\VANthebibliography\thebibliography
\def\thebibliography{\DeclareRobustCommand{\VAN}[3]{##3}\VANthebibliography}

\usepackage{epsfig}
\usepackage{graphicx}
\usepackage{float}
\usepackage[caption =false]{subfig}
\usepackage{graphicx}
\usepackage{multirow}
\usepackage{longtable}
\usepackage{multicol}
\usepackage{lipsum}
\usepackage{supertabular}
\usepackage{enumerate}

 \usepackage{times}
\renewcommand{\thefootnote}{\fnsymbol{footnote}}

\def\lsim{\mathrel{\rlap{\lower4pt\hbox{\hskip1pt$\sim$}}
    \raise1pt\hbox{$<$}}}                
\def\gsim{\mathrel{\rlap{\lower4pt\hbox{\hskip1pt$\sim$}}
    \raise1pt\hbox{$>$}}}                

\usepackage{graphicx}	
\usepackage{amsmath}	

\title[]
{Spectroscopy of CASSOWARY gravitationally-lensed galaxies in SDSS: characterisation of an extremely bright reionization-era analog at $z=1.42$} 

\begin{document}

\author[Mainali et al.] 
{Ramesh Mainali$^{1,2,3}$\footnotemark[1], 
Daniel P. Stark$^2$,
Tucker Jones$^{4}$\footnotemark[2], 
Richard S. Ellis$^{5}$,
Yashar D. Hezaveh$^{6,7}$,\newauthor 
 \&  Jane R. Rigby$^{1}$ \\
$^{1}$ Observational Cosmology Lab, NASA Goddard Space Flight Center, Greenbelt, MD 20771, USA \\
$^{2}$ Steward Observatory, University of Arizona, 933 N Cherry Ave, Tucson, AZ, USA \\
$^{3}$Department of Physics, The Catholic University of America, Washington, DC 20064, USA \\
$^{4}$ Department of Physics, University of California Davis, 1 Shields Avenue, Davis, CA 95616, USA \\  
$^{5}$ Department of Physics and Astronomy, University College London, Gower Street, London, WC1E 6BT, UK \\
$^{6}$Center for Computational Astrophysics, Flatiron Institute, 162
Fifth Avenue, New York, NY 10010, USA\\
$^{7}$D{\'e}partement de Physique, Universit{\'e} de Montr{\'e}al,
Montreal, Quebec, Canada H3T 1J4\\
}


\date{Accepted XXX. Received YYY; in original form ZZZ}

\pubyear{2022}


\label{firstpage}
\pagerange{\pageref{firstpage}--\pageref{lastpage}}
\maketitle

\label{firstpage}
\begin{abstract}  
We present new observations of sixteen bright ($r=19-21$) gravitationally lensed 
galaxies at $z\simeq 1-3$ selected from the CASSOWARY survey. Included in our sample is the $z=1.42$ galaxy CSWA-141, one of the brightest known reionization-era analogs at high redshift (g=20.5), with a large sSFR (31.2 Gyr$^{-1}$) and an [OIII]+H$\beta$ equivalent width (EW$_{\rm{[OIII]+H\beta}}$=730~\AA) that is nearly identical to the 
average value expected at $z\simeq 7-8$.  In this paper, we investigate the rest-frame UV nebular line emission in our sample with the goal of understanding the factors that regulate strong CIII] emission. Whereas most of the sources in our sample show weak UV line emission, we find elevated CIII]  in the spectrum of CSWA-141 (EW$_{\rm{CIII]}}$=4.6$\pm1.9$~\AA) together with detections of other prominent emission lines (OIII], Si III], Fe II$^\star$, Mg II).  We compare the rest-optical line properties of high redshift galaxies with strong and weak CIII] emission, and find that systems with the strongest UV line emission tend to have young stellar populations and nebular gas that is moderately metal-poor and highly ionized, consistent with trends seen at low and high redshift. The brightness of CSWA-141 enables detailed 
investigation of the extreme emission line galaxies which become common at $z>6$.
We find that gas traced by the CIII] doublet likely probes higher densities than 
that traced by [OII] and [SII]. Characterisation of the spectrally resolved Mg II emission line and several low ionization absorption lines suggests neutral gas around the young stars is likely optically thin, potentially facilitating the escape of ionizing radiation.  

\end{abstract} 


\renewcommand{\thefootnote}{\fnsymbol{footnote}}
\footnotetext[1]{E-mail: ramesh.mainali@nasa.gov}

\section{Introduction}

Over the last decade, much progress has been made in our understanding  of galaxies in 
the first billion years of cosmic time (for a review, see \citealt{Stark2016}).   Deep infrared
imaging has uncovered thousands of photometrically-selected star forming systems thought to lie in
the redshift range $6<z<9$ (e.g.,
\citealt{McLure2013,Finkelstein2014,Bouwens2015,Bouwens2021,Livermore2017,Ishigaki2018}), providing 
a census of UV-selected galaxies throughout the reionization era.    
The spectral energy distributions (SEDs) point toward a population 
undergoing rapid stellar mass growth with blue UV continuum spectral slopes (e.g., \citealt{Rogers2013,Bouwens2014a}), low
stellar masses and large specific star formation rates
\citep{Stark2013,Gonzalez2014,Grazian2014,Salmon2015,Curtislake2016}.   

In the last five years, our first constraints on the nebular emission properties 
of galaxies at these early epochs have emerged. {\it Spitzer}/IRAC
photometry suggests that nearly half of the UV-selected galaxies at $z\simeq 7$ have extremely large [OIII]+H$\beta$ equivalent widths (EWs) \citep{Labbe2013, Smit2014a, Smit2015,Robertsborsani2015,deBarros2017, Endsley2020a}, indicating that very recent 
($\lsim 50$ Myr) activity powers the UV and optical luminosity, as expected for 
galaxies undergoing rapidly rising star formation histories. Roughly 20\% of the 
population have yet more intense rest-optical nebular emission ([OIII]+H$\beta$ EW $>$ 1000~\AA), indicating an extremely young stellar population ($<$10 Myr) is dominating 
the light, as expected for systems that have undergone a recent burst of 
star formation. Since such extreme emission line galaxies (EELGs) are very rare 
at lower redshifts \citep{Boyett2022}, we lack a detailed understanding of the gas and ionizing agents in typical reionziation-era systems.

Ground-based near-infrared spectrographs offer a path toward progress at the highest redshifts, providing access to the rest-frame ultraviolet where a suite of valuable diagnostic lines  are 
situated. The first deep spectra have revealed strong nebular emission from high ionization metal species ([CIII],CIII]$\lambda\lambda$1907,1909~\AA,
OIII]$\lambda\lambda$1660,1666~\AA, CIV$\lambda\lambda$1548,1550~\AA), a significant 
departure from what is commonly seen at lower redshifts 
\citep{Stark2015a,Stark2015b,Stark2017,Mainali2017,Laporte2017,Mainali2018,Hutchison2019,Topping2021}. The detection of these lines reveals 
gas under extreme ionization conditions and points to a 
population of intense ionizing agents, potentially AGN in 
some cases \citep{Laporte2017,Mainali2018} and low metallicity 
massive stars in others \citep{Mainali2017,Stark2017,Berg2019,Berg2021}. The detection of CIV$\lambda\lambda$1548,1550~\AA\ and [CIII],CIII]$\lambda\lambda$1907,1909~\AA\ may further indicate a higher fraction of ionizing photons escape from a galaxy\citep{Schaerer2022}. While
the equivalent width distribution in the total population 
is still subject to limited statistics \citep{Mainali2018}, it 
appears that both the gas and ionizing sources at $z\gsim 6$ are  often significantly different from that in galaxies at 
$z\simeq 2-3$. 

The presence of strong rest-UV nebular emission in a subset of $z\gsim 6$ galaxies bodes well for future studies of the 
reionization era. For the next-generation 25-30m ground based optical/infrared observatories (which are limited to constraints on the rest-frame UV at $z>6$), these  lines may provide the only way in which early galaxies can be studied spectroscopically.  While typically fainter than the strong lines in the rest-optical, the suite of emission lines in the far-UV provide unique diagnostic power of the ionizing spectrum and gas physical conditions (i.e, density, temperature, ionization). Meanwhile in the near-UV, the resonant nature of the nebular Mg II emission line makes it an ideal probe of the neutral gas opacity in early galaxies, potentially providing an indirect indicator of LyC leakage at $z\gsim 6$ (e.g., \citealt{Henry2018,Chisholm2020,Xu2022,Izotov2022,Seive2022}). The advantage of Mg II relative to  Ly$\alpha$ is that it is not obscured by the neutral IGM at very high 
redshifts. 
 
To ensure that the faint rest-UV spectra provide reliable physical diagnostics from these future facilities, it is important that we understand the gas 
conditions and stellar populations which support prominent UV line emission.  
With current facilities, this is most easily done at lower redshifts where rest-frame optical emission lines which constrain the gas-phase metallicity and ionization parameter are observable with ground-based facilities. 
Over the last few years, a wide range of observations have been conducted 
with the goal of better understanding the physics regulating rest-UV 
emission line spectra. These studies have demonstrated that prominent  CIII] appears to be a fairly ubiquitous feature in the rest-UV spectra of dwarf star forming galaxies \citep{Erb2010,Christensen2012,Stark2014,Rigby2015, Vanzella2016,Vanzella2017, Senchyna2017, Berg2018, Mainali2020, Du2020, Schmidt2021,Tang2021a,Llerena2021,Berg2022,Mainali2022}, reflecting the large electron temperatures associated with metal 
poor gas and the hard ionizing spectrum of young, low metallicity 
massive stars. 

Here we seek to build on this progress, improving our 
understanding of the connection between UV line
emission and the physical properties of the massive stars and ionized gas.    
Over the last decade, we have utilized a wide range of 
ground-based facilities (LBT, Keck, Magellan) to obtain deep spectra and imaging of some of the brightest known (g=19-21)  
$z\simeq 1.5-3$ gravitationally lensed galaxies within Sloan Digital Sky Survey (SDSS). This sample contains galaxies with a range of physical properties, including two of the brightest known systems with the large specific star formation rates (and extreme emission lines) that are 
typical in the reionization era. Central to this observational campaign are newly-acquired optical spectra from the Echellette
 Spectrograph and Imager (ESI; \citealt{Sheinis2002}) on the Keck II telescope (see \citealt{Jones2018} for more details).
The ESI spectra provide moderate resolution (R= 6300) rest-UV spectra, 
enabling a unique exploration of the nature of the massive stars and the physical conditions of the ionized gas. We supplement the Keck observations (9 galaxies) with optical spectra from LBT (1 galaxy) and MMT (6 galaxies), near-infrared spectra from Magellan, 
and new optical and near-infrared imaging.

The large continuum brightness of  the SDSS lensed galaxies offers several distinct advantages with respect to the much fainter low mass galaxies that are typical of cluster fields imaged by {\it HST}
(e.g., \citealt{Christensen2012, Stark2014,Vanzella2017}). First, it enables improved constraints on the ionized gas physical conditions (metallicity, density, ionization parameter) through detection of the full suite of rest-optical strong emission lines as well as occasionally allowing detection of multiple temperature and density-sensitive lines, providing a 
more comprehensive view of the conditions in the gas and the most important factors regulating the rest-UV line spectra. Second, the large continuum S/N in the rest-UV spectra makes it possible to detect very weak rest-UV nebular lines. This enables the rest-UV lines to be characterized in individual 
galaxies for a wide range of gas conditions, not only in the most extreme 
and metal-poor systems. This will provide a valuable control sample for our analysis, offering insight into what factors are most important in 
regulating rest-UV emission line spectra. This insight will be critical 
in assessing the feasibility of detecting and characterizing lines in the rest-frame far and near-UV with future facilities. In this paper, we 
will focus primarily on CIII] emission, comparing measured line strengths to 
other empirical and model-based quantities (i.e., gas-phase metallicity, optical line ratios with the goal of understanding the factors regulating the CIII] strength.

The paper is organized as follows.  We present the new imaging and spectra and discuss population synthesis modeling of broadband SEDs in \S2.   In \S3, we provide our results on individual galaxies. We discuss implications of our spectra of a reionization-era analog in \S4 and summarize our findings 
in \S5.

Throughout the paper, we adopt a $\Lambda$-dominated, flat universe
with $\Omega_{\Lambda}=0.7$, $\Omega_{M}=0.3$ and
$\rm{H_{0}}=70\,{\rm km\,s}^{-1}\,{\rm Mpc}^{-1}$. All
magnitudes in this paper are quoted in the AB system and equivalent widths (EW) are given in rest-frame, unless stated otherwise. 
 
\begin{table*}
\begin{tabular}{lccccccccccc}
\hline   Object & RA & DEC &  $z_{spec}$& $\rm m_{AB}$&  Dates & Rest-UV Coverage & $\rm{t_{exp}}$  &  PA & $\rm M_{UV}$ & $\mu$ & Instrument \\
 &  &  &  &  & & (\AA) & (ksec)  &  (deg) &  & &  \\
\hline \hline
CSWA-165 & 01:05:19.65 & +01:44:56.4 & 2.127 &21.1& 12 Dec 2012 & 1020-2560& 3.6 & 5.0 & -22.0 & 5.4$^{a}$ &MMT/BCS\\
CSWA-116  & 01:43:50.13 & +16:07:39.0 &1.499 &20.8& 29 Sep 2011& 1280-3200 & 2.7 & 105 & -20.9 & 10.7$^{a}$&MMT/BCS \\
CSWA-103  &01:45:04.18 & -04:55:42.7 & 1.959 & 22.1& 8-10 Nov 2012 &1350-3425    & 25 &115  & -20.9 & 4.7$^{a}$ & Keck/ESI\\
CSWA-164  &02:32:49.97 & -03:23:29.3 & 2.512  & 19.9 &8-10 Nov 2012 & 1140-2880   & 28 &158 & -22.0 & 20.8$^{a}$ & Keck/ESI\\
CSWA-11 & 08:00:12.37 & +08:12:07.0 & 1.409 & 21.1& 24 March 2012 &1330-3280 & 1.8 & 60 & -22.3 & 1.9$^{c}$ &MMT/BCS\\
 CSWA-139 & 08:07:31.51 & +44:10:48.5 & 2.536 & 22.1& 23 Mar 2012 & 905-2260 & 3.6 & 80 & -20.8 & 3.8$^{a}$ &MMT/BCS\\
 CSWA-141 &    08:46:47.53 & +04:46:09.3 & 1.425 & 20.5 & 8 Nov 2012  & 1655-4180  & 5.2  & 190 & -22.1 & 5.5$^{a}$ & Keck/ESI\\
  \ldots &    \ldots & \ldots & \ldots &  \ldots &11 Nov 2015  & 1440-2140  & 3.6  & 190 & \ldots & \ldots & LBT/MODS\\
  CSWA-19  & 09:00:02.64 & +22:34:04.9 & 2.032 & 20.9& 9-10 Nov 2012  & 1320-3330  & 20 & 86 & -21.9 & 6.5$^{a}$ & Keck/ESI\\
  CSWA-40 & 09:52:40.22 & +34:34:46.1 & 2.189 & 21.4& 5-6 Mar 2013 & 1255-3175  & 16.2 & 70 & -22.3 & 3.2$^{a}$ & Keck/ESI\\
CSWA-2  & 10:38:43.58 & +48:49:17.7 & 2.196 & 20.9& 5 Mar 2013 & 1250-3125  & 7.2 & 17 & -22.3 & 8.4$^{a}$ & Keck/ESI\\
CSWA-16  & 11:11:03.68 & +53:08:54.9 &1.945 & 22.1 &24 Mar 2012 &1085-2715 &2.7& 267 & -20.8 & 3.9$^{d}$ &MMT/BCS\\
 CSWA-38  & 12:26:51.69 & +21:52:25.5 & 2.925 & 21.3 &6 Mar 2013 & 1020-2580  & 10.8 &130 & -20.2 & 40$^{b}$ & Keck/ESI\\
CSWA-13 & 12:37:36.20 & +55:33:42.9 &  1.864 & 20.3 & 24 Mar 2012 &1120-2790 & 1.8 &  320& -23.0 & 1.9$^{d}$ &MMT/BCS\\ 
CSWA-39 & 15:27:45.02 & +06:52:33.9 & 2.762 & 20.9 & 5-6 Mar 2013 & 1065-2695   & 18 & 105 & -21.5 & 15$^{b}$ & Keck/ESI\\
 CSWA-128  & 19:58:35.65 & +59:50:53.6&  2.225 & 20.6 & 8-10 Nov 2012 & 1240-3140   & 16.3 & 60 & -21.9 & 10$^{c}$ & Keck/ESI\\
 CSWA-163 & 21:58:43.68 & +02:57:30.2 & 2.081& 22.4 &30 Sep 2011 & 1035-2600 & 1.8 & 20 & -20.4 & 6.5$^{a}$ &MMT/BCS\\ 
\hline \hline
\end{tabular}
$^{a}$\citet{Stark2013b}, $^{b}$\citet{Koester2010}, $^{c}$\citet{Nicha2016}, $^{d}$This work. \\
\caption{Summary of  rest-frame UV observations of our bright gravitationally lensed CASSOWARY galaxy sample.   
New ultra-deep moderate resolution optical spectra have been obtained for nine of these sources using ESI on Keck.   
We also include an additional seven sources from \citet{Stark2013b} with high quality MMT blue channel spectra.  
From left to right, we present the object ID in the CASSOWARY catalog, the RA and DEC, the spectroscopic redshift, apparent magnitude, the date the optical spectra 
 were obtained, the rest-UV wavelength coverage provided by the optical spectra, the exposure time and position 
angle of the optical spectra, the absolute magnitude of the arc in the rest-UV, the magnification factor provided 
by gravitational lensing, and the optical spectrograph utilized.    Further details are presented in \S2.}

\label{table:sample}
\end{table*}

\section{Observations and Analysis}

\subsection{Sample Selection}

The galaxies studied in this paper were originally 
identified using a search algorithm that identifies blue arcs surrounding red early type galaxies 
in Sloan Digital Sky Survey (SDSS) imaging as part of the Cambridge And Sloan 
Survey Of Wide ARcs on the skY (CASSOWARY; e.g., \citealt{Belokurov2007,Belokurov2009}).  
The most recent catalog of CASSOWARY lensed galaxies was presented in \citet{Stark2013b} 
following a large spectroscopic campaign aimed at obtaining redshifts of the source and lens galaxies. 
The catalog of galaxies released in \citet{Stark2013b} contains more than 
50 gravitationally-lensed galaxies in SDSS.  Optical magnitudes of this sample are in the range $19.6<r<22.3$.
 
\begin{figure*}
\centering
\hbox{\hspace{-0.8 cm}\subfloat{\includegraphics[scale=0.85]{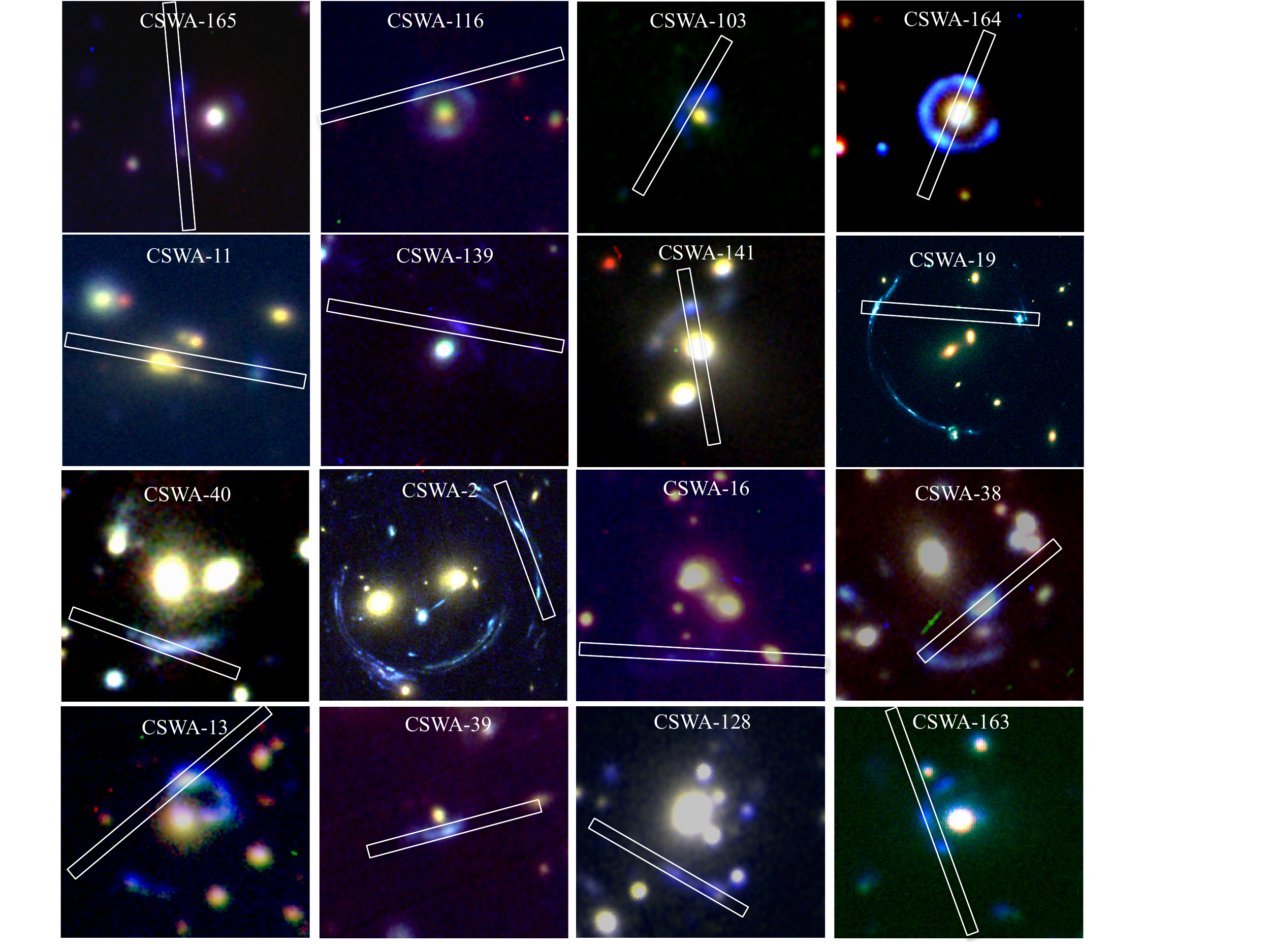}}}
\caption{Color images of sixteen gravitationally lensed galaxies. The name of each sources is presented at the top of each image, along with the slit position used for the observations. For seven galaxies (CSWA-141, CSWA-13, CSWA-139, CSWA-116, CSWA-16, CSWA-165, CSWA-164) the color images are created from g,r and i band images from LBT/MODS. The color images for four galaxies (CSWA-11,CSWA-163,CSWA-128,CSWA-103) are made with g,r and i bands images from LBT/LBC. Three color images (CSWA-38, CSWA-39, CSWA-40) are made using Keck/ESI images in V,R and I bands. For other two galaxies (CSWA-2, CSWA-19), color images are created using archival HST images.  North is up and east is to the left. Each postage stamp is $25"\times25"$ in size.  The orientation and centroid of the ESI or MMT slit is overlaid in each image.}
\label{fig:ciiimosaic}
\end{figure*}

Here we present the results from a large observational investment targeting sixteen of these 
bright lensed sources with Keck, Magellan, LBT, and MMT (see Table~\ref{table:sample}).  Object identifiers 
from the CASSOWARY survey are abbreviated as ``CSWA" in the table and subsequent discussion.  
A mosaic of the  galaxies considered in this paper is shown in Figure~\ref{fig:ciiimosaic}.
 In the following subsections, we describe 
the imaging and spectral datasets that we have obtained for this paper.
The galaxies considered here were selected from the larger sample of \citet{Stark2013b} 
based on the brightness of the continuum.  We also preferentially targeted sources at 
redshifts which place rest-optical lines 
in regions of significant atmospheric transmission in the near-IR.  

\subsection{Spectroscopy}
\subsubsection{Optical Spectroscopy}

We have acquired deep Keck/ESI optical spectra of 
nine galaxies from the \citet{Stark2013b} sample (see Table~\ref{table:sample} for details), enabling robust 
constraints to be placed on the strength of nebular UV metal line emission.
 The positioning of the ESI slit on the lensed galaxies is shown in Figure~\ref{fig:ciiimosaic}.   The spectra were obtained in two observing runs between November 2012 and March 2013.   The ESI spectra cover observed wavelengths between 4000 to 10100 \AA, providing rest-frame spectral coverage that typically ranges between 1300 and 2000~\AA\  (see Table~\ref{table:sample}).   A slit width of 0\farcs75 was used, providing a resolving power of R=6300 (FWHM=48 km s$^{-1}$).     The data were reduced using the ESIRedux code written by J. X. Prochaska.  Sky subtraction is performed following the bias subtraction and flat fielding.   The 1D spectra are then extracted using a boxcar aperture  matched to the 
spatial extent of the arc.  When multiple lensed images appear on the slit, the traces are extracted separately 
and then combined to maximize the S/N.   The continuum is generally well detected, with S/N $\simeq 10$ 
per resolution element at 6200~\AA.   Example spectra are shown in Figure~\ref{fig:esispectra}.  A full description of the ESI 
observations and spectral reduction is presented in \citet{Jones2018}.

One source in the ESI sample, CSWA-141, was also observed with the Multi Object Double Spectrograph 
(MODS, \citealt{Pogge2010}) on the LBT.   MODS provides bluer wavelength coverage than ESI, allowing constraints 
to be placed on emission from CIV, OIII], and He II.  We used a long slit of width 0\farcs8 with the 400 lines/mm grating, 
providing spectral resolution of $\sim$ 3 $\rm \AA$. 

We also include seven other galaxies from \citet{Stark2013b} for which the MMT Blue Channel Spectrograph (BCS) discovery spectra have sufficient continuum S/N to characterize the rest-UV metal emission lines.  The MMT BCS data 
were obtained with the 300 lines/mm grating, providing total spectral coverage of $\sim5300$ $\rm \AA$. For the  slit width of 1\farcs0  used in the observations, the typical spectral resolution is $\sim$ 6.5 $\rm \AA$. More details of the MMT observations are provided in \citet{Stark2013}. 

In total, the Keck, LBT, and MMT data provides rest-UV spectra for 16 galaxies. 
Our aim is to characterize CIII] emission feature, typically the strongest rest-UV 
emission line, along with other rest-UV emission features in the whole sample. Using redshift presented in \citet{Jones2018} for Keck/ESI data and \citet{Stark2013b}
 for MMT data, we first searched for CIII] emission in individual galaxy spectrum. The line is spectrally resolved in the Keck spectra, whereas it remained unresolved 
 in the MMT spectra. For the resolved CIII] doublet, we measured emission line fluxes and associated errors (for error spectrum) in individual
 components by directly integrating the flux levels within $\pm$1~\AA\ of individual line centers (rest-frame).  When the line is unresolved, we computed total CIII] emission line fluxes and associated line 
 flux errors directly from the flux spectra and error spectra, respectively. If the line remained undetected at 3$\sigma$ level, we calculated upper limits by 
 integrating error spectrum from 1905~\AA\ to 1912~\AA. The emission line equivalent width is then computed by dividing the line fluxes (or upper limits) by
 median continuum level measured on either side of the CIII] line.  We then searched for any other rest-UV emission features in the spectra and characterized them when detected.
 Our sample includes 10 galaxies with CIII] detection where the equivalent widths range from 0.4~\AA\ to 4.6~\AA\ (see Table~\ref{table:uvlines}).  We will 
 come back to interpret the CIII] equivalent widths in \S3.2, comparing the 
 line strengths to optical line ratios and gas physical properties.

\begin{table}
\begin{tabular}{lcccc}
\hline Object & &EW (\AA)   & &  \\
 & Ly$\alpha$   & [CIII]$\lambda1907$ & CIII]$\lambda1909$ &  CIII]$\lambda1908$$^{a}$   \\
 \hline 

CSWA-141 & \ldots  & 2.3(1.3) & 2.3(1.4)&4.6(1.9)\\
CSWA-13 & 5.4(2.1)& \ldots& \ldots & 4.4(0.9) \\
CSWA-139 & \ldots& \ldots& \ldots  & 3.4(2.6)  \\
CSWA-2  & \ldots &  2.0(1.6) &1.1(0.9) & 3.1(1.6)\\
CSWA-39  & 11.8(6.2) & 0.6(0.1) & 0.5(0.1) &1.1(0.2)\\
CSWA-19  & \ldots &0.4(0.1) & 0.3(0.1) & 0.7(0.1)\\
CSWA-38  & 2.6(1.1) & \ldots & 0.4(0.1)  &\\
CSWA-128  & \ldots & 0.3(0.1) & 0.3(0.1) &0.6(0.1)\\
CSWA-103  & \ldots & 0.3(0.1) & 0.2(0.1) & 0.5(0.1)\\
CSWA-164  & 2.0(0.5) & 0.2(0.1) & 0.2(0.1)& 0.4(0.1) \\
CSWA-163 & \ldots & \ldots& \ldots& $<3.1$ \\
CSWA-16 & \ldots& \ldots& \ldots & $<2.3$ \\
CSWA-165 & \ldots& \ldots& \ldots  & $<1.9$  \\
CSWA-40  & \ldots  & $<1.2$ & $<1.2$ &$<1.7$\\
CSWA-11 & \ldots & \ldots& \ldots & $<1.4$ \\
CSWA-116 & \ldots & \ldots& \ldots &$<1.1$  \\
 \hline
\end{tabular}
$^{a}$ Total CIII] doublet.\\
\caption{Equivalent width measurements of rest-UV emission lines. The numbers within parentheses represent uncertainty. The upper limits are 3$\sigma$. }
\label{table:uvlines}
\end{table}

\subsubsection{Near-Infrared Spectroscopy}

\begin{table}
\begin{tabular}{llccc}
\hline Object & Dates  &$\rm{t_{exp}}$  &  PA  & Observatory/ \\ 
 &  &(ksec) & (deg) & Instrument \\ \hline 
 CSWA-165   & 2014 June 22 &  7.2 & 120  & Magellen/FIRE \\  
  CSWA-164    & 2015 Nov 03 &  4.8 & 58   & Magellen/FIRE  \\ 
 CSWA-11   & 2015 Nov 04 &  2.7 & -10  & Magellen/FIRE  \\
  CSWA-141     & 2012 Feb 15 &  1.2 & 280 & Magellen/FIRE   \\ 
  CSWA-128 & 2012 Nov  7 & 3.6 & 190  & LBT/LUCI  \\
   CSWA-163   & 2014 June 22 &  9.0 & 90  & Magellen/FIRE  \\   
\hline
\end{tabular}
\caption{Details of near-infrared spectroscopic observations obtained for this paper.  From 
left to right, the columns denote the object ID, observation dates, exposure time, position angle of slit, and the observatory and instrument used to acquire near-IR spectroscopy. Further details are provided in \S2.2.
 }
\label{table:nir}
\end{table}

Near-infrared (NIR) spectra supplement the optical spectra described above, 
providing constraints on the metallicity, ionization parameter, and electron 
density of the ionized gas.  Near-infrared spectroscopic analysis is limited 
to the subset of sources from Table~\ref{table:sample} which are located 
at redshifts  placing strong  rest-optical emission lines in spectral windows 
in which atmospheric transmission is near-unity.    One of the sources in our 
sample (CSWA-2) was targeted with Keck near-infrared spectroscopy in \citet{Jones2013}.
In observing runs between 
November 2012 and November 2015, we have obtained near-infrared spectroscopic 
observations of five additional sources 
(CSWA-141, CSWA-164, CSWA-165, CSWA-163, CSWA-11) 
using the Folded-port InfraRed Echellette (FIRE; \citealt{Simcoe2008}) on the 
Magellan Baade Telescope. We used FIRE in its echelle mode, providing 
continuous spectral coverage spanning 0.82-2.51 $\mu$m.   We adopted 
a slit width of 0\farcs75, delivering a spectral resolution of 2.6 $\rm \AA$ in 
the J band, 3.4  $\rm \AA$ in the H band and 4.7  $\rm \AA$ in the K band. 
Spectroscopic data reduction was performed using standard routines in 
the FIREHOSE data reduction pipeline\footnote{wikis.mit.edu/confluence/display/FIRE/FIRE+Data+Reduction}.

One additional galaxy (CSWA-128) was observed on 2012 Nov 7 with the LUCI near-IR spectrograph on the Large Binocular 
Telescope (LBT).  We used the N1.8 camera and 200$\_$H+K grating in longslit mode.  
We first observed the lensed source with the HKSpec filter centered at 1.93 microns, providing spectral coverage from 
1.50 to 2.30 $\mu$m.  We also observed CSWA-128 with the zJSpec filter centered at 1.1 microns, providing spectral coverage 
between 0.95 and 1.40$\mu$m.   A slit width of 1\farcs0 was used, resulting in a spectral resolution of 16 $\rm \AA$. Spectroscopic data reduction was performed using standard IDL long slit reduction packages (see \citealt{Bian2010} for details).   
We summarize details of near-infrared spectroscopy of CASSOWARY galaxies in Table~\ref{table:nir}. Example spectra are shown in Figure~\ref{fig:cswa_nir}.

Emission line fluxes in the NIR spectra are measured using the IDL routine MPFITPEAK which computes line fluxes after fitting a gaussian model. In cases where the 
emission lines are partially affected by a skyline, we mask the contaminated region 
before fitting the emission line.  Additionally, if one of the components of [OIII]$\lambda\lambda$4959,5007 is strongly 
affected by an emission line, we assume the theoretical line ratio of
 [OIII]$\lambda$5007/[OIII]$\lambda$4959=2.98 (Storey \& Zeippen 2000) in calculating the total [OIII] line flux.
 
We calculate the impact of dust on the nebular lines using the Balmer decrement flux ratio of H$\alpha$/H$\beta$. The observed line ratio is compared to the  line ratio  expected in absence of dust (H$\alpha$/H$\beta$=2.86; \citealt{Osterbrock2006}) for case B recombination assuming T$_{e}$=10,000 K. In the two cases where the 
Balmer line ratios are not available, 
we use the stellar reddening inferred from the broadband data to estimate the nebular attenuation (see \S2.5). We note that using stellar reddening for nebular attenuation correction doesn't impact our results. We assume that the nebular gas attenuation is similar to the stellar continuum attenuation. 
 We presented the observed and dust-corrected emission line measurements in Table~\ref{table:opticallines}.

For one object (CSWA-141) where auroral [OIII]$\lambda$4363 line is detected, we calculated electron temperature using PyNeb (version 1.1.8; \citealt{Luridiana2015}) 
 using emission line flux ratio of OIII]$\lambda$4363/[OIII]$\lambda$5007.  For our calculations, we adopted the default PyNeb atomic data sets. We assumed electron density of  
250 cm$^{-3}$, which is typical to $z\sim2$ galaxies \citep{Sanders2016}. However, we note that our assumed electron density value has negligible effect on the derived electron temperature in the low density regime ($<$10$^{3}$ cm$^{-3}$). 
Since we don't have T$_{e}$([OII]) sensitive emission line measurements, we followed the relation given by \citet{Izotov2006} for low 
metallicity to estimate the electron temperature in the $\rm{O^{+2}}$ region.
 We use our electron temperature measurements to infer the direct oxygen abundance. We only use $\rm{O^{+}/H^{+}}$ 
and $\rm{O^{+2}/H^{+}}$ to compute oxygen abundance since the ionization states higher than $\rm{O^{+2}}$ contributes significantly low
at less than 1 per cent \citep{Izotov2006}. $\rm{O^{+}/H^{+}}$ and $\rm{O^{+2}/H^{+}}$  are calculated from PyNeb using 
our T$_{e}$([OIII], T$_{e}$([OII], $n_{e}$ and emission line fluxes of [OII], H$\beta$ and [OIII]. 

We use PyNeb to calculate the electron density using the flux ratio of the [OII], [SII], and [CIII], CIII] doublets.
The typical error in measurements is then calculated using the errors in emission line fluxes.
When an electron temperature measurement is not available, we assumed electron temperature of 10,000K following \citet{Sanders2016}
 to adopt temperature dependent effective collision strengths. 
Adopting electron temperature of 7000K (15000K) instead would overestimate (underestimate)
electron density by 15-20$\%$ which is effectively lower than density error from our line flux measurements.

\label{sec:nir_spec}

 \begin{table}
\begin{tabular}{l|ccccc}
\hline Line  &$\lambda_{\rm{rest}}$(\AA)& $\lambda_{\rm{obs}}$ (\AA) & F($\lambda$)/F(5007)   &   I($\lambda$)/I(5007) \\
\hline \hline
  \multicolumn{5}{|c|}{CSWA-141} \\ \hline
 $\rm{[OII]}$  & 3727.13 & 9039.2 & $0.060\pm0.002$ & $0.088\pm0.003$\\
 $\rm{[OII]}$ & 3729.92 &9045.9 &$0.075\pm0.004$ &$0.110\pm0.006$\\
   $\rm[NeIII]$ & 3869.66  & 9384.9 & $0.054\pm0.017$&$0.075\pm0.023$ \\
  H$\delta$ & 4102.90& 9950.5  & $0.030\pm 0.004$ &$0.039\pm0.005$\\
  H$\gamma$ & 4341.58& 10529.4 & $0.063\pm 0.003$ &$0.076\pm0.004$\\
 $\rm{ [OIII]}$& 4365.31  & 10586.9 & $0.019\pm0.002$ &$0.023\pm0.002$\\
 H$\beta$  &4862.55& 11792.9 & $0.13\pm0.003$&$0.135\pm0.003$\\
 $\rm{ [OIII]}$ &4960.25 &  12029.8 & $0.338 \pm 0.004$ &$0.342\pm0.004$\\
  $\rm{[OIII]}$& 5008.27& 12146.3 &  1.000&1.000\\
 $\rm{[SIII]}$ & 6310.48& 15304.4 & $0.007\pm 0.001$&$0.005\pm0.001$\\
  $\rm{[NII]}$  & 6549.84 &\ldots  &$<0.004$& $<0.002$\\
  H$\alpha$  &6564.61& 15920.7 & $0.529 \pm 0.003$ &$0.386\pm0.002$\\
  $\rm{[S II]}$  &6717.96&  16292.7 & $0.017\pm 0.001$ &$0.013\pm0.001$ \\
 $\rm{ [S II]}$  &6732.34&  16327.5 &  $0.016\pm0.001$ &$0.012\pm0.001$\\
\hline 
  \multicolumn{5}{|c|}{CSWA-165} \\ \hline
 $\rm{[OII]}$  & 3727.13 & 11657.1 & $0.56\pm0.06$& $1.01\pm0.11$\\
 $\rm{[OII]}$ &  3729.92 & 11665.9 &$0.63\pm0.08$& $1.13\pm0.14$\\
   $\rm[NeIII]$  & 3869.66 & \ldots & $<0.23$& $<0.38$\\
  H$\beta$ &4862.55& 15208.9 & $0.46\pm0.08$&$0.49\pm0.08$ \\
 $\rm{[OIII]}$ & 5008.27& 15665.2 &  1.00 &1.00\\
 H$\alpha$  & 6564.61& 20532.4 & $2.25 \pm 0.15$ &$1.40\pm0.09$\\
  $\rm{[NII]}$  & 6585.28& 20597.5  &$0.55\pm0.09$ & $0.34\pm0.06$\\
\hline 
  \multicolumn{5}{|c|}{CSWA-163} \\ \hline
 $\rm{[OII]}$ & 3727.13& 11482.5 & $0.28\pm0.01$ &$0.53\pm0.02$\\
 $\rm{[OII]}$&  3729.92& 11491.2 &$0.35\pm0.03$ &$0.66\pm0.06$\\
  $\rm[NeIII]$  & 3869.66& \ldots & $<0.04$& $<0.07$\\
 H$\gamma$  & 4341.58& 13375.6 & $0.06 \pm 0.03$&$0.08\pm0.04$ \\
  H$\beta$  &4862.55 & 14981.5 & $0.22\pm0.03$ & $0.23\pm0.03$\\
 $\rm{ [OIII]}$  &4960.25&  15282.5 & $0.35\pm0.02$ &$0.35\pm0.02$\\
 $\rm{[OIII]}$ & 5008.27& 15429.9 &  1.00 &1.00\\
  H$\alpha$  & 6564.61& 20225.5 & $1.12 \pm 0.04$ &$0.67\pm0.02$\\
  $\rm{[NII]}$ & 6585.28 & 20289.2  &$0.18\pm 0.06$ &$0.11\pm0.04$\\
\hline 
  \multicolumn{5}{|c|}{CSWA-128} \\ \hline
  $\rm{ [O II]}$   & 3729.01& 12019.6 &  $0.28\pm0.08$ &$0.40\pm0.11$\\
 $\rm{ [Ne III]}$    &3869.66& 12476. 4 &  $0.10\pm0.03$ &$0.14\pm0.04$\\
 H$\beta$  &4862.55& 15681.7 & $0.24\pm0.06$ &$0.25\pm0.06$\\
 $\rm{ [OIII]}$  &4960.25 &  15997.7 & $0.32\pm0.04$ & $0.33\pm0.04$\\
 $\rm{[OIII]}$ & 5008.27& 16151.8 &  1.00 & 1.00\\
 $\rm{[NII]}$ & 6549.84&21121.3  &$0.05\pm0.01$ & $0.04\pm0.01$\\
 H$\alpha$  & 6564.61& 21170.3 & $0.96 \pm 0.05$ & $0.71\pm0.04$\\
 $\rm{[NII]}$  & 6585.28 &  21237.2 &$0.13\pm0.02$ &$0.09\pm0.01$\\
 $\rm{[S II]}$  & 6717.96&  21664.9 & $0.10\pm0.02$ & $0.07\pm0.01$\\
 $\rm{ [S II]}$   &6732.34 & 21711.1  &  $0.08\pm0.04$& $0.06\pm0.02$\\
\hline 
  \multicolumn{5}{|c|}{CSWA-164} \\ \hline
 $\rm{[OII]}$  & 3727.13& 13090.9 & $0.52\pm0.11$&$0.80\pm0.17$ \\
 $\rm{[OII]}$&  3729.92  & 13100.4 &$0.64\pm0.04$& $0.99\pm0.0.07$\\
  $\rm{[OIII]}$& 5008.27 & 17590.4 &  1.00&1.00\\
  H$\alpha$  & 6564.61& 23064.8 & $1.91\pm0.09$&$1.34\pm0.06$\\
 \hline 
   \multicolumn{5}{|c|}{CSWA-11} \\ \hline
  $\rm{[OII]}$ & 3727.13 & 8979.9 & $0.67\pm0.13$ &$0.85\pm0.16$ \\
  $\rm{[OII]}$  & 3729.92 & 8986.6 & $0.66\pm0.09$ &$0.83\pm0.11$\\
  $\rm{ [Ne III]}$  &3869.66  & \dots &  $<0.43$ &$<0.53$\\
  $\rm{[OIII]}$ & 5008.27 & 12068.7 &  1.00&1.00\\
  H$\alpha$ & 6564.61&  15816.5 & $1.08\pm0.06$&$0.89\pm0.05$\\
\hline \hline

 \end{tabular}
\caption{Rest-optical emission line measurements of six bright lensed Cassowary galaxies.  Measurements are 
from  Magellan/FIRE (CSWA-141, CSWA-165, CSWA-163, CSWA-164, CSWA-11) and LBT/LUCI (CSWA-128). Line flux measurements are quoted relative to  [OIII]$\lambda$5007. We adopt [OIII]$\lambda$5007 as a reference line due to its high S/N.  Three-sigma upper limits are reported for emission lines that are not detected.}
\label{table:opticallines}
\end{table}

 \subsection{Imaging}
 
Each of the CASSOWARY galaxies was discovered in SDSS imaging.  In many cases, 
the photometric constraints from SDSS are unreliable owing to blending with neighbors and 
low S/N detection of diffuse emission associated with the arcs.   We have obtained deeper 
optical multi-band imaging for each of the  galaxies discussed in this paper using 
cameras on the LBT and Keck. In order to better characterize the stellar populations, we have also obtained near-IR imaging sampling across the Balmer break for a subset of our targets using the LBT, MMT and Keck.   In the following subsections, we describe 
the optical and near-infrared imaging observations and analysis.  Details of the new imaging 
observations are summarized in Table~\ref{table:imaging}.

\begin{table}
\begin{tabular}{lcccc}
\hline   Object & Observatory / & Filters &  Dates  &$\rm{t_{exp}}$\\  
&  Instrument & Observed & & (sec) \\ \hline \hline 
CSWA-165 & LBT/MODS  & g  & 2014 Jan 27 &  720   \\ 
   & LBT/MODS  &  r, z & 2014 Jan 27 &  360   \\ 

  & MMT/MMIRS & J & 2015 Sep 30 &   900 \\ 
  & MMT/MMIRS & K & 2015 Sep 30 &   1200 \\
  CSWA-116 & LBT/MODS  & g  & 2014 Jan 27&  720  \\ 
                   & LBT/MODS  & r, z & 2014 Jan 27&  360   \\
       & MMT/MMIRS & J,K & 2015 Oct 01 &   600 \\ 
  CSWA-103    & LBT/LBC & g  & 2015 Nov 15 & 600   \\
          & LBT/LBC & r, i  & 2015 Nov 15 & 300   \\
   & MMT/MMIRS & J,K & 2015 Oct 01 &   600 \\ 
    CSWA-164 & LBT/MODS  & g  & 2014 Jan 27&  720  \\ 
                   & LBT/MODS  & r, z & 2014 Jan 27&  360   \\
  CSWA-11    & LBT/LBC & g  & 2015 Nov 15 & 600   \\
       & LBT/LBC & r, z  & 2015 Nov 15 & 300   \\
       & Keck/ MOSFIRE  & K$_{s}$&  2015 Nov 30  &  300  \\ 
       CSWA-139 & LBT/MODS  & g  & 2014 Jan 27&  720  \\ 
                   & LBT/MODS  & r, z & 2014 Jan 27&  360   \\
                     & MMT/MMIRS & H & 2017 Jan 11 &   900 \\               
 CSWA-141 & LBT/ MODS  & g & 2014 Jan 27 &  720 \\
& LBT/ MODS  &   r, z & 2014 Jan 27 &  360  \\  
     & Keck/ MOSFIRE  & K$_{s}$&  2014 Apr 11  &  300  \\ 
     & Keck/ MOSFIRE  & Y &  2015 Nov 30  &  300  \\
     & Keck/ MOSFIRE  & J2 &  2015 Nov 30  &  300  \\ 
      CSWA-40 & Keck/ESI & V, R, I& 2013 Mar 06 & 200 \\
      
      CSWA-16 & LBT/MODS  & g, r & 2014 Jan 27&  360  \\
       CSWA-38 & Keck/ESI & V, R, I& 2013 Mar 06 & 200 \\ 
CSWA-13 & LBT/MODS  & g  & 2014 Jan 27 &  720  \\ 
               & LBT/MODS & r, z & 2014 Jan 27 &  360  \\ 
      & LBT/LUCI & K$_{s}$ &  2014 Apr 13 &  1200  \\ 
      CSWA-39 & Keck/ESI & V, R, I& 2013 Mar 06 & 200 \\
 
        CSWA-128    & LBT/LBC & g  & 2015 Nov 15 & 600   \\
          & LBT/LBC & r, i  & 2015 Nov 15 & 300   \\
        & Keck/ MOSFIRE  & K$_{s}$&  2014 Apr 11  &  300  \\ 
        
            CSWA-163    & LBT/LBC & g  & 2015 Nov 15 & 600   \\
          & LBT/LBC & r, i  & 2015 Nov 15 & 300   \\
& MMT/MMIRS & J & 2015 Sep 30 &   900 \\   
    & MMT/MMIRS & K & 2015 Sep 30 &   1200 \\

\hline
\end{tabular}
\caption{New optical and near-infrared imaging of CASSOWARY lensed galaxies obtained with 
Keck, LBT, and MMT.  From left to right, columns denote the object ID, observatory and instrument, filters utilized, dates of observations, and exposure time.}
\label{table:imaging}
\end{table}

 \subsubsection{Optical imaging}  
We secured images of seven galaxies from Table~\ref{table:sample} using the Multi Object Double Spectrograph 
(MODS, \citealt{Pogge2010}) on LBT in January 2014.   The dual  channel mode of MODS provides images in 
two separate filters simultaneously over a field of view of $6\times6$ arc minutes.  We used the blue and red channel to obtain 
g, r, and z-band images for  CSWA-13, CSWA-16, CSWA-116,  CSWA-139, CSWA-141, CSWA-164 and CSWA-165.  
Each arc was first observed in the g and r-band for 360 seconds and then in the g and z-band for an additional 360 seconds.   For one source (CSWA 16), we obtained imaging only in the g and r-bands.  
The sky was partly covered by clouds throughout the MODS observations and the seeing varied between 1\farcs0 and 1\farcs5. Optical images of three other galaxies (CSWA-39, CSWA-38, CSWA-40) were taken with ESI on Keck, providing a field of view of $2\times3.5$ arc minutes. These three sources were observed with V, R and I filters.  There was some cloud during the observations, and the seeing was between 0\farcs8 and 1\farcs2. For the two other sources in Table~\ref{table:sample}, CSWA-2 and CSWA-19, we make use of archival HST images  (program 11974, PI: Allam). 
We summarize the details of the imaging observations in Table~\ref{table:imaging}.   

The optical images were reduced using standard routines for flat fielding and image combination. The 5$\sigma$ typical limiting magnitude in the g, 
r and z bands of MODS images are 25.5, 24.7 and 23.2 respectively. Similarly, for the ESI images the typical 5$\sigma$ limiting magnitude
 in V, R and I bands are 22.7, 22.6 and 22.6 respectively.  For the archival HST images, the typical 5$\sigma$ limiting magnitudes in F450W, F606W, and F814W filters are 26.3, 26.1 and 25.5 respectively.  Each image was flux calibrated using stars that are isolated and well detected in 
 SDSS imaging.   The mosaic  in Figure~\ref{fig:ciiimosaic} shows the multi-color optical imaging obtained for each source.  

Before performing photometry on the lensed galaxies, we modeled the foreground lens using GALFIT \citep{Peng2002} and subtracted its contribution to the lensed source.       
To calculate the absolute magnitude for the galaxies in our sample, we first measured the integrated flux 
 in the g-band.  After applying the appropriate magnification 
 corrections (see \S2.4), we arrive at the absolute UV magnitudes shown in Table~\ref{table:sample}.  The values 
 span M$_{\rm{UV}}=-23.1$ to M$_{\rm{UV}}=-20.4$, corresponding to  0.8-9 L$^\star_{\rm{UV}}$ at $z\simeq 2-3$  \citep{Reddy2009,Parsa2016}. 
    
  \subsubsection{Near-infrared imaging}  
We obtained near-infrared images of CSWA-141, CSWA-11 and CSWA-128 in the $\rm K_{s}$-band using the Multi-Object Spectrometer for Infra-Red Exploration (MOSFIRE, \citealt{Mclean2012}) on Keck I. Conditions were photometric with seeing of 0\farcs5. We obtained 10 dithered frames having $6.1\times6.1$ arc minute field of view, each with exposure time of 30 seconds. The 5$\sigma$ AB limiting magnitude  in the MOSFIRE $\rm K_{s}$-band images is 23.1. With the goal of constraining the equivalent width of [OIII] and H-beta emission, we observed CSWA-141 with the MOSFIRE J2 medium band filter.  The filter spans 1.11-1.25 $\mu$m, covering H-beta and [OIII]. 

We secured a $\rm K_{s}$-band image of CSWA-13 using the LUCI near-IR Spectrograph on the LBT when the seeing was 1\farcs2. We utilized the N3.75 camera providing $4\times4$ arc minutes of field of view with a plate scale of  0.12 arcseconds per pixel. The near-IR imaging reduction was performed using standard IDL routines designed to flat field, sky-subtract and stack the dithered frames. Finally, we used isolated and unsaturated stars that are in the 2MASS catalog to  flux calibrate the images.   
The  5$\sigma$ AB limiting magnitude in the LUCI $\rm K_{s}$-band image is 23.4. For another four galaxies (CSWA-116, CSWA-165, CSWA-103, CSWA-163),  
near-IR imaging was obtained using the MMIRS instrument 
on the MMT.  Each of the four sources was observed in the J and K bands.  MMIRS provides imaging over a field of view of  $6.9\times6.9$ arc minutes with a plate scale of 0.20 arcsec per pixel. The seeing was between 0\farcs6-0\farcs9 throughout the observations. The typical 5$\sigma$ AB limiting magnitude in MMIRS J and K-band images are 
22.9 and 22.8, respectively. 

 \subsection{Lensing Magnification}
 
Derivation of stellar mass and intrinsic luminosity of lensed systems requires magnification correction. In \citet{Stark2013}, magnification factors were presented for eleven out of sixteen sources given in Table~\ref{table:sample} and typically range between $\mu$=5 and 10. Three out of the five remaining sources have a published lens model. We adopted magnifications for CSWA-38 and CSWA-39 from \citet{Koester2010} and CSWA-11 from \citet{Nicha2016}. 
 
For the two remaining  sources (CSWA-13 $\&$ CSWA-16), we follow a similar approach to \citet{Hezaveh2013}. In brief, we assumed a symmetric Gaussian light distribution for the lensed source. We also tested the impact of adding 
a second Gaussian component to the source.  The foreground lens is modeled as a Singular Isothermal Ellipsoid (SIE). Multiple lenses are allowed in the modeling procedure. To obtain the best fit model, we perform a Markov Chain Monte Carlo (MCMC) analysis to minimize $\chi^2$ in the image plane. In the case of CSWA-16, the observed data are better fit after introducing a second Gaussian component to the background source. 
The magnification factor is then calculated as the ratio of the lensed to unlensed flux. 
We derive a magnification of $\mu=3.9\pm0.3$ for CSWA-16 and $\mu=1.9\pm0.2$ for CSWA-13.  

\subsection{Stellar Population Synthesis Modeling} 

\label{sec:sedfitting}

Thirteen of the galaxies shown in Figure \ref{fig:ciiimosaic} have the necessary 
optical and near-IR imaging to derive 
physical properties from broadband SED fitting.   For these systems, 
we infer the stellar mass, 
specific star formation rate and dust attenuation  using
 the Bayesian galaxy SED modeling and interpreting  tool  BEAGLE  tool (version
 0.20.3; \citealt{Chevallard2016}). BEAGLE is based on the photoionization models of star-forming galaxies in \citet{Gutkin2016}, combining the latest version of  the \citet{Bruzual2003} stellar 
population synthesis models with the photoionization code CLOUDY \citep{Ferland2013}.

We fit the broadband photometric fluxes as well as CIII] equivalent widths. When relative fluxing between rest-UV and optical emission lines is possible, we include all emission lines in the fitting procedure. The models assume constant star formation where the maximum stellar age is allowed to vary freely between 5 Myr to the Universe age at the given redshift.  We use \citet{Chabrier2003} initial mass function and the \citet{Calzetti2000} extinction  curve.  The metallicity is allowed to vary in the range of -2.2$\leq$log(Z/Z)$_{\odot}$$\leq$0.25 assuming
 equivalent stellar and nebular metallicity (Z$_{\star}$=Z$_{ISM}$). The redshift of all objects is fixed to their spectroscopic redshift given in Table~\ref{table:sample}.  The  ionization parameter (U$\rm_{S}$; here defined as the ratio of ionizing-photon
 to gas densities at the edge of the Str\"{o}mgren sphere)  is  varied in the range of -4.0$\leq$$U_{\rm S}$$\leq$-1.0, and the dust-to-metal mass ratio spanned within the range of $\xi_{d}$=0.1-0.5. We adopt models with hydrogen density (n$\rm_{H}$=100 $\rm cm^{-3}$) and C/O abundance of 0.5 of solar value [(C/O)$_{\odot}\approx$0.44]. Finally, the prior on the V-band dust attenuation optical depths ($\hat{\tau}_V$) is taken as an exponential distribution after fixing the fraction of attenuation optical depth arising from the ambient ISM ($\mu$) to be 0.4.  We note that our assumption of constant star formation may underestimate the stellar mass by as high as 0.5 dex when the galaxy is dominated by a young stellar population, but this would not affect the primary conclusions of this paper.  We discuss the main results of this SED fitting in \S 3.2.

\begin{figure*}
\centering

\subfloat{ \includegraphics[angle=90,width=0.52\textwidth]{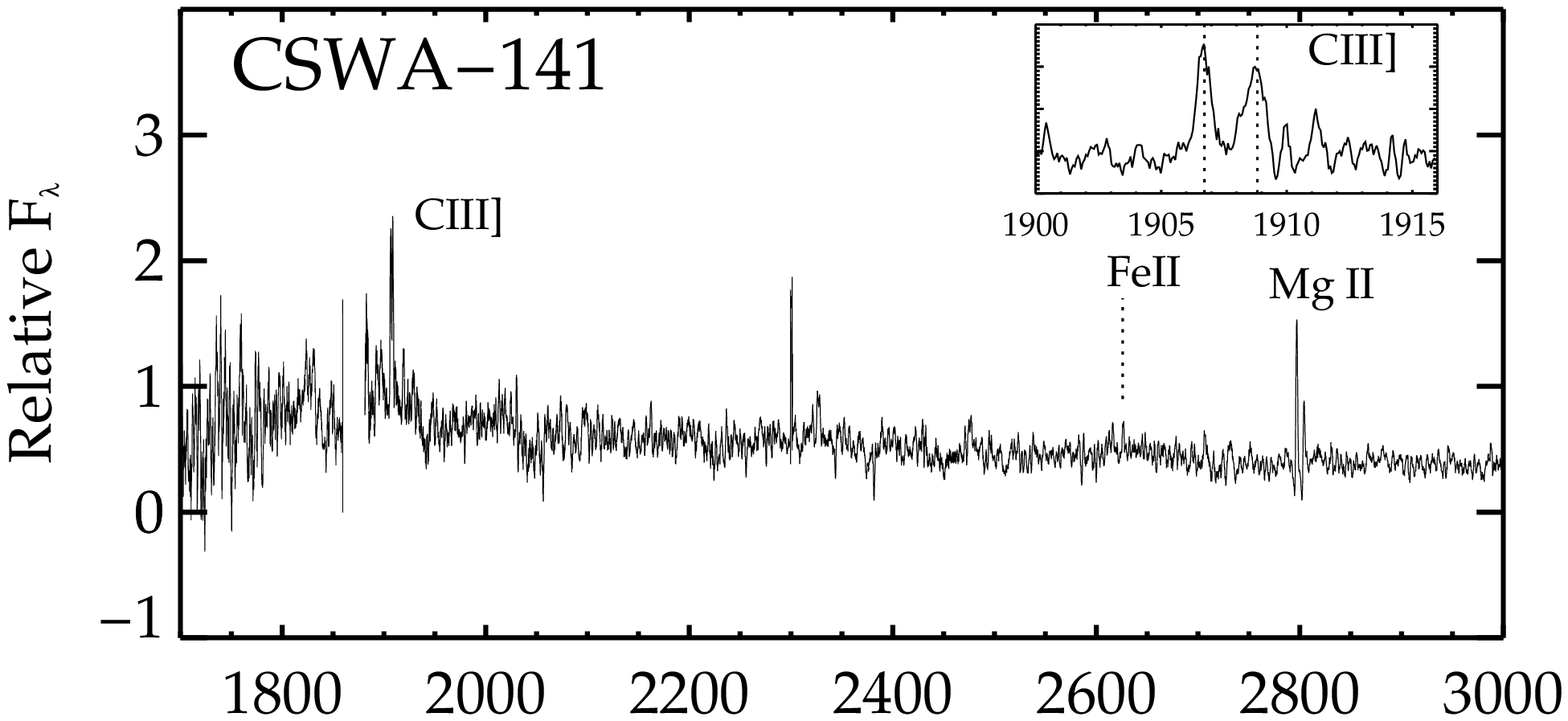}} 
\hbox{\hspace{-0.8 cm} \subfloat{ \includegraphics[angle=90,width=0.52\textwidth]{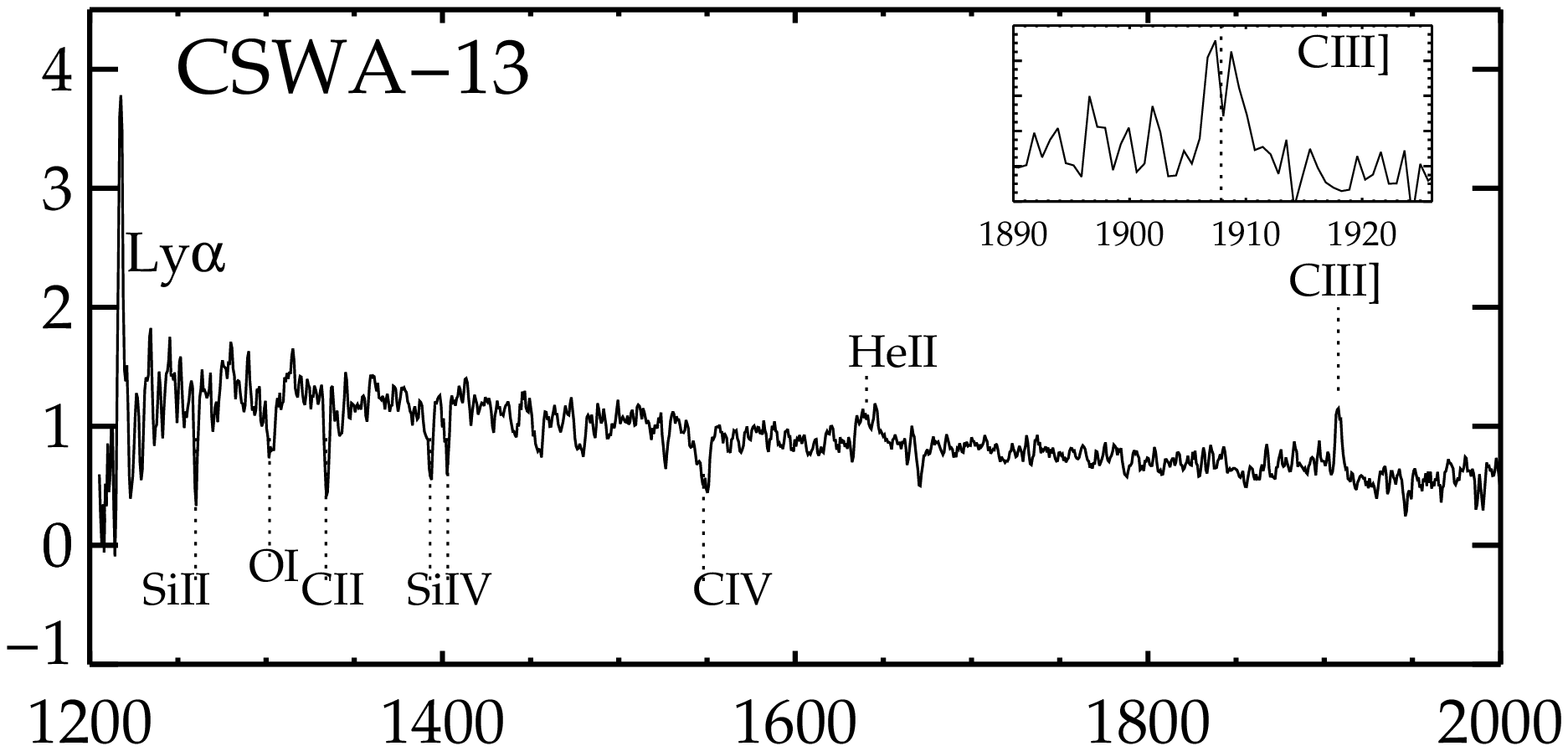}}}\\
\vspace{-7 mm}
\subfloat{ \includegraphics[angle=90,width=0.52\textwidth]{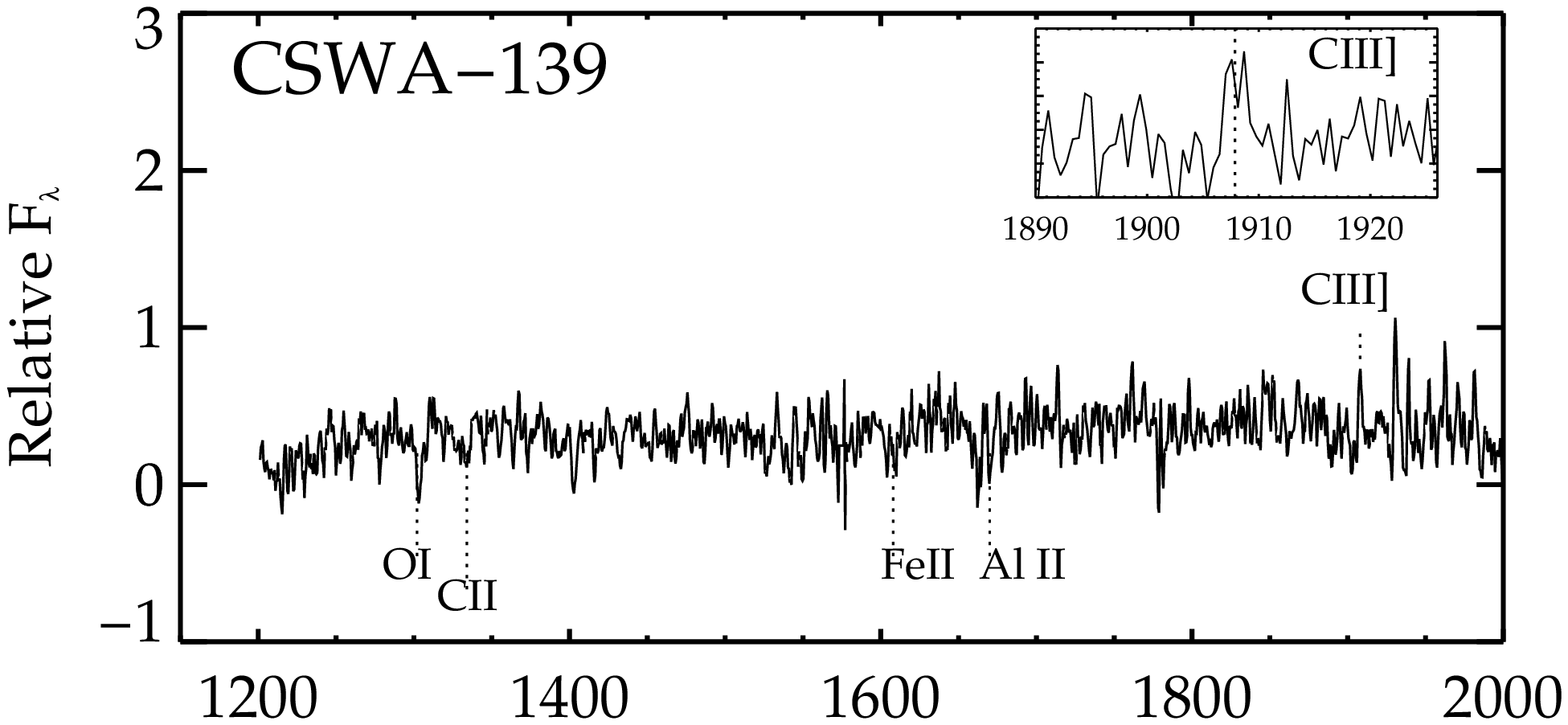}} 
\hbox{\hspace{-0.8 cm} \subfloat{ \includegraphics[angle=90,width=0.52\textwidth]{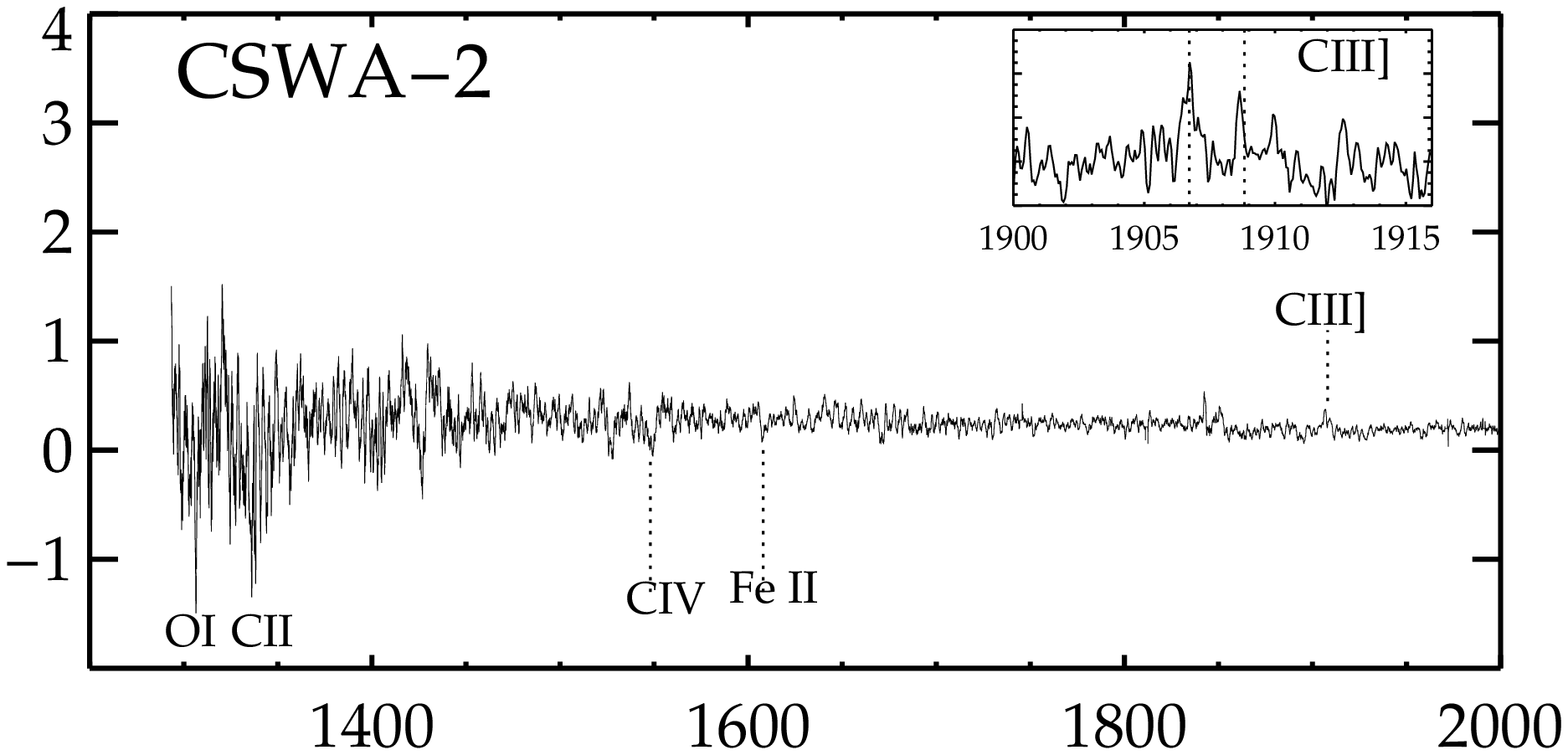}}}\\
\vspace{-7 mm}
\subfloat{ \includegraphics[angle=90,width=0.52\textwidth]{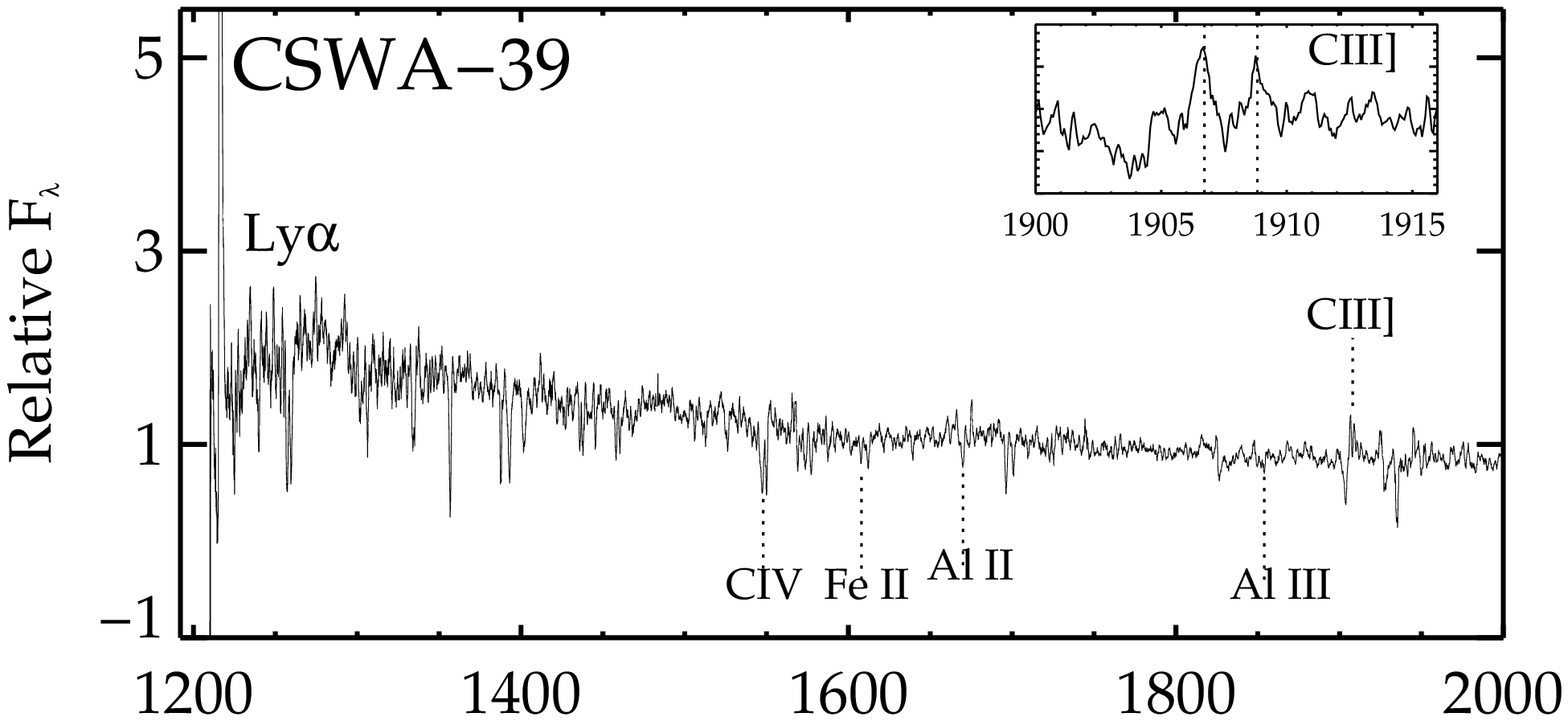}} 
\hbox{\hspace{-0.8 cm} \subfloat{ \includegraphics[angle=90,width=0.52\textwidth]{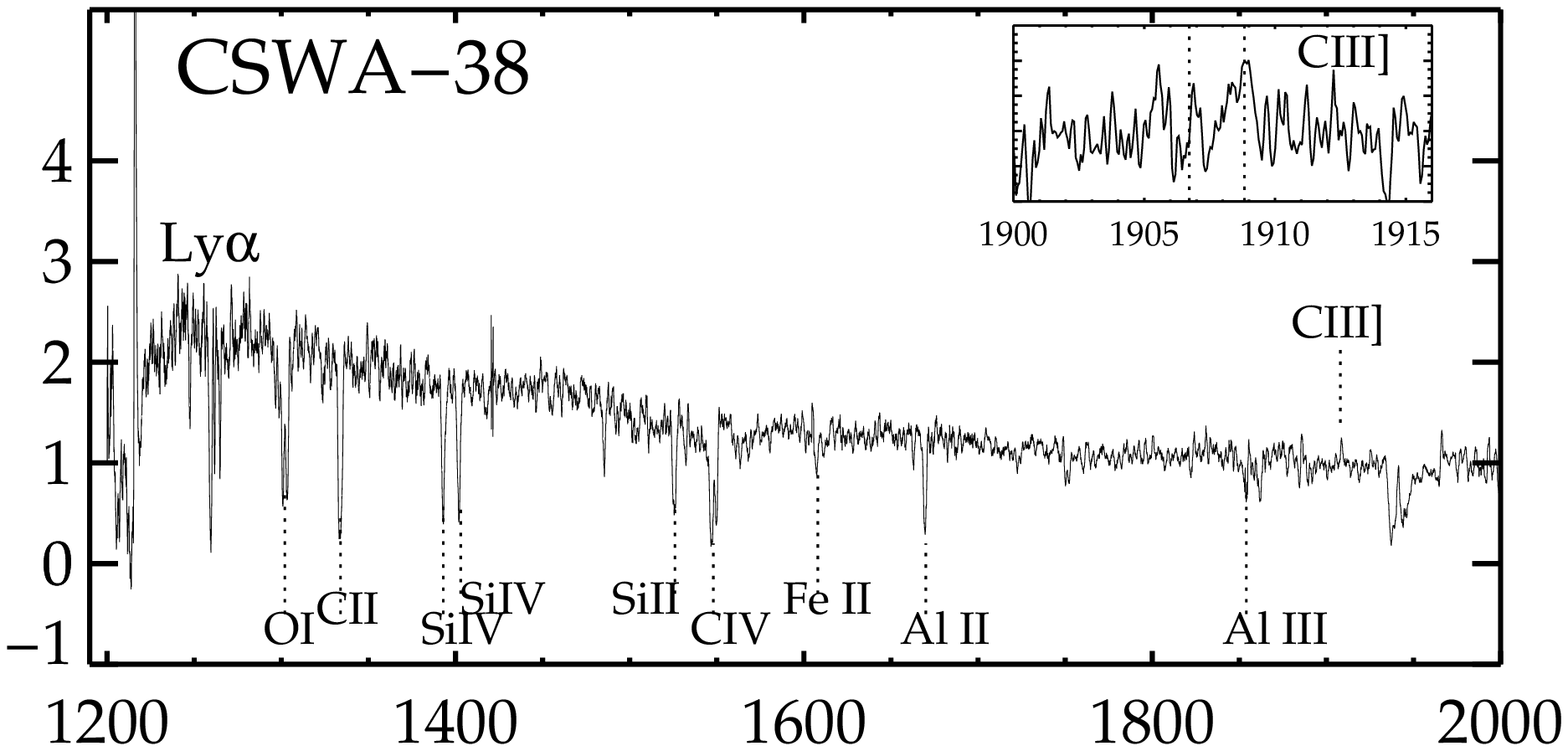}}}\\
\vspace{-7 mm}
\subfloat{ \includegraphics[angle=90,width=0.52\textwidth]{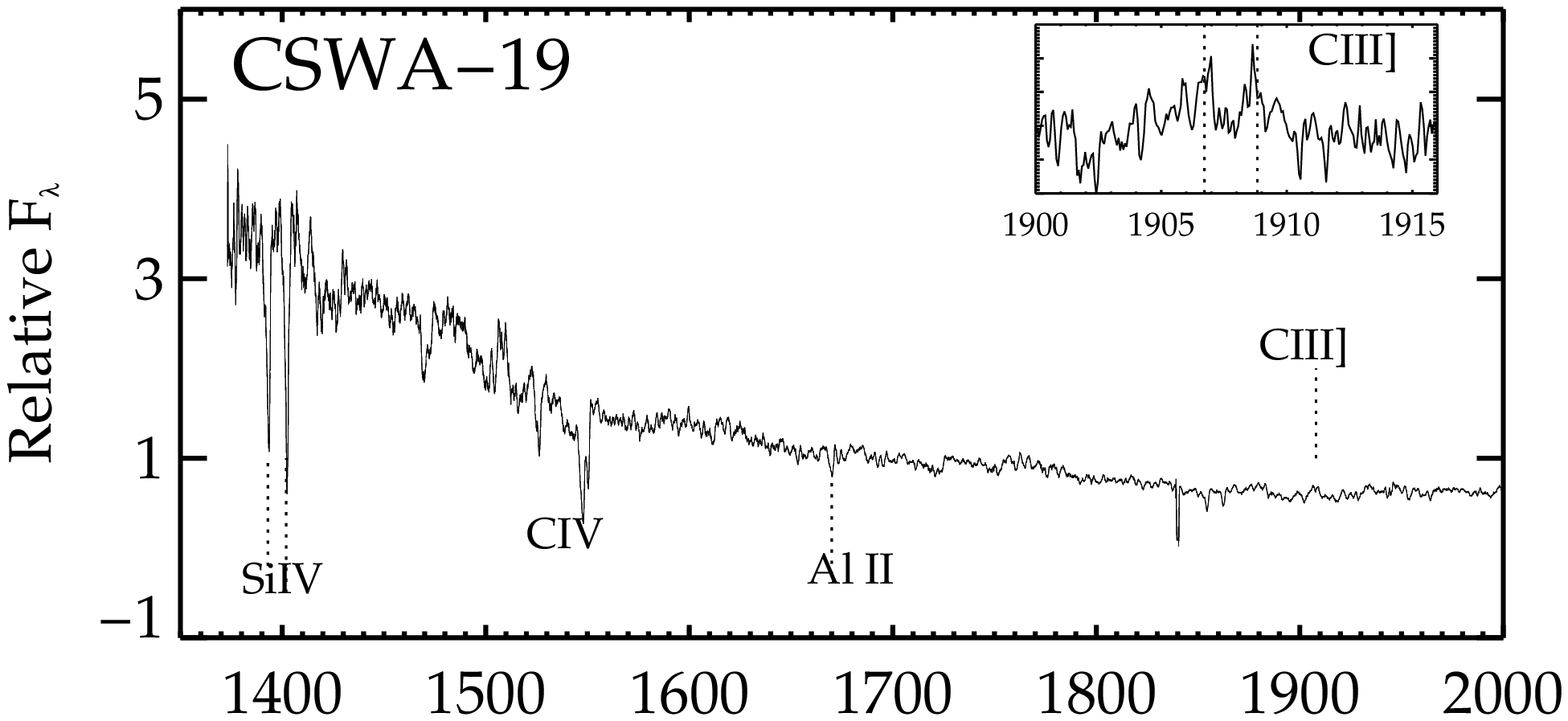}} 
\hbox{\hspace{-0.8 cm} \subfloat{ \includegraphics[angle=90,width=0.52\textwidth]{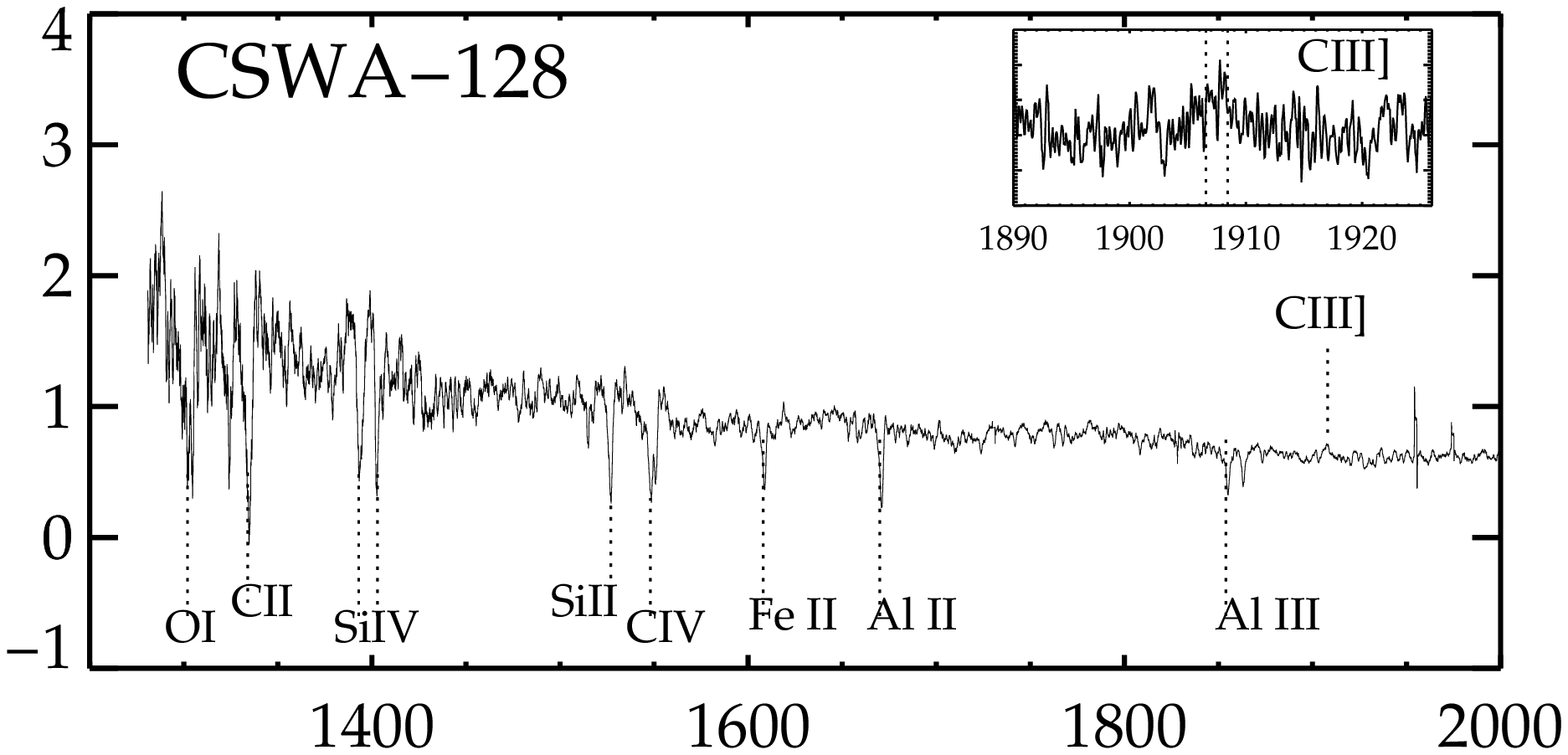}}}\\
\vspace{-7 mm}
\subfloat{ \includegraphics[angle=90,width=0.52\textwidth]{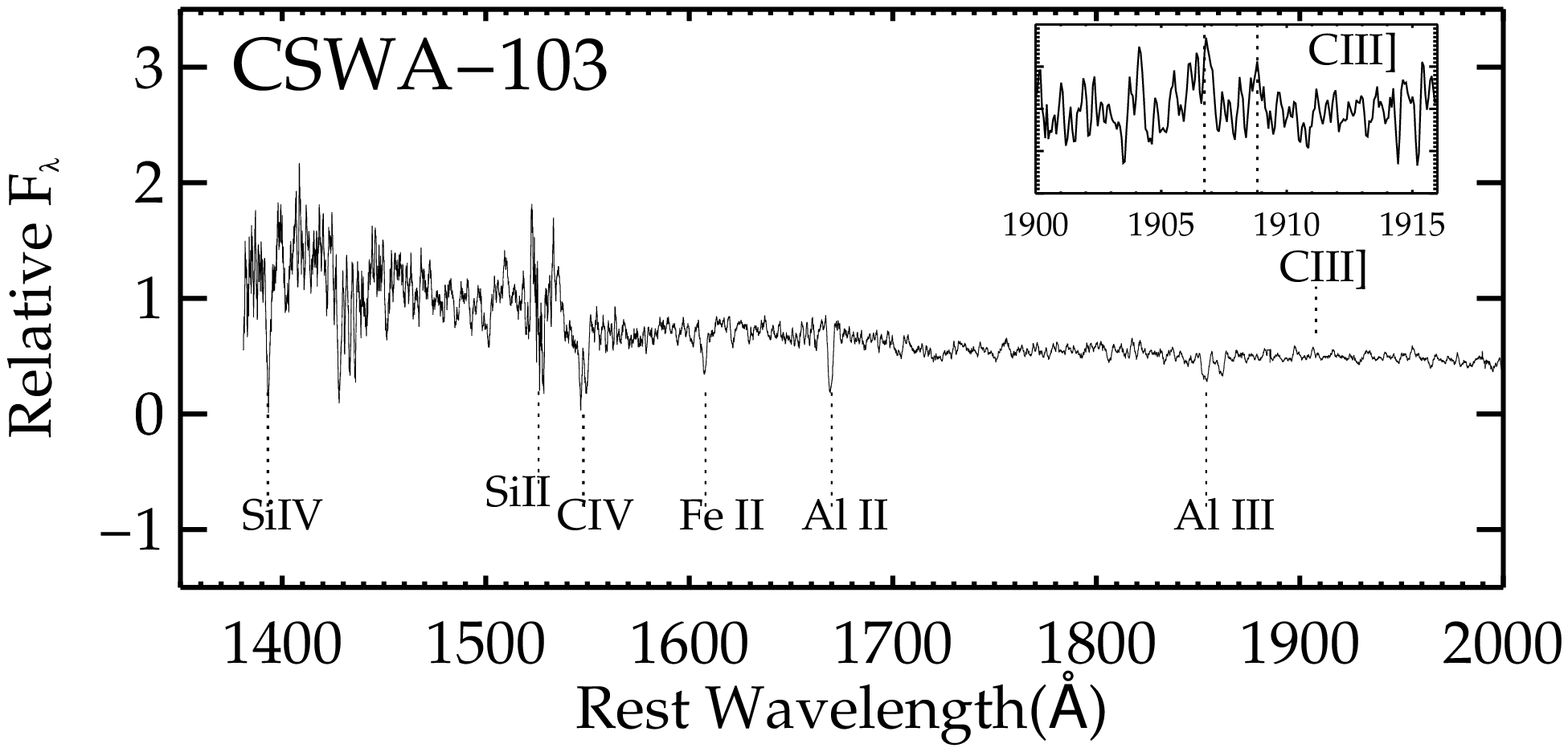}} 
\hbox{\hspace{-0.8 cm} \subfloat{ \includegraphics[angle=90,width=0.52\textwidth]{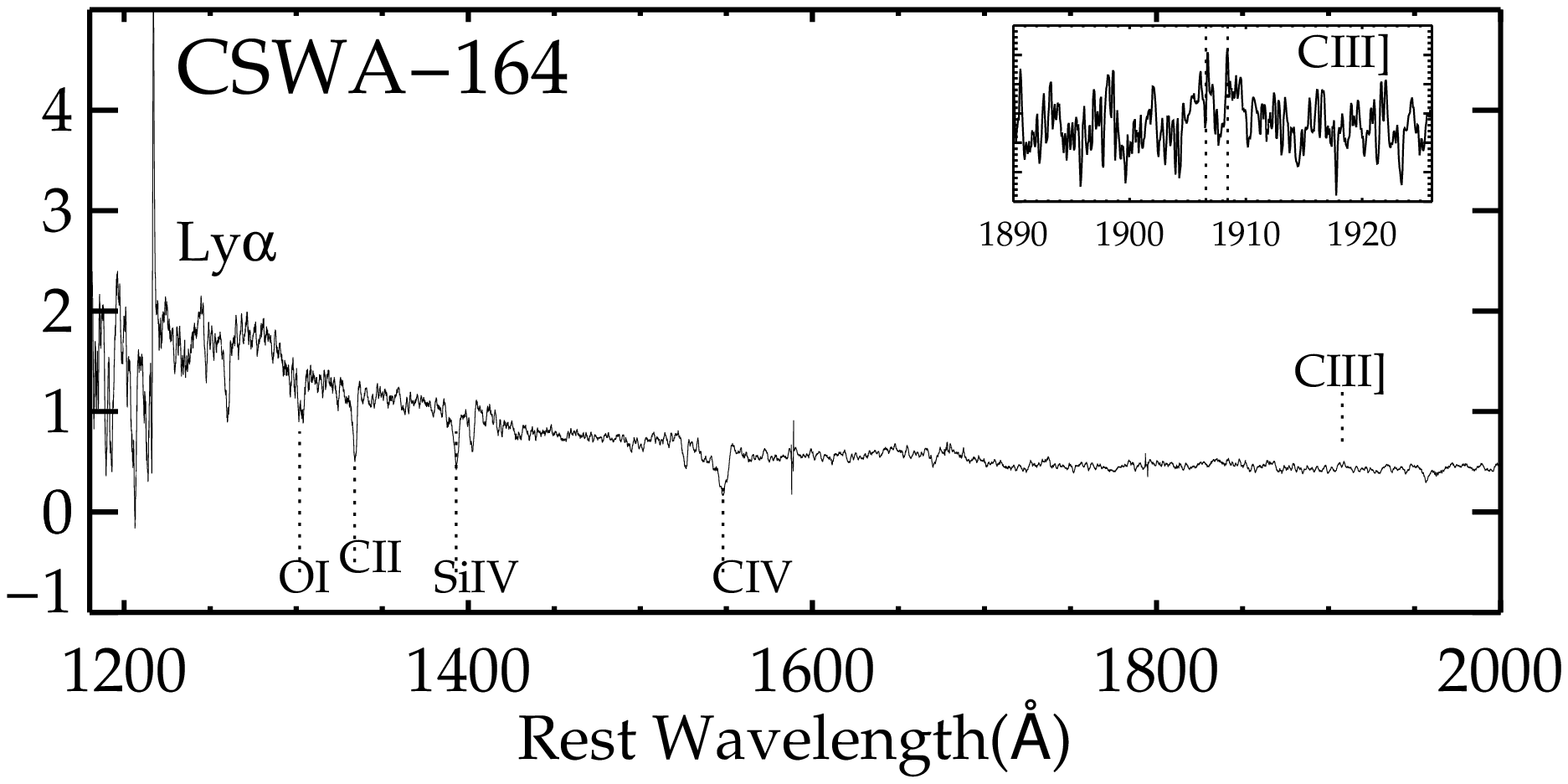}}}\\

\caption{Rest-UV spectra of gravitationally lensed galaxies presented in Table~\ref{table:sample} with CIII] detections. The upper left side of each image contains the CASSOWARY-ID. The dotted dashed line below and above the UV continuum represents absorption and emission features identified in the spectra. The upper right of each panel shows zoom in spectral coverage near CIII]$\lambda\lambda$1907,1909. The vertical dashed lines in the inset show the location of CIII]$\lambda\lambda$1907,1909 doublet. The CIII] doublet remain unresolved in the MMT/BCS spectra of CSWA-13 and CSWA-139. }
\label{fig:esispectra}
\end{figure*}

\begin{figure*}
\centering
\subfloat{\includegraphics[angle=90,width=0.32\textwidth]{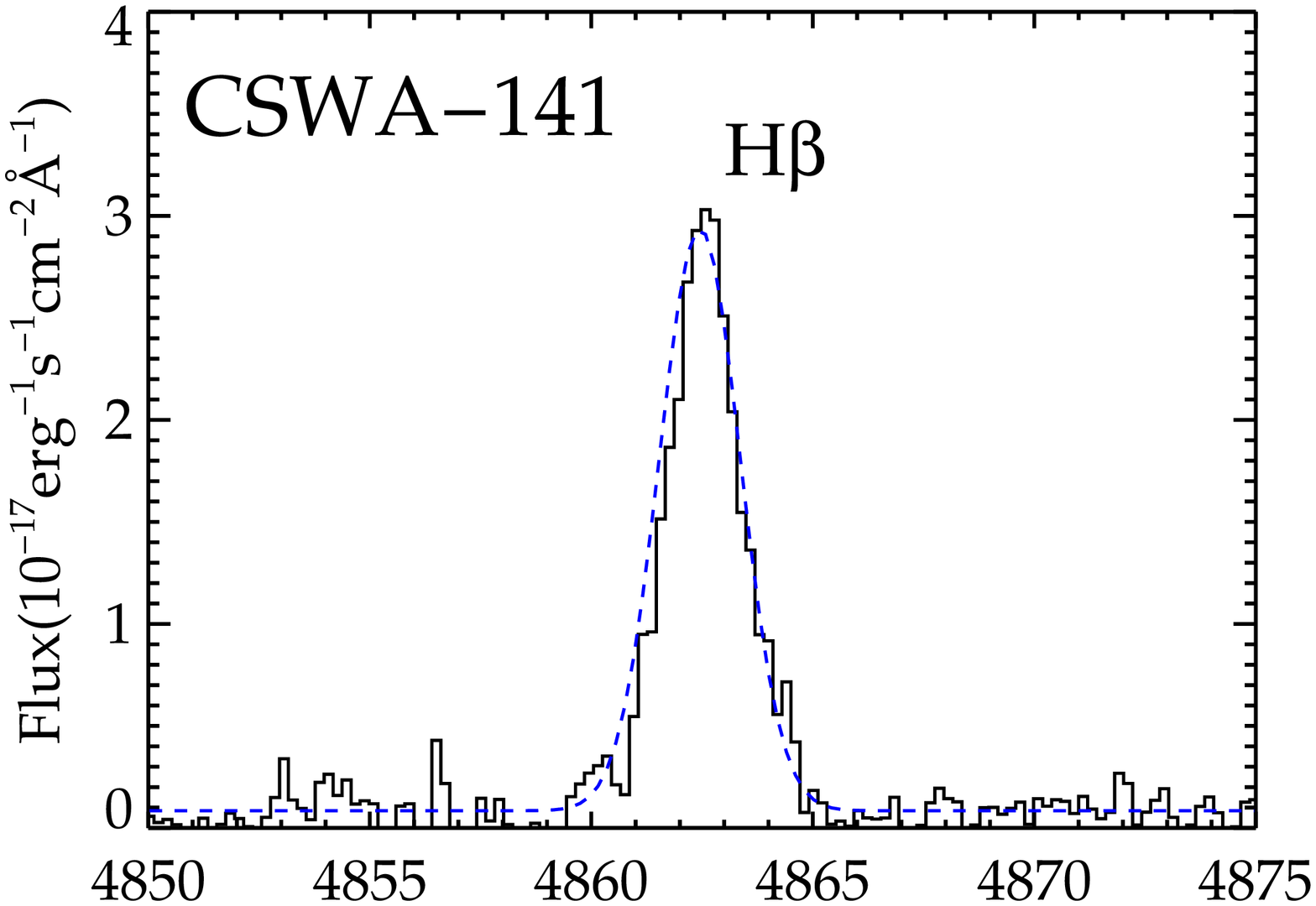}}
\subfloat{\includegraphics[angle=90,width=0.32\textwidth]{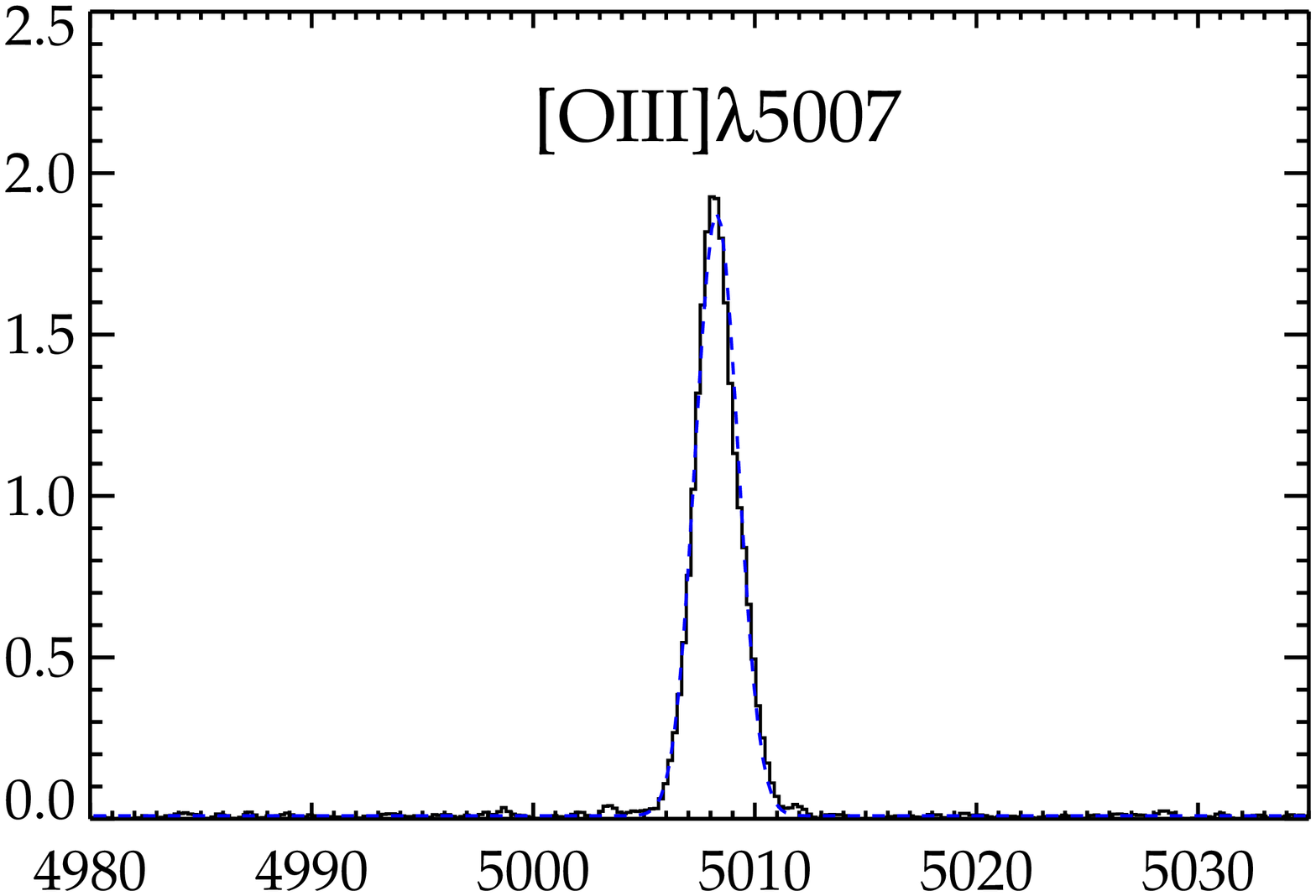}}
\subfloat{\includegraphics[angle=90,width=0.32\textwidth]{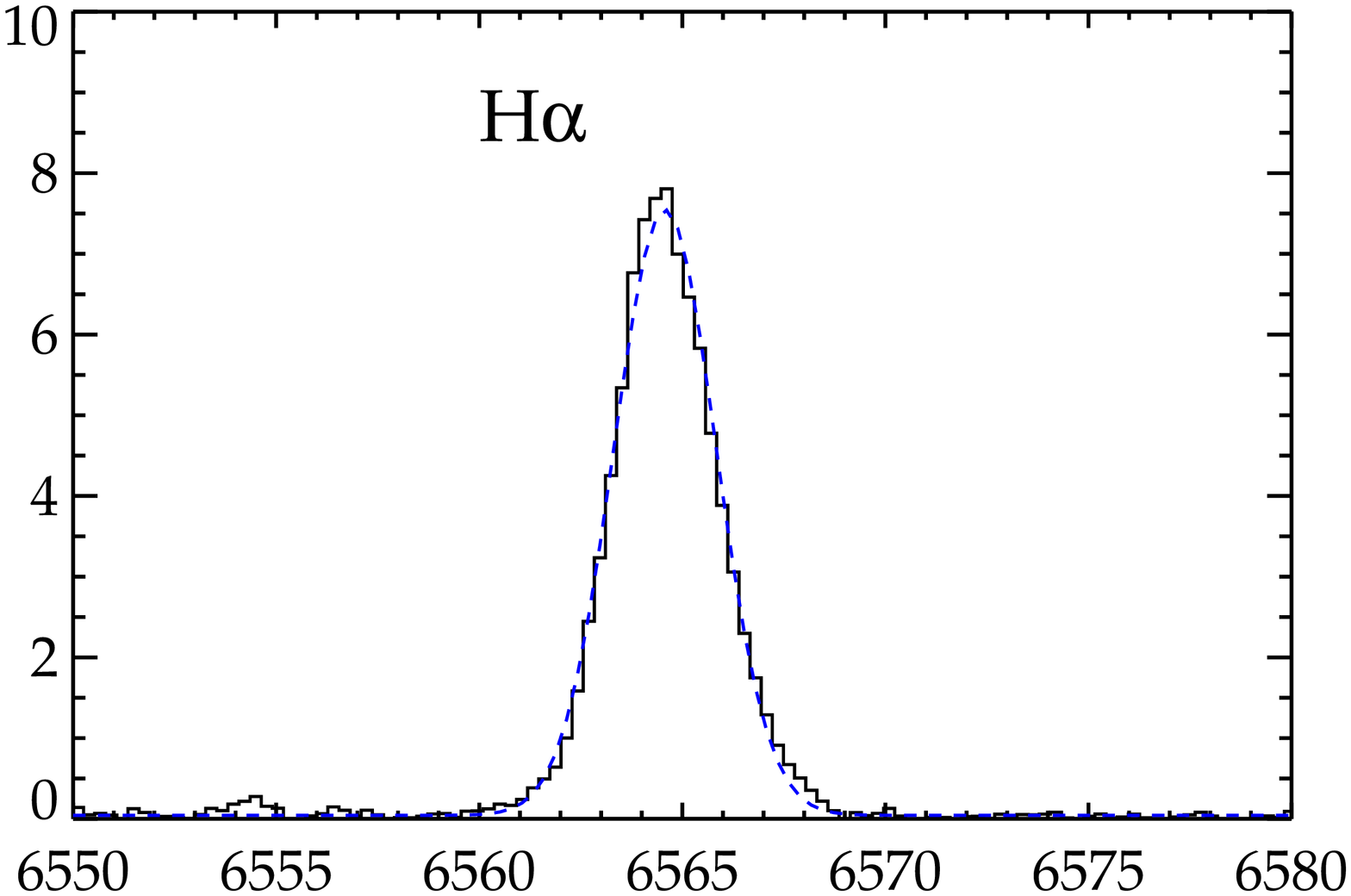}}\\
\vspace{-4 mm}
\subfloat{\includegraphics[angle=90,width=0.32\textwidth]{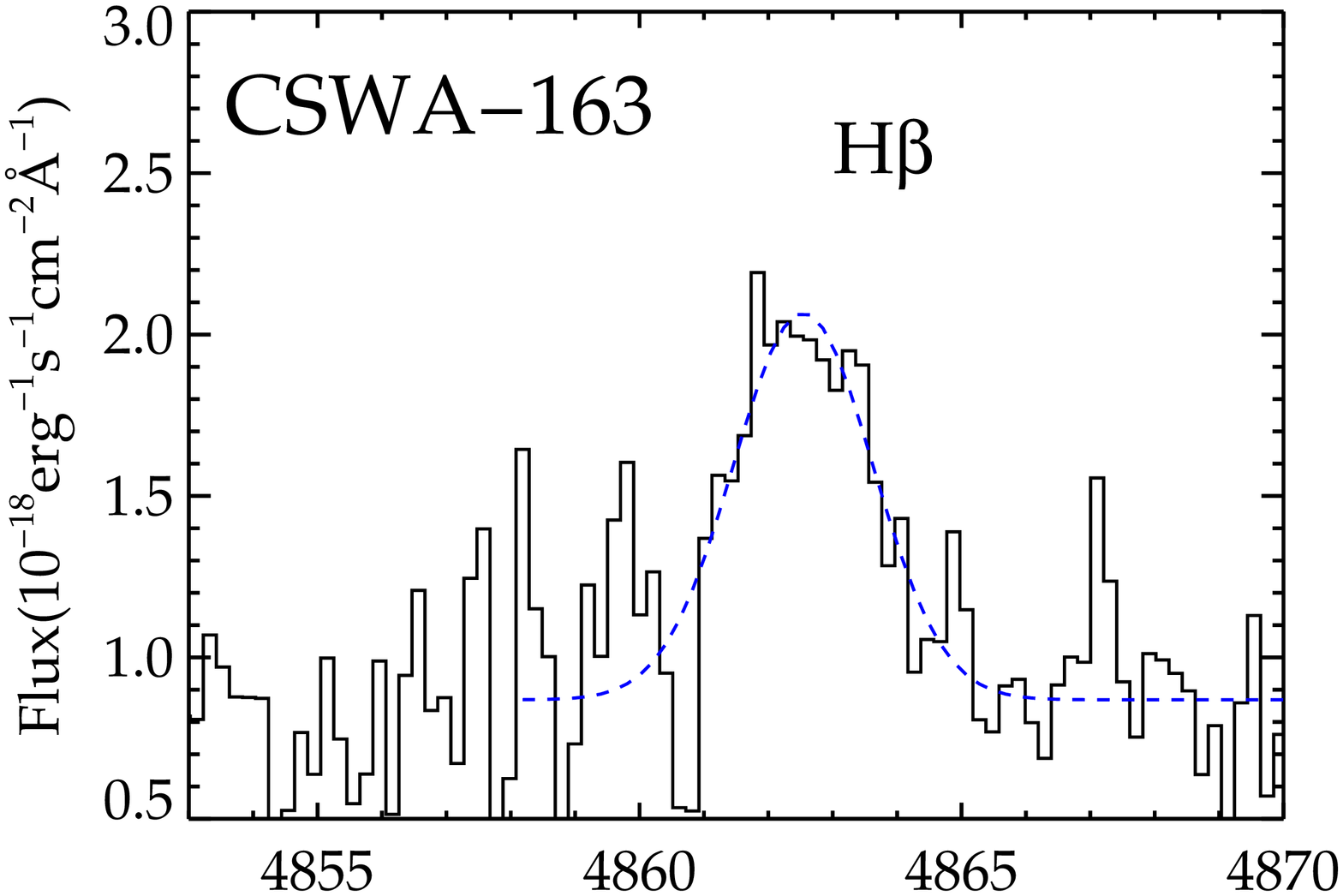}}
\subfloat{\includegraphics[angle=90,width=0.32\textwidth]{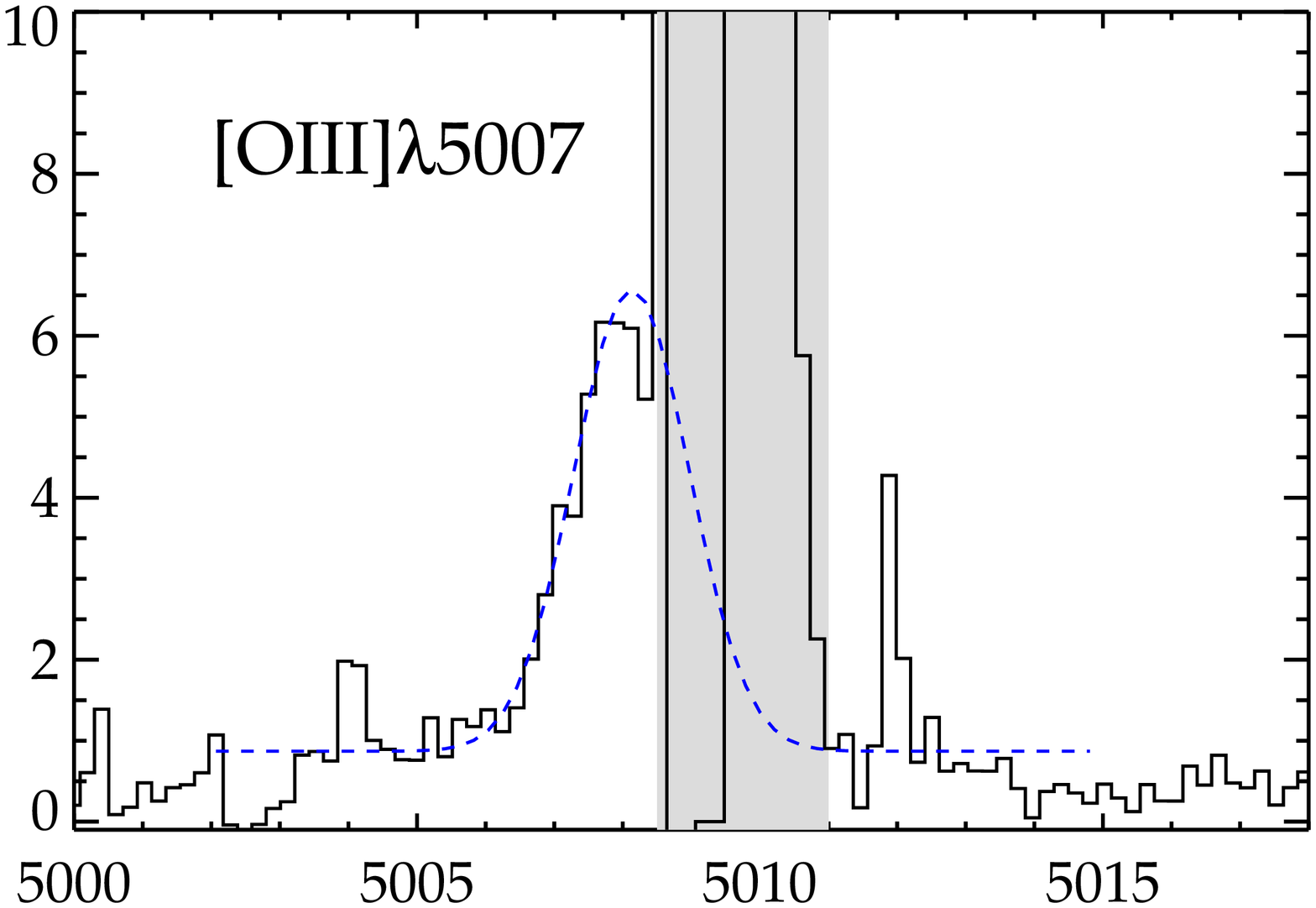}}
\subfloat{\includegraphics[angle=90,width=0.32\textwidth]{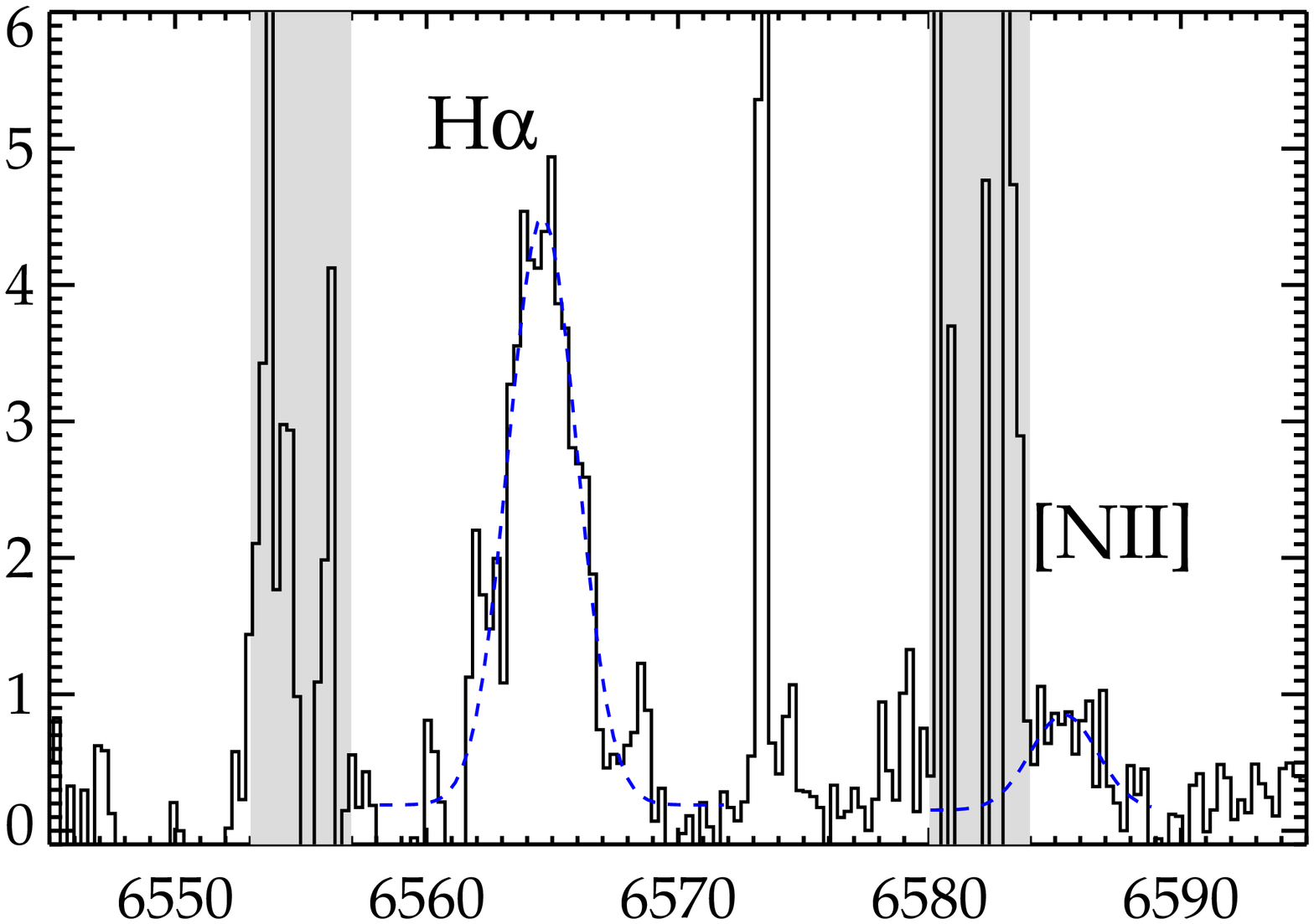}}\\
\vspace{-4 mm}
\subfloat{\includegraphics[angle=90,width=0.32\textwidth]{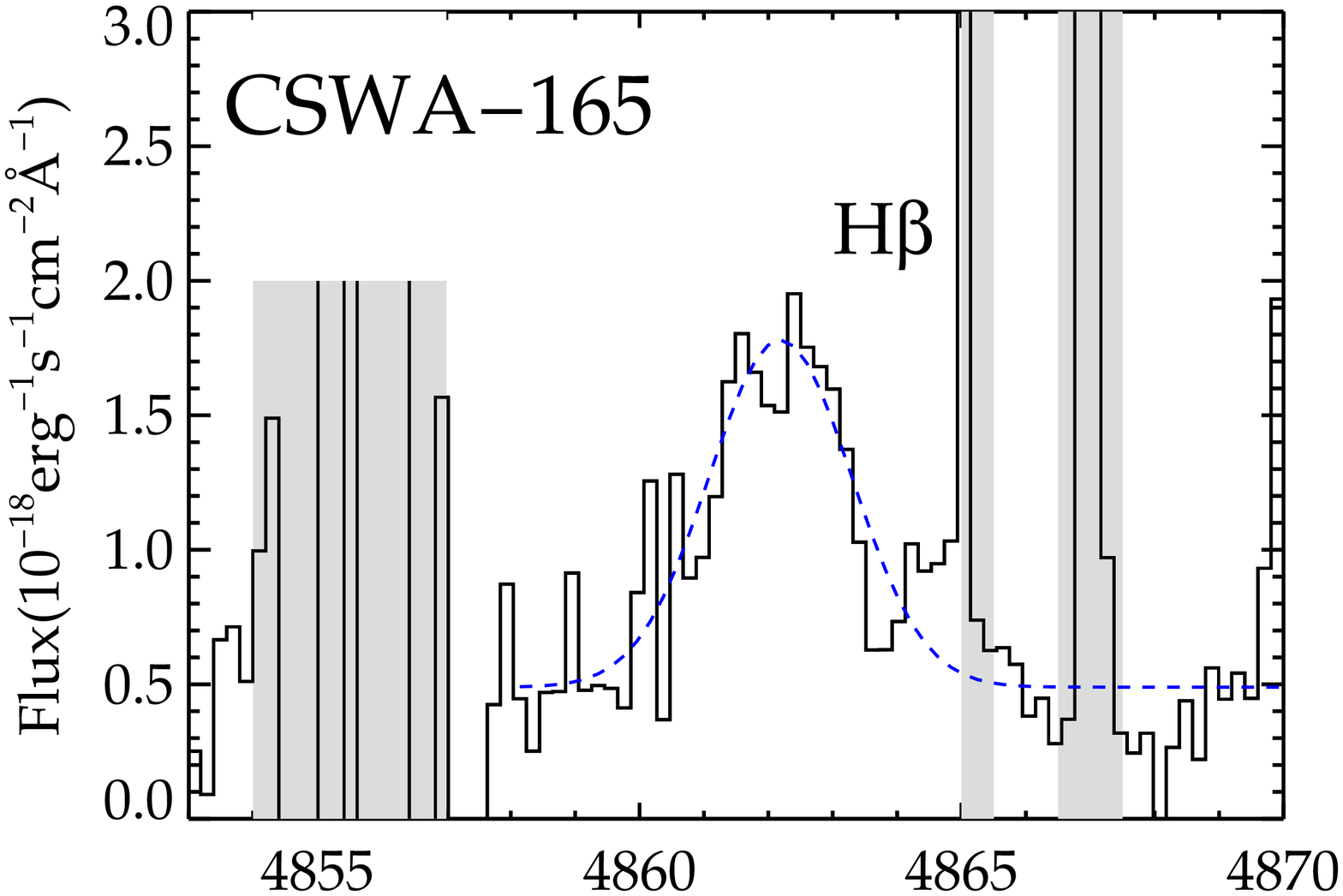}}
\subfloat{\includegraphics[angle=90,width=0.32\textwidth]{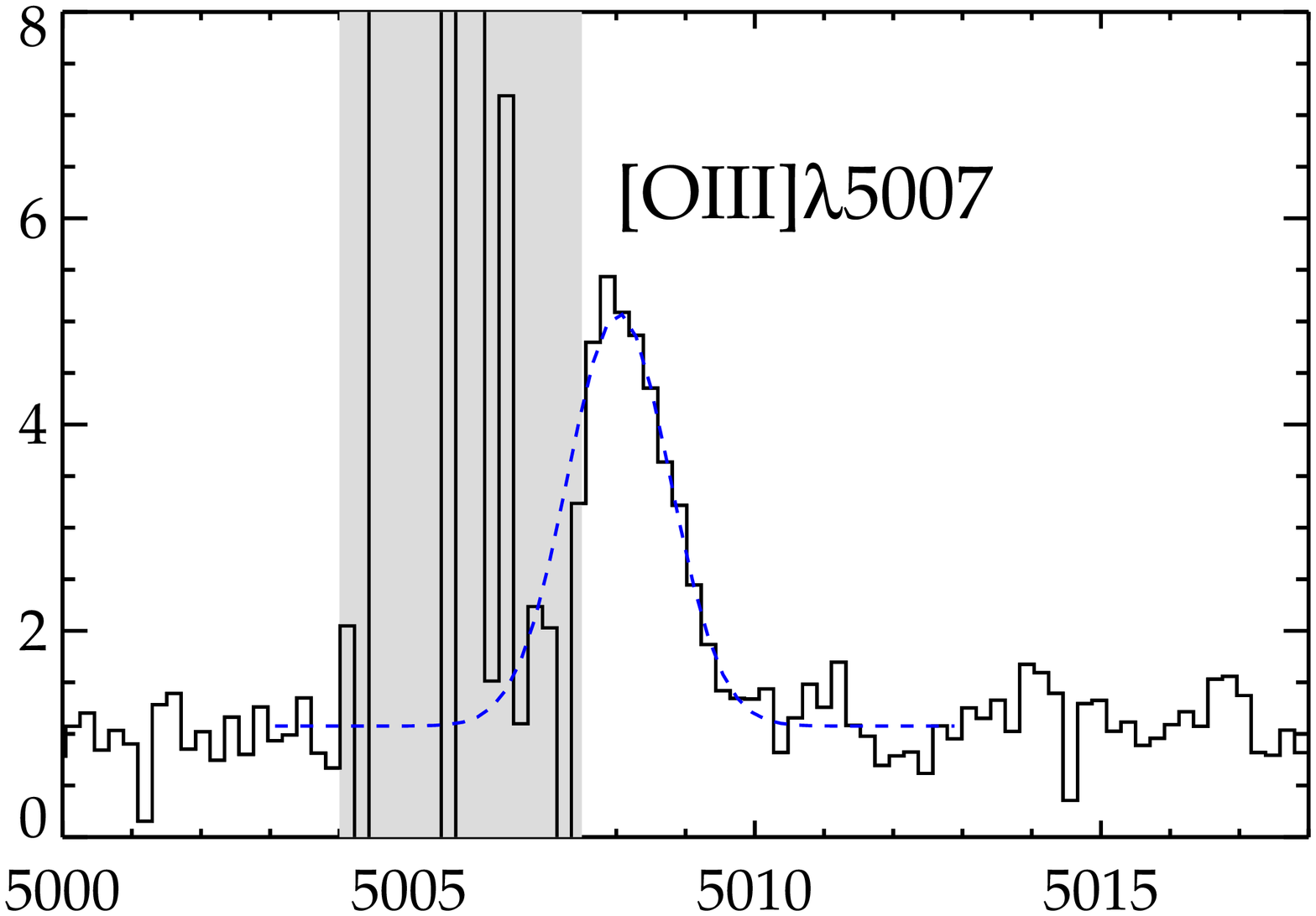}}
\subfloat{\includegraphics[angle=90,width=0.32\textwidth]{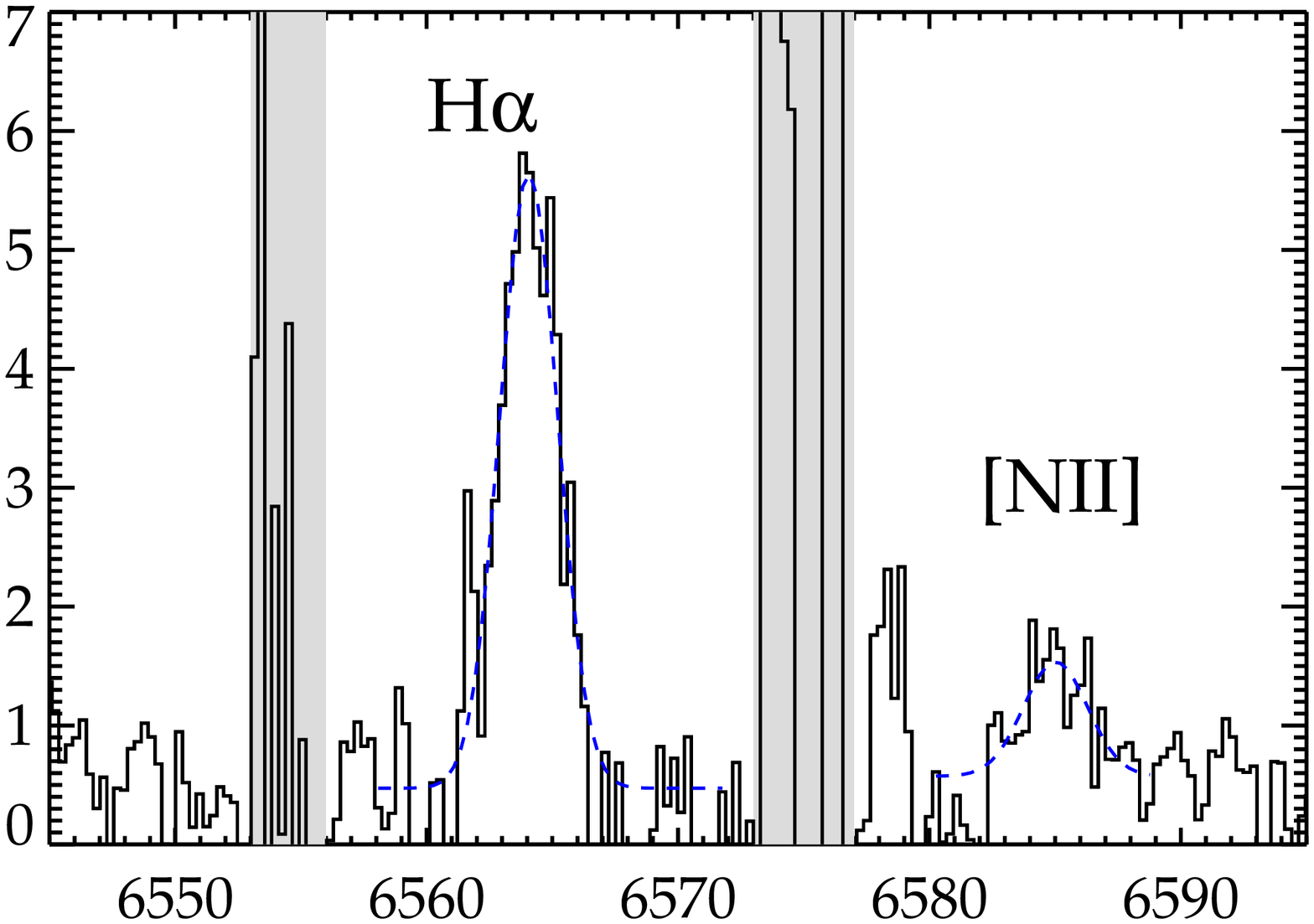}}\\
\vspace{-4 mm}
\subfloat{\includegraphics[angle=90,width=0.32\textwidth]{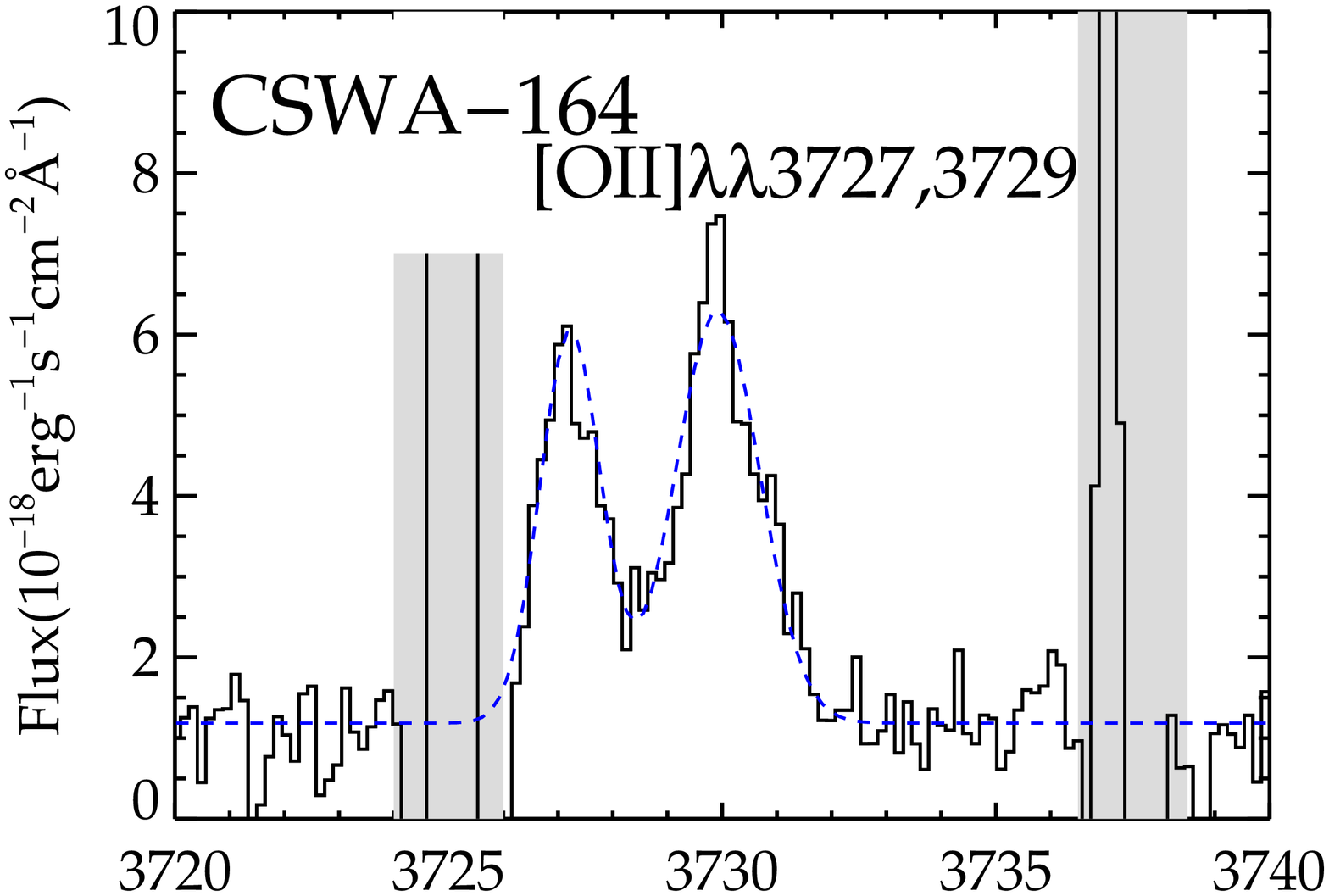}}
\subfloat{\includegraphics[angle=90,width=0.32\textwidth]{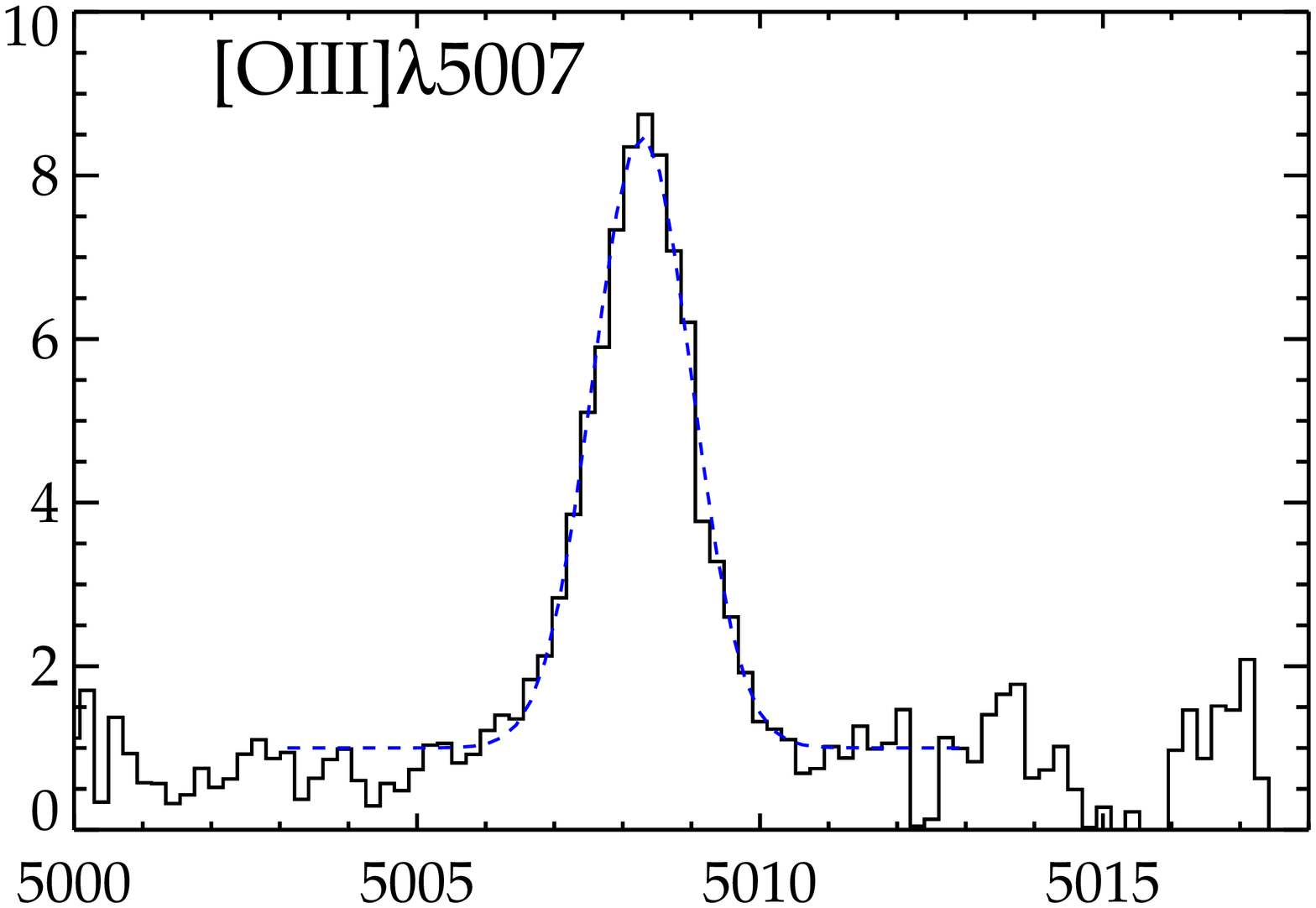}}
\subfloat{\includegraphics[angle=90,width=0.32\textwidth]{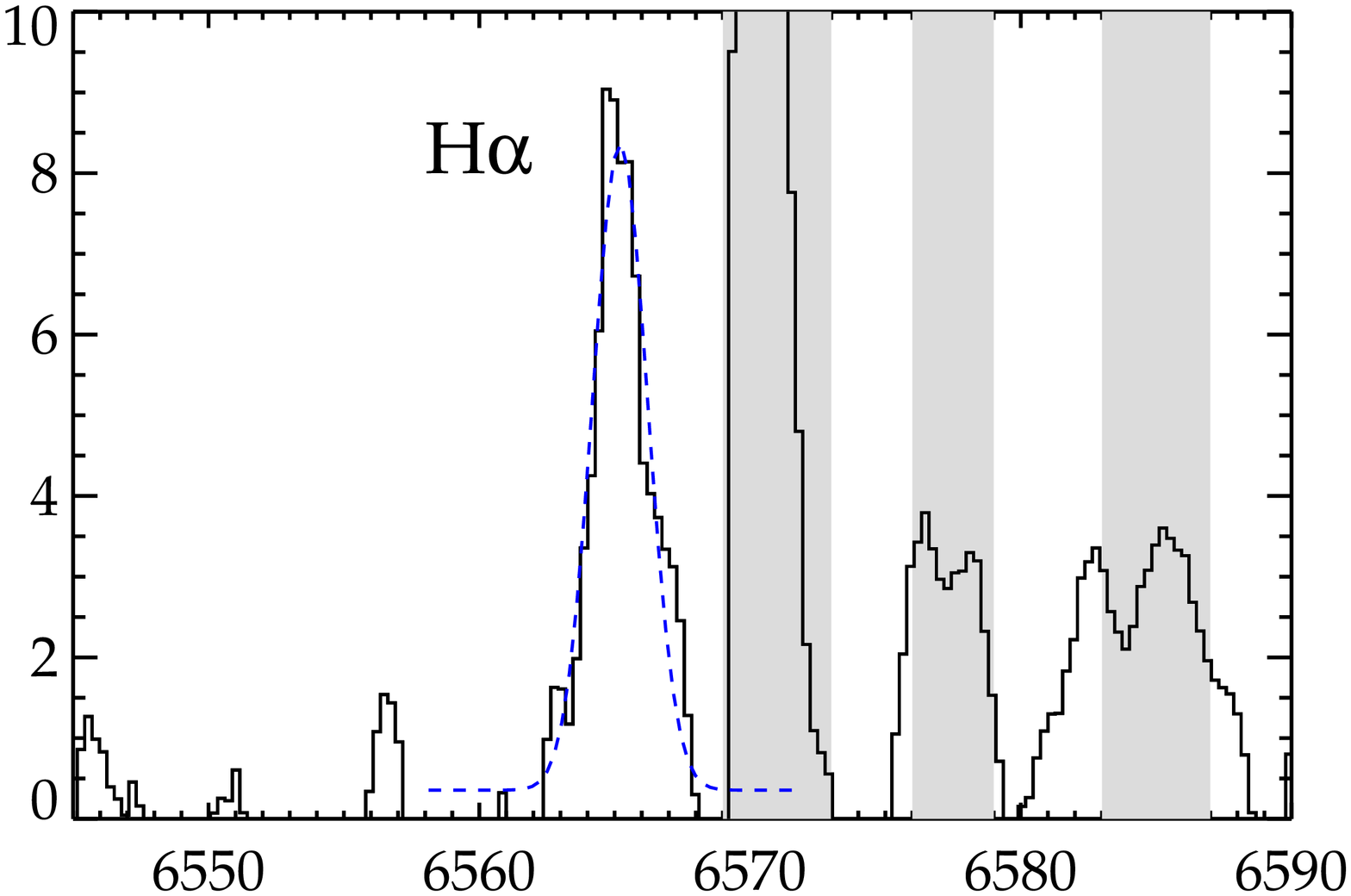}}\\
\vspace{-4 mm}
\subfloat{\includegraphics[angle=90,width=0.32\textwidth]{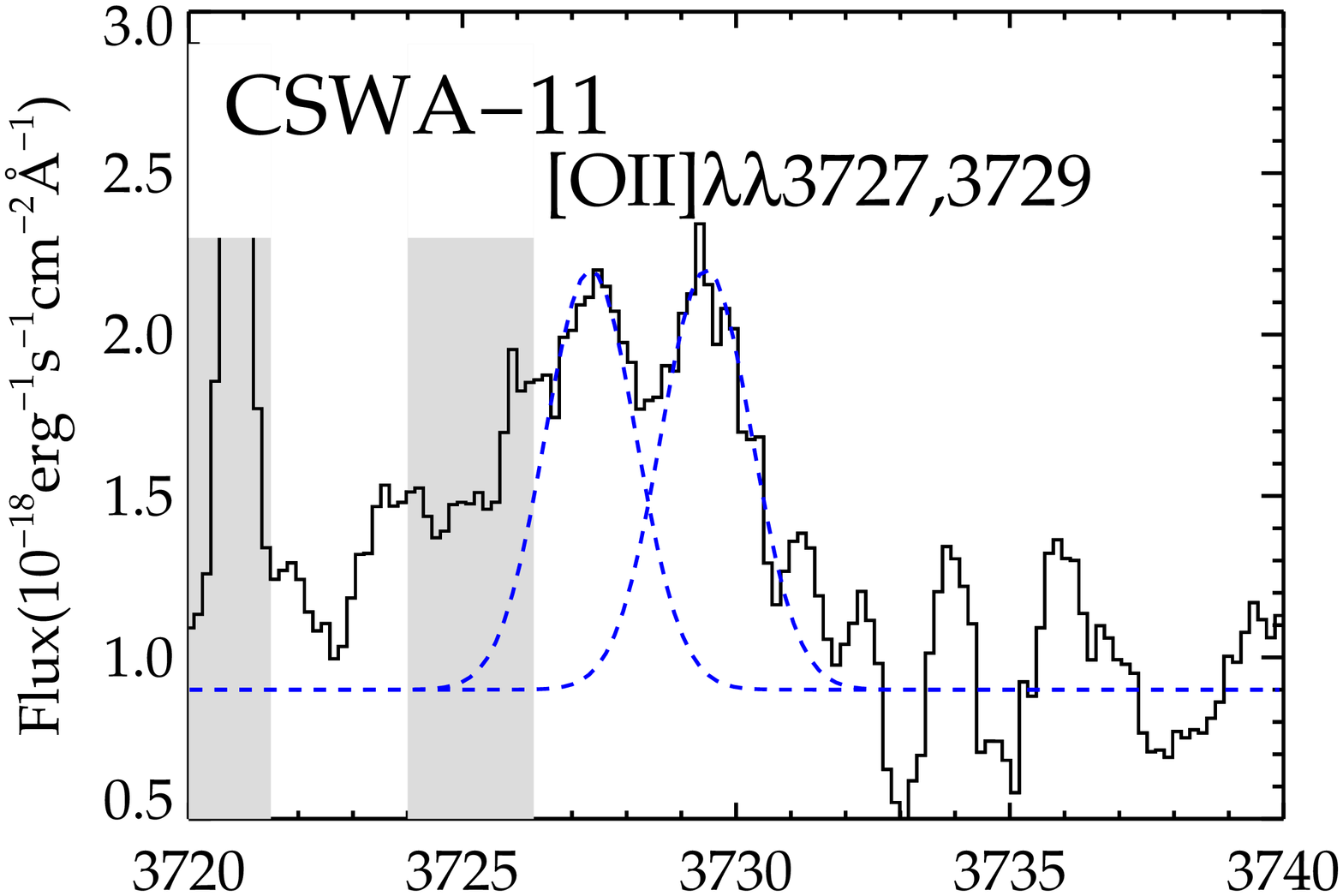}}
\subfloat{\includegraphics[angle=90,width=0.32\textwidth]{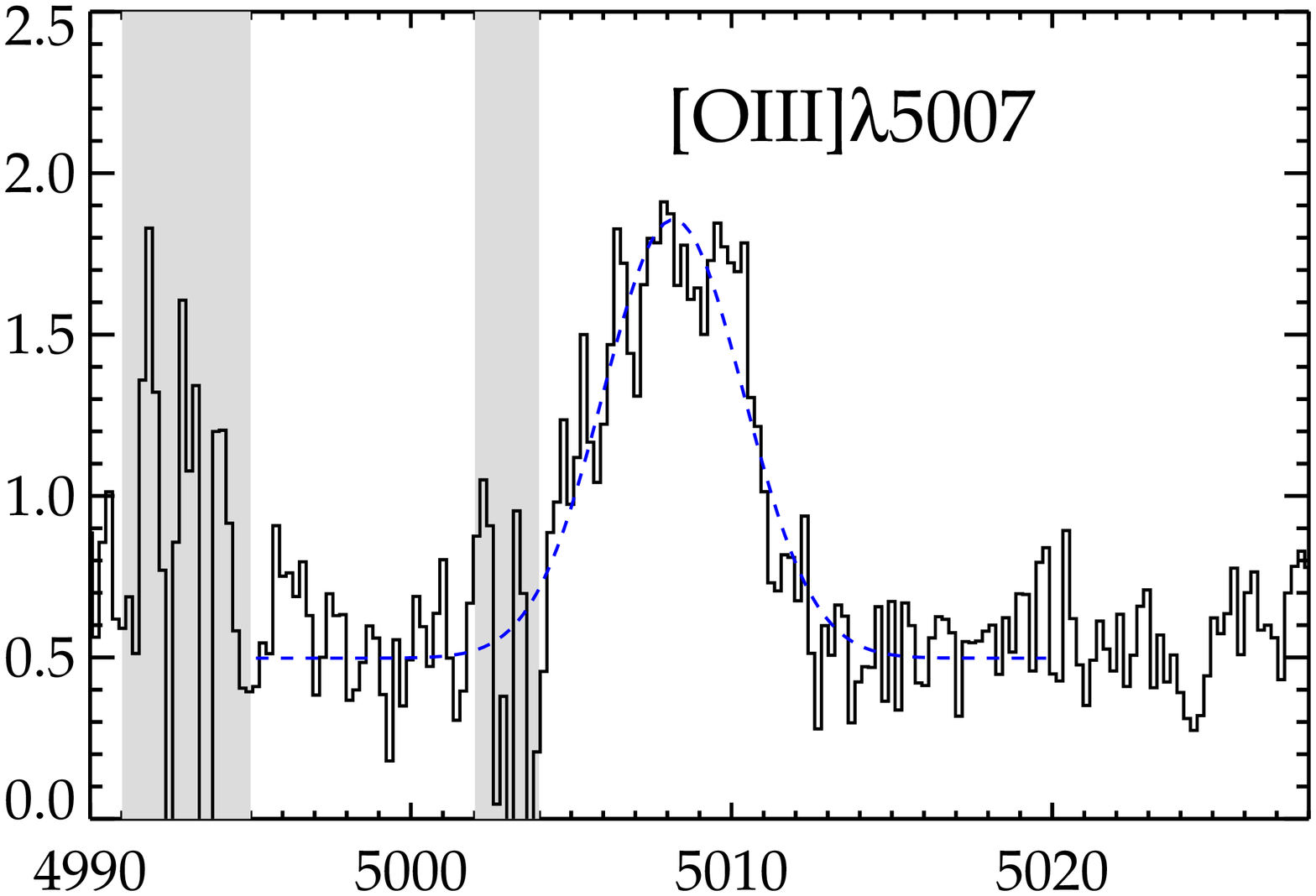}}
\subfloat{\includegraphics[angle=90,width=0.32\textwidth]{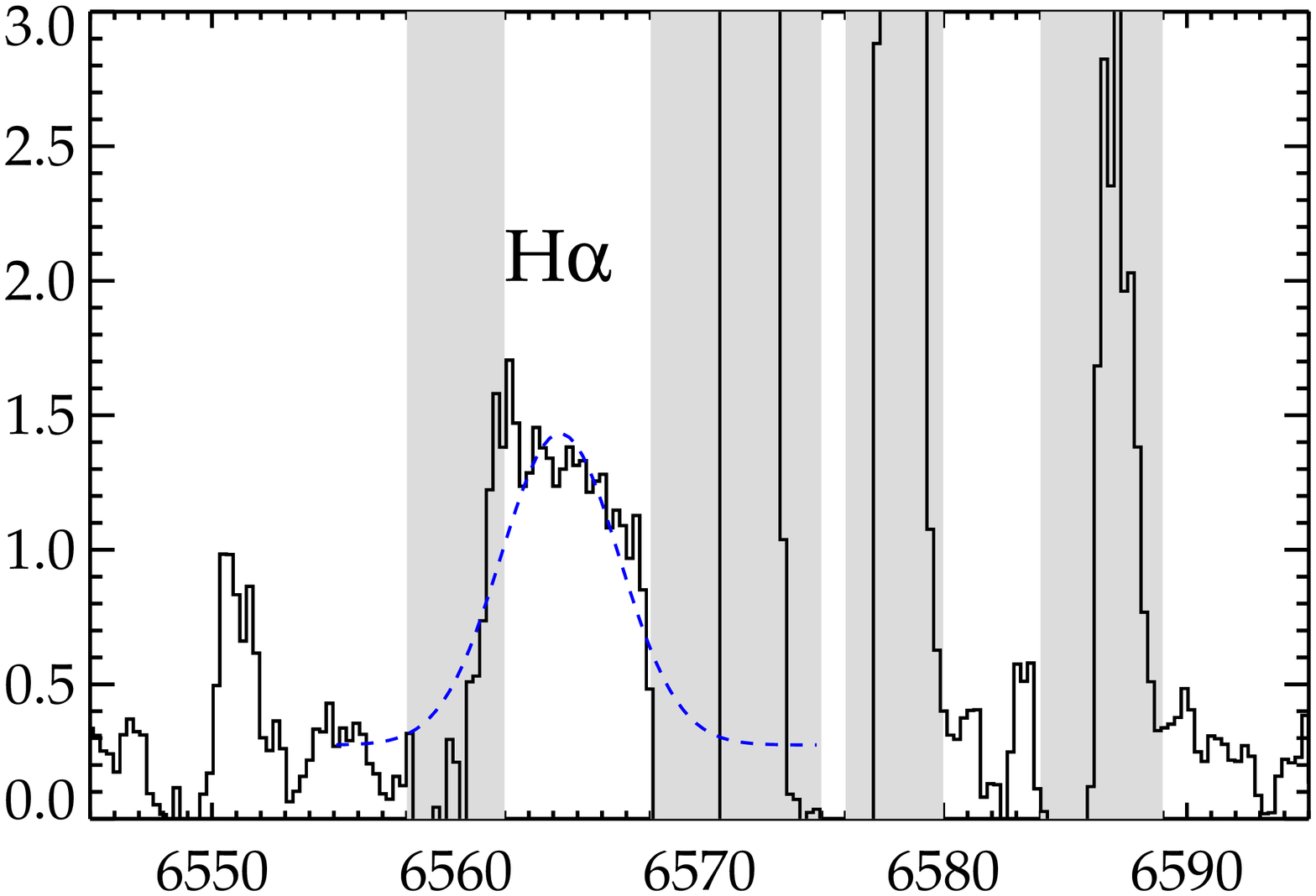}}\\
\vspace{-4 mm}
\subfloat{\includegraphics[angle=90,width=0.32\textwidth]{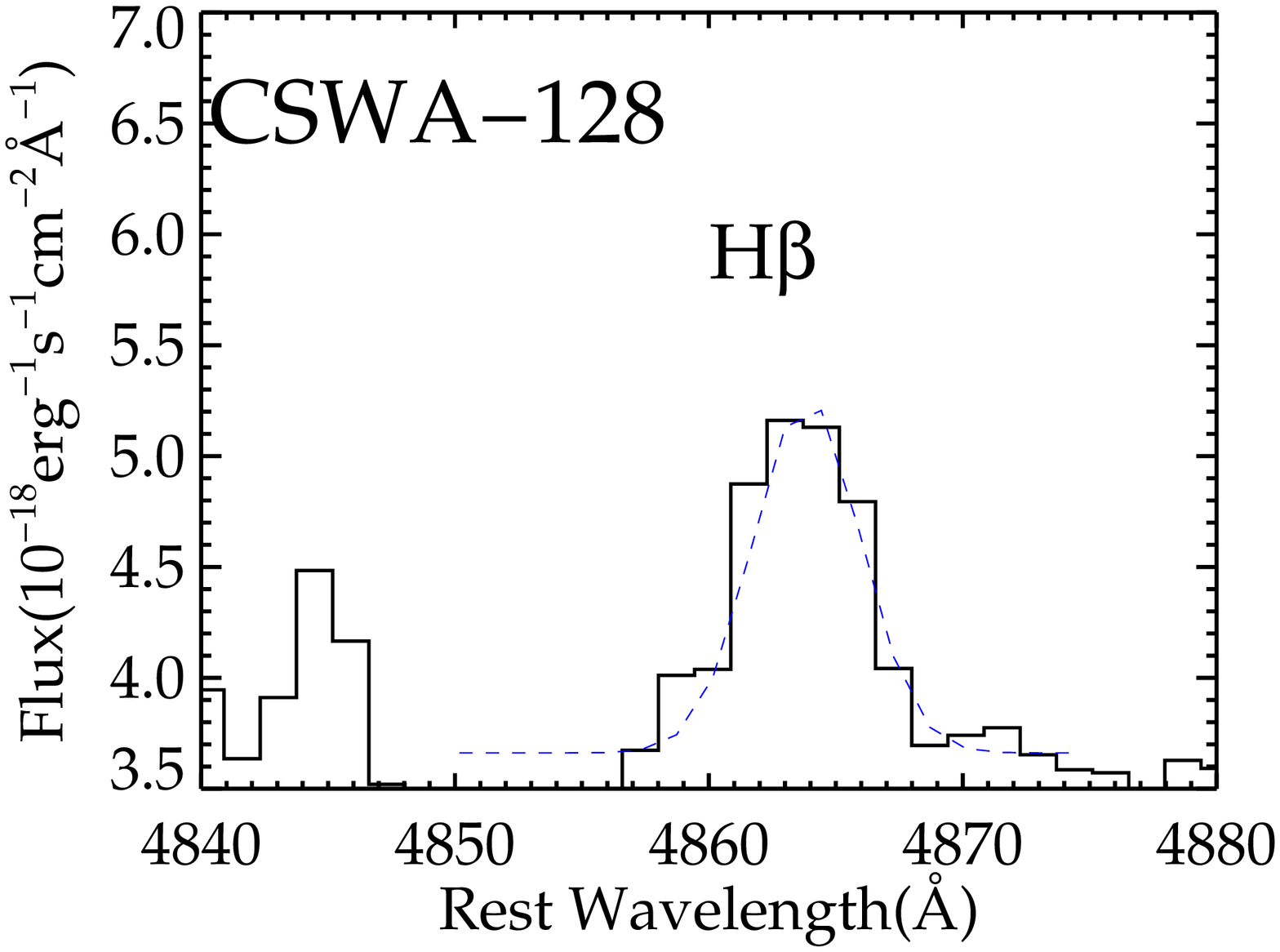}}
\subfloat{\includegraphics[angle=90,width=0.32\textwidth]{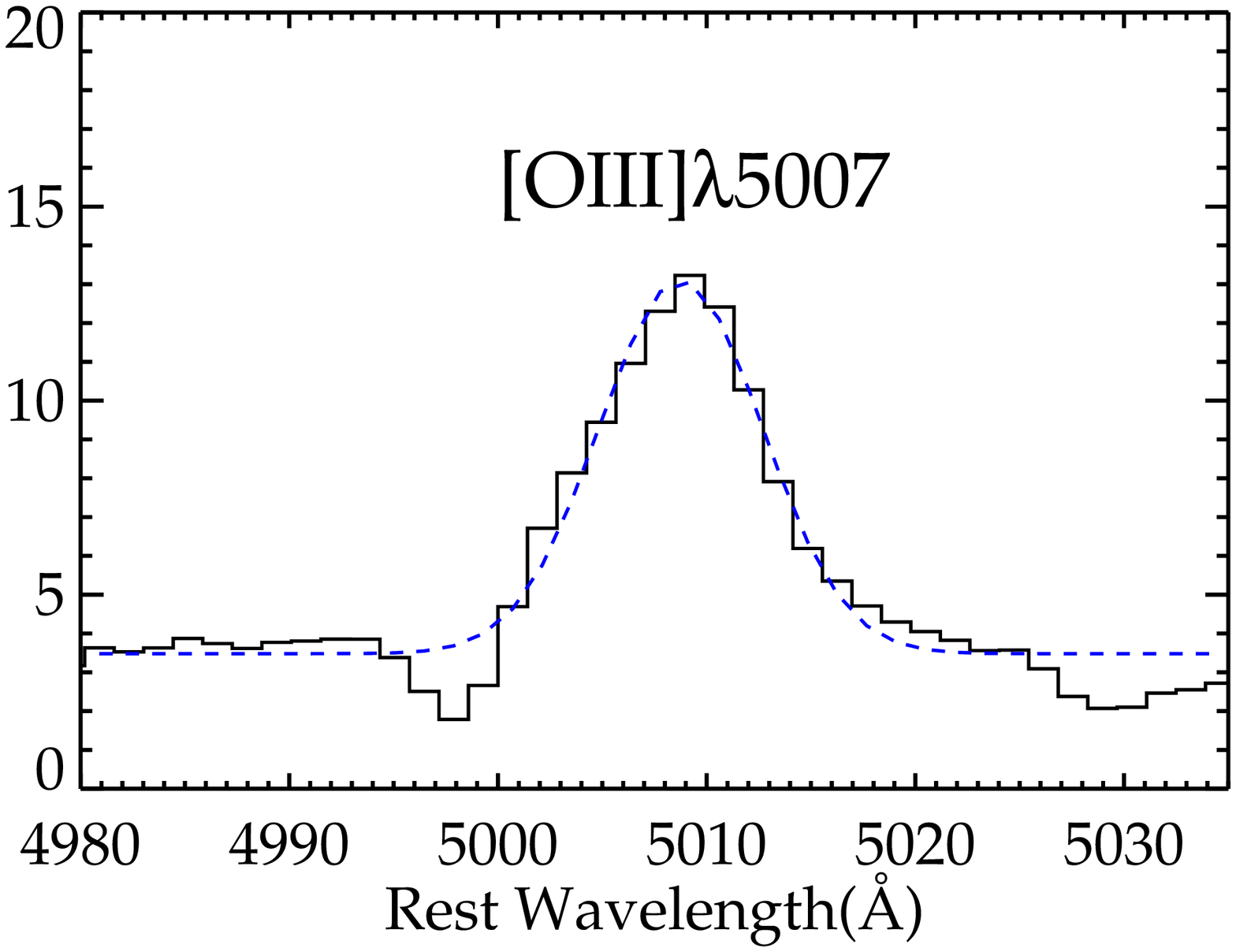}}
\subfloat{\includegraphics[angle=90,width=0.32\textwidth]{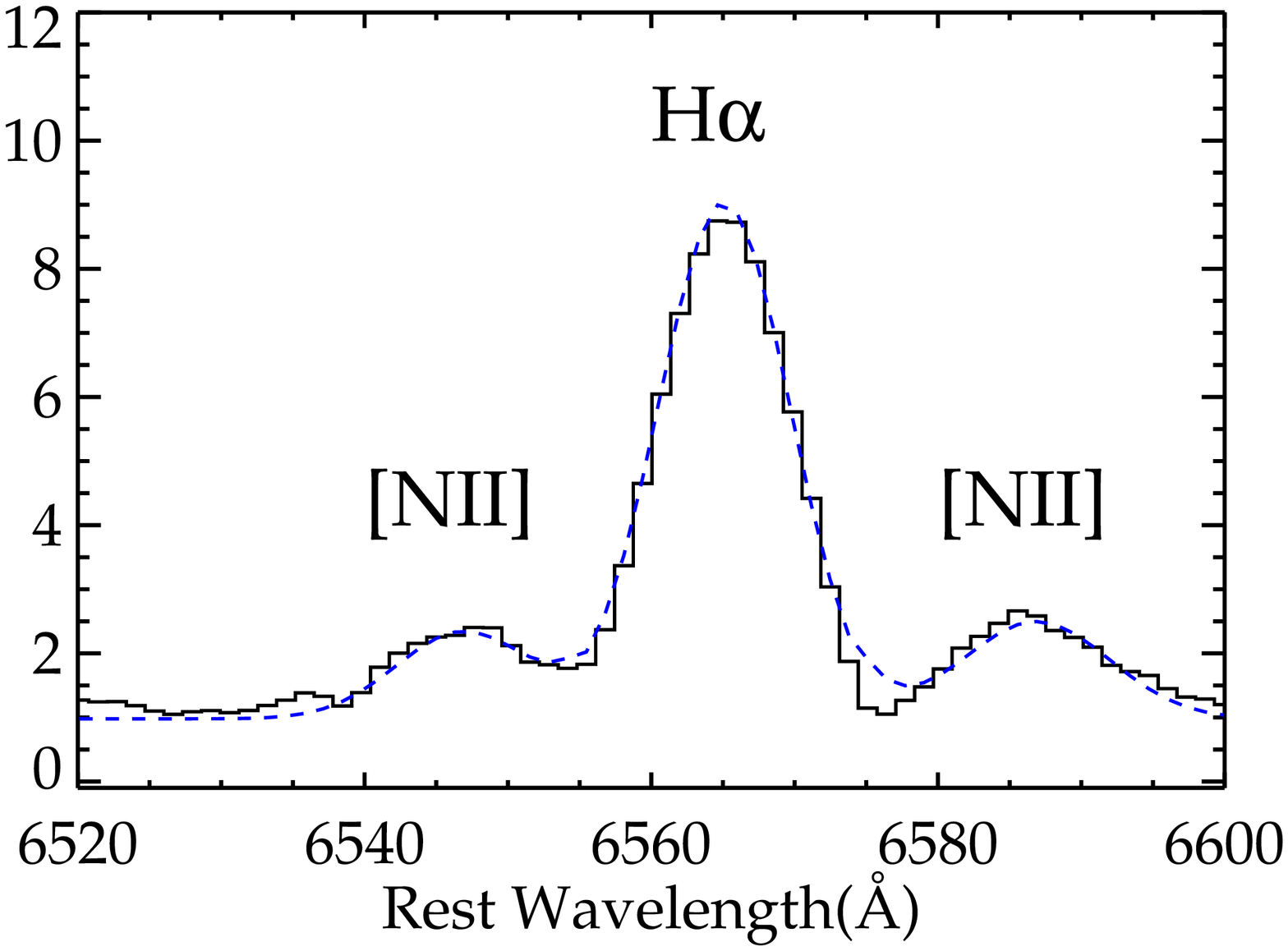}}
\vspace{-4 mm}
\caption{Optical emission lines in six gravitationally lensed galaxies discussed in this paper. Each row shows optical emission lines from a single source. The object ID is given in the leftmost panel of each row. The black line represents observed flux whereas blue line shows gaussian fit to the line profile calculated using IDL routine MPFITPEAK. The region affected by skylines is shown as a grey swath.}
\label{fig:cswa_nir}
\end{figure*}   

\begin{table}
\begin{tabular}{lccc}
\hline Object & log (M$_{\star}$/M$_{\odot})$ & sSFR    & $\hat{\tau}_V$  \\ 
  & & (Gyr$^{-1})$ & \\ \hline 

 CSWA-141      & $8.6^{+0.2}_{-0.3}$  & $31.2^{+60.8}_{-16.3}$  & $0.17^{+0.18}_{-0.12}$     \\  
  CSWA-13     & $9.8^{+0.3}_{-0.3}$  &  $43.1^{+56.6}_{-28.7}$& $0.53^{+0.13}_{-0.27}$    \\ 
    CSWA-139     &  $10.3^{+0.3}_{-0.3}$  & $2.9^{+14.3}_{-2.0}$  & $0.77^{+0.41}_{-0.33}$     \\ 
    CSWA-2   &  $9.1^{+0.3}_{-0.3}$   & $19.9^{+26.1}_{-10.4}$  & $0.96^{+0.31}_{-0.27}$     \\ 
    CSWA-39   &  $9.9^{+0.3}_{-0.4}$   & $11.7^{+13.8}_{-8.1}$  & $0.96^{+0.24}_{-0.31}$     \\ 
    CSWA-128      &  $9.9^{+0.1}_{-0.1}$  & $1.0^{+0.4}_{-0.3}$  & $0.13^{+0.06}_{-0.07}$     \\ 
   CSWA-19      &  $10.5^{+0.1}_{-0.1}$  & $0.5^{+0.4}_{-0.1}$  & $1.09^{+0.21}_{-0.16}$     \\
    CSWA-103  & $10.4^{+0.1}_{-0.2}$ &$0.7^{+1.1}_{-0.3}$ & $0.48^{+0.31}_{-0.23}$   \\
      CSWA-163   & $9.9^{+0.1}_{-0.1}$  & $1.8^{+0.9}_{-0.9}$  & $0.67^{+0.09}_{-0.13}$     \\ 
       CSWA-40   &  $10.8^{+0.2}_{-0.2}$   & $2.1^{+3.3}_{-1.3}$  & $0.43^{+0.29}_{-0.25}$     \\ 
 CSWA-165   &$10.0^{+0.2}_{-0.2}$  & $3.9^{+3.0}_{-1.9}$   & $0.62^{+0.12}_{-0.13}$   \\  
   CSWA-11      & $10.0^{+0.4}_{-0.3}$   & $7.4^{+14.0}_{-5.5}$  & $0.28^{+0.31}_{-0.20}$     \\ 
  CSWA-116  & $9.4^{+0.2}_{-0.2}$ & $1.3^{+2.4}_{-0.7}$  & $0.27^{+0.34}_{-0.18}$    \\ 
\hline
\end{tabular}
\caption{Physical properties inferred from BEAGLE fitting for subset of CASSOWARY galaxies with optical and 
near-IR photometry. From left to right, the columns give the object name, C~III] equivalent widths, magnification corrected stellar mass, specific star formation rates and V-band optical depth.
}
\label{table:sed}
\end{table}

\begin{figure*}
\centering
\subfloat{\includegraphics[angle=90,width=0.99\textwidth]{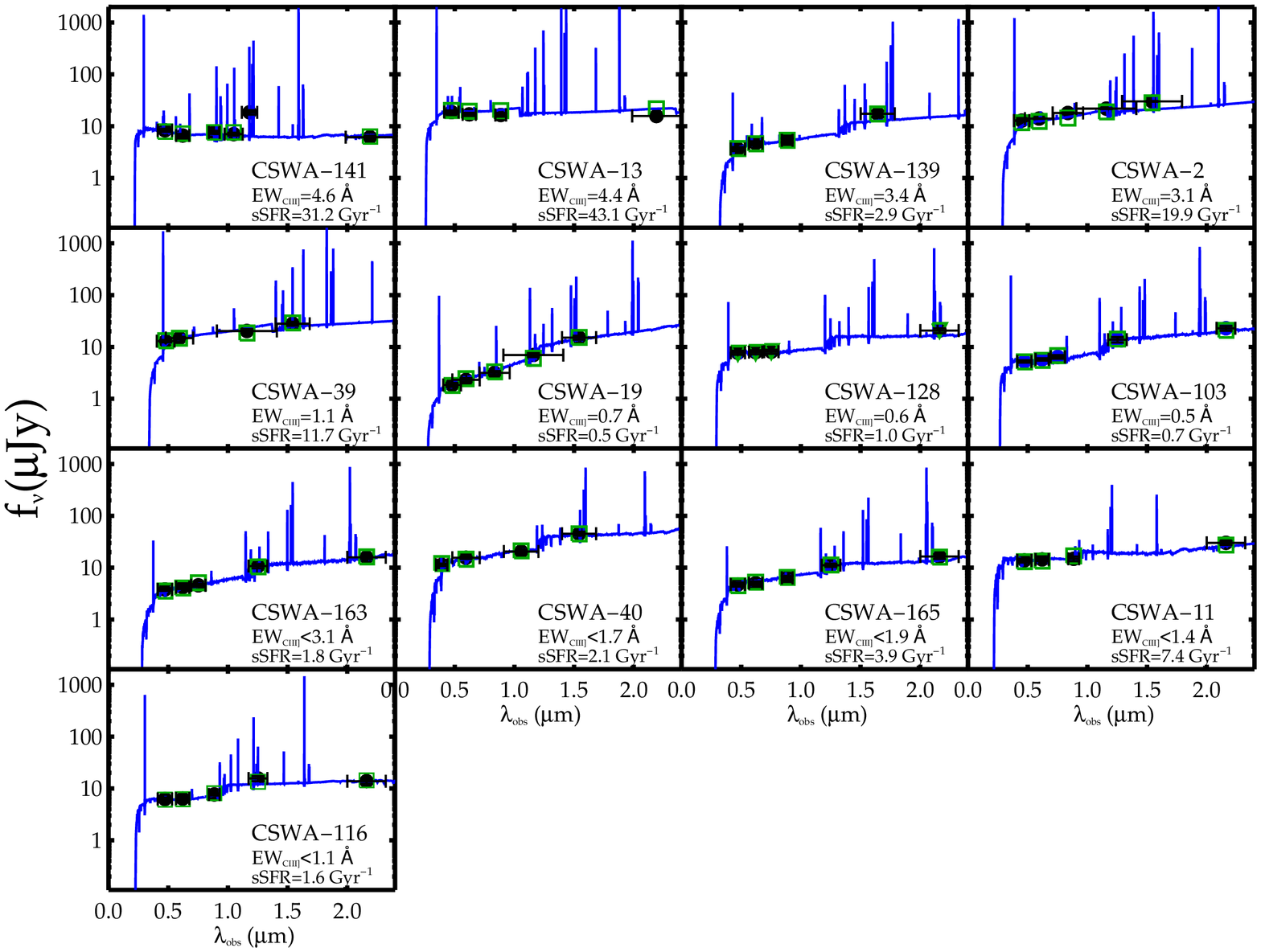}}
\caption{SEDs of galaxies in the  sample. The black circle represents broad band photometry data, blue line represents best fit population synthesis model to the observed data (see \S 2.5) and green square shows synthetic broad band flux from the best fit model. The lower right of each panel shows object ID, sSFR and CIII] equivalent widths.}
\label{fig:sed_cswa}
\end{figure*}

 \section{Results}  
    
All sixteen sources presented in this paper have optical spectra covering CIII] to quantify equivalent widths. 
We report these values (or $3\sigma$ upper limits) in Table~\ref{table:uvlines}. Of the sixteen sources, six galaxies have near-IR spectra enabling rest optical line flux ratios (Table~\ref{table:nir}).
 Ten sources have multi-wavelength photometry with SED information enabling 
 stellar mass and sSFR measurements.  Below we briefly comment on individual galaxy properties
where sources are ordered by descending CIII] equivalent width.

 \subsection{Notes on Individual Sources}

 \begin{table}
\begin{tabular}{lcccc}

\hline  Line & $\lambda_{\rm{rest}}$& $\lambda_{\rm{obs}}$  &  $\rm F_{line}$ & EW  \\
  & (\AA)  &  (\AA)  & $\rm \overline{F_{CIII]1909}}$ &  (\AA)   \\
 \hline \hline
 \multicolumn{5}{|c|}{Keck/ESI} \\
\hline \hline
 [CIII]  & 1906.73 & 4624.32  & 1.02(0.10) &  2.3(1.3) \\
  CIII]  & 1908.68 & 4628.84 &1.00 & 2.3(1.4) \\
  Fe II* & 2626.64  & 6370.24  & 0.22(0.07) & 0.9(0.5) \\
   Mg II & 2796.36  & 6783.42 & 1.30(0.07) & 9.7(6.8)  \\
   & 2803.53 & 6800.79 & 0.65(0.04) & 4.8(2.6) \\
  $\rm{[OII]}$ & 3727.10  &  9039.02 & 5.10(0.04) & 26.3(12.4)    \\
  &  3729.86  & 9045.68 & 6.62(0.09) & 39.2(19.9)   \\
 $\rm[NeIII]$ &  3870.16  & 9385.28 & 4.79(0.23) & 19.5(13.8)   \\
 \hline 
  \multicolumn{5}{|c|}{LBT/MODS} \\
  \hline  Line & $\lambda_{\rm{rest}}$& $\lambda_{\rm{obs}}$  &  $\rm F_{line}$ & EW  \\
  & (\AA)  &  (\AA)  & $\rm \overline{F_{CIII]1908}}$ &  (\AA)   \\
 \hline\hline
  [CIII]  & 1908 & 4622.13 & 1.00 &  3.5(1.5) \\
            He II & 1640.52 & 3974.16 & $<0.08$ &  $<0.7$ \\
            CIV & 1549 & 3758.75 & $<0.11$ &  $<0.8$ \\
            OIII] & 1660.81 & 4027.93 & 0.09(0.02) & 0.3(0.2) \\
             & 1666.15 & 4040.15 & 0.23(0.03) & 0.7(0.2) \\
            Si III] & 1882.98 & 4566.25 & 0.24(0.03) & 0.7(0.1) \\
            Si III] & 1892.03 & 4588.16 & 0.11(0.02) & 0.3(0.1) \\
 \hline
\end{tabular}
\caption{Rest-UV emission line measurements of CSWA-141. Emission line fluxes are presented relative to the CIII]$\lambda$1909 line flux for the Keck/ESI data, while the line fluxes are presented relative to unresolved combined CIII] flux for LBT/MODS data. The upper limits are 3$\sigma$. }
\label{table:c141_uvlines}
\end{table}

{\it CSWA-141}  is an extreme equivalent width optical line 
emitting galaxy at $z=1.425$ with an integrated apparent magnitude of $r=20.7$.  The  Magellan/FIRE 
spectrum reported in \citet{Stark2013b} shows a large number 
of rest-frame optical emission lines, including the temperature sensitive [OIII]$\lambda$4363 auroral line and 
the density sensitive [SII]$\lambda\lambda$6717,6731 and [OII]$\lambda\lambda$3727,3730 lines  (see Figure~\ref{fig:cswa_nir}).  
We have since obtained deep Keck/ESI and LBT/MODS optical spectra and  optical and near-IR imaging, providing 
constraints on the rest-UV metal lines and the SED. 

The optical to near-infrared SED of CSWA-141 (Figure~\ref{fig:sed_cswa}) implies a very large specific 
star formation rate (sSFR=$31^{+61}_{-16}$ Gyr$^{-1}$).  After correcting for magnification 
due to lensing ($\mu$=5.5), the stellar mass and star formation rate of the best fitting model 
are 3.98$\times$10$^{8}$ M$_\odot$  and 12$^{+24}_{-7}$ M$_\odot$ 
yr$^{-1}$, respectively (Table~\ref{table:sed}).  The flux in the medium-band J2 MOSFIRE filter (covering 1.117 to 1.246 $\mu$m)
is significantly in excess of that in adjacent filters, as expected given the 
contamination by extremely strong [OIII] and H$\beta$ lines.  We use the J2 flux excess to calculate the 
equivalent width from [OIII] and H$\beta$, following the methodology adopted  at 
higher redshift (e.g.,\citealt{Stark2013, Smit2015}).  This approach yields a rest-frame equivalent width of 
W$_{\rm{[OIII]+H\beta}}$ =730~\AA, consistent with the very young stellar populations (32 Myr for constant star 
formation) implied by 
the population synthesis modeling. While the optical line EW is significantly in excess of what is seen in 
typical star forming galaxies at $z\simeq 2$ (e.g.,\citealt{Boyett2022}), it is nearly identical to the average [OIII]+H$\beta$ EW at 
$z\simeq 7-8$, as implied by flux excesses in {\it Spitzer}/IRAC bandpasses \citep{Labbe2013,Smit2015, deBarros2017,Endsley2020a}.  
The continuum brightness of CSWA-141 enables a unique and detailed view of this population.

\begin{figure*}
\centering
\subfloat{ \includegraphics[angle=90,width=0.48\textwidth]{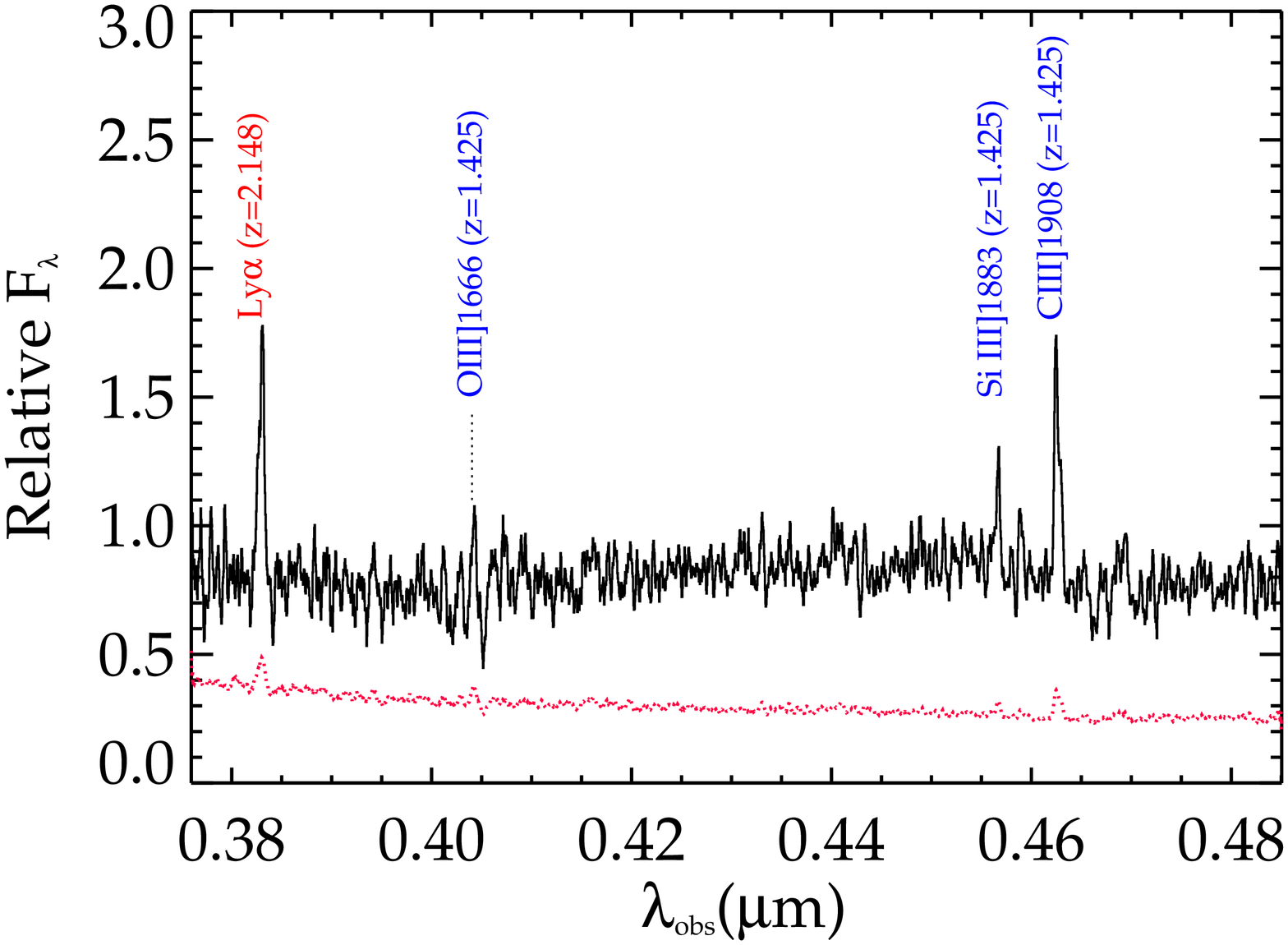}}
\subfloat{ \includegraphics[angle=90,width=0.48\textwidth]{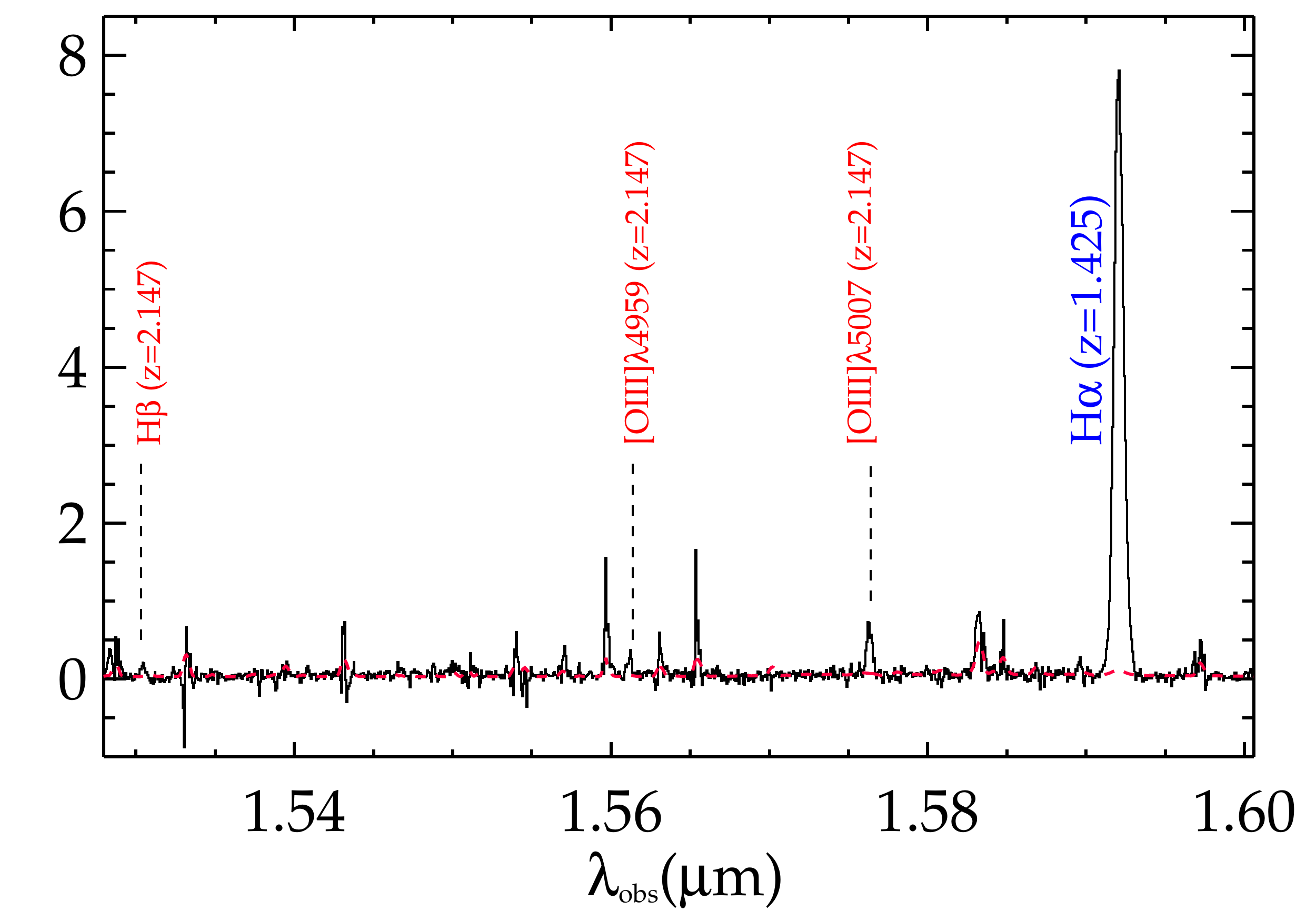}} \\ 
\caption{Optical and NIR spectra of CSWA-141. A secondary source (CSWA-141b) is identified in the Optical+NIR spectra (see \S 3) (Left:) Ly$\alpha$ emission at $z=2.148$ from CSWA-141b in the 
 LBT/MODS spectrum alongside [SiIII]$\lambda$1883 and CIII]$\lambda$1908 from CSWA-141 at 
 $z=1.425$. (Right:)  Rest optical lines ([OIII]$\lambda\lambda$4959,5007, H$\beta$) from CSWA-141b at $z=2.147$ observed in the FIRE spectrum alongside the H$\alpha$ emission line 
 from CSWA-141.}
\label{fig:c141_second}
\end{figure*}

The ESI spectrum covers 4100-10000~\AA, revealing prominent emission from 
[CIII]$\lambda$1907, CIII]$\lambda$1909, Fe II$^\star\lambda$2626, 
Mg II$\lambda\lambda$2797,2804, [OII]$\lambda\lambda$3726,3729, 
[Ne IIII]$\lambda$3869, and H$\delta$ (Table~\ref{table:c141_uvlines}).  
The [CIII], CIII] doublet is easily resolved by ESI (Figure~\ref{fig:esispectra}) with a total 
rest-frame equivalent width of W$_{\rm{CIII]}}$=4.6\AA, the largest of the 
sixteen galaxies considered in this paper.   Nebular Mg II emission also exhibits 
a very large rest-frame equivalent width (W$_{\rm{Mg II}}$ = 15.0~\AA), as is common 
in lower mass galaxies (e.g.,\citealt{Erb2012,Guseva2019}).  The MODS spectrum provides better 
blue sensitivity than ESI, yielding detection of 
OIII]$\lambda$1661,1666 and  Si III]$\lambda\lambda$1883,1892.  The CIII] doublet is also detected, but it is unresolved at the resolution of MODS (Figure~\ref{fig:c141_second}). The summed equivalent widths of the OIII] and Si III] doublet (1.0~\AA\ and 1.0~\AA, 
respectively) are considerably lower than CIII] (see Table~\ref{table:c141_uvlines}).  
The high ionization lines He II$\lambda$1640 and CIV$\lambda\lambda$1548,1550 are not detected, implying rest-frame equivalent widths below 0.3 and 0.5~\AA\ (at 3$\sigma$), respectively. 
For emission lines detected by both ESI and MODS, we will adopt whichever instrument provides the 
higher S/N EW measurement for our subsequent analysis and discussion.  

The rest-frame optical emission lines detected in the FIRE spectrum provide an array of constraints on the 
nebular gas physical conditions.   The detection of [OIII]$\lambda$4363 in the FIRE spectrum enables a 
measure of the nebular electron temperature (T$_{\rm{e}}=1.5\pm 0.1\times10^{4}$ K ) and 
the oxygen abundance via the direct T$_e$ method.  Following the process discussed in \S2.2.3, we
find that $\rm{12 +log(O/H)} = 7.95\pm0.08$, implying a gas-phase metallicity of 0.18 Z$_{\odot}$. Our measurement is very similar to \citet{Sanders2016b} who reported oxygen abundance ($\rm{12 +log(O/H)}$) of CSWA-141 based on our line flux measurements given in Table~\ref{table:opticallines}. The derived metallicity is also consistent with the metallicities derived from the N2 and O3N2 indices with the calibration presented in \citet{Bian2018} ($\rm{12 +log(O/H)} <8.0$).   

The O32 (6.6) and R$_{23}$ indices (11.4) point to a large ionization parameter and gas excitation.  The exact values depend on the input ionizing spectrum. Photoionization modeling using BEAGLE resulted in a large ionization parameter of log $U_{s}$ = -2.1$\pm$0.1, which is broadly consistent with O$_{32}$ vs ionization parameter relationships in the literature (e.g., \citealt{Sanders2016, Berg2019}). The photoionization modeling further suggested a high ionizing photon production efficiency of log ($\rm \xi_{ion}$/Hz erg$^{-1}$) = 25.5$\pm$0.1. This is higher than the canonical value typically assumed for galaxy-driven reionization model \citep{Robertson2010} but consistent with other strong CIII] emitters in the literature (e.g., \citealt{Nakajima2018}). While the measured O$_{32}$ and R$_{23}$ are rare among more massive star-forming galaxies at $z\simeq 2$, 
they are consistent with the O$_{32}$-sSFR and O$_{32}$-R$_{23}$ trends that are observed at high redshift (e.g., \citealt{Sanders2016,Strom2016}). 
 
The electron densities inferred from the flux ratios of the [OII] and [SII] doublets using PyNeb  
(160$^{+76}_{-74}$ cm$^{-3}$ and 350$^{+294}_{-206}$ cm$^{-3}$, respectively) are  consistent with the median density of 
more massive star forming galaxies at $z\simeq 2$ (250 cm$^{-3}$; e.g., \citealt{Sanders2016}).   In contrast, the 
resolved [CIII], CIII] doublet suggests gas at very high density ($16500^{+12100}_{-7800}$cm$^{-3}$).  
Such an offset between densities derived from CIII] and those inferred from [OII], [SII] have been seen 
in other galaxies at high redshift (e.g., \citealt{James2014}). We will come back to discuss this in more detail in \S4.

Finally, we note that the MODS and FIRE spectra reveal emission lines at 3827, 15304, 15612, 15763, 20663~\AA\ 
which appear nearly spatially 
coincident with CSWA-141 but are not  associated with the $z=1.425$ galaxy (see Figure~\ref{fig:c141_second}).  
We identify these as Ly$\alpha$, H$\beta$, [OIII]$\lambda\lambda$4959,5007, and H$\alpha$ in a second fainter 
gravitationally lensed source at $z=2.148$.  This higher redshift galaxy is  unresolved from the $z=1.425$ source in 
existing ground-based optical and near infrared images.   Given that the H$\alpha$ flux of the $z=1.425$ galaxy is 7.4$\times$ 
larger than the $z=2.148$ source, we expect that the newly-discovered higher redshift source contributes negligibly (at the 5\% level) 
to the broadband flux and the rest-UV continuum in the MODS and ESI spectra.  Higher resolution imaging will be required to disentangle the two sources.  

{\it CSWA-13} is a bright galaxy at $z=1.87$ that was first confirmed in \citet{Stark2013b} through the detection of 
Ly$\alpha$ emission and numerous interstellar absorption lines in an MMT blue channel spectrum.  
CIII] emission is confidently detected (W$_{\rm{CIII]}}$=$4.4\pm 0.9$~\AA) 
in the discovery spectrum.   He~II is also detected (W$_{\rm{He II}}$=$4.3\pm 0.9$~\AA) with 
a broad FWHM (2430 km s$^{-1}$) that is indicative of a stellar wind origin.  We do not detect 
the OIII] doublet, implying individual components with equivalent widths less than  
0.8~\AA.    While this is among the strongest CIII] emitters in our sample, the redshift 
places the strong rest-optical lines in regions of poor atmospheric transmission.   
We have obtained multi-wavelength imaging in the 
optical and near-IR (Figure \ref{fig:sed_cswa}).   The SED reveals  a large sSFR (43.1 Gyr$^{-1}$), little 
dust attenuation ($\hat{\tau}_V$=0.53), and a magnification-corrected stellar mass of log (M$_{\star}$/M$_{\odot}$) = 9.8.

 {\it CSWA-139} was confirmed to have a redshift of $z=2.54$ in \citet{Stark2013b} 
 based on the presence of Ly$\alpha$ absorption and interstellar metal absorption lines in an MMT spectrum. 
 The spectrum also shows emission from the [CIII],CIII]$\lambda\lambda$1907,1909 doublet 
 with a total equivalent width of 
W$_{{\rm CIII]}}=3.4\pm 2.6$\AA.   Here we present new optical and near-IR imaging from the LBT and MMT.  
The SED is best fit with an sSFR of 2.9 Gyr$^{-1}$, $\hat{\tau}_V$=0.77, and a stellar mass of log (M$_{\star}$/M$_{\odot}$) = 10.3 after 
magnification correction.

 {\it CSWA-2} (SDSS J1038+4849) was first reported in \citet{Belokurov2009}, and the source redshift ($z=2.20$) was subsequently confirmed in \citet{Jones2013} through detection of rest-optical emission lines.   
 The lens reconstruction described in  \citet{Jones2013} shows that CSWA-2 is a merger of two 
 systems with a stellar mass ratio ($6\pm3$):1.   
log (M$_{\star}$/M$_{\odot}$)  = $9.1^{+0.2}_{-0.1}$ and one of the largest sSFR (19.9 Gyr$^{-1}$). 
The oxygen abundance 12+log(O/H) of the system as 
calculated by using N2 index is 8.25 ($\sim$0.4 Z$_{\odot}$). The Balmer decrement ratio (H$\alpha$/H$\beta$=3.47) suggests little nebular extinction 
whereas O3 of 1.80 imply relatively larger excitation from ionized gas. Using the metallicity calibration of \citet{Bian2018}, we estimate the oxygen abundance 
12+log(O/H) of the system using the O3 index as 8.4 ($\sim$0.5 Z$_{\odot}$)
 In this paper, we present new  optical spectroscopy of this system obtained with ESI.  
 The spectrum is dominated by continuum 
 emission from the lower mass source (denoted J1038 North in \citet{Jones2013} which is considerably 
 brighter in the optical.   The spectrum shows  emission from   [CIII],CIII]$\lambda\lambda$1907,1909 with a total equivalent width of W$_{\rm CIII]}$=$3.1\pm1.8$~\AA.

{\it CSWA-39} (SDSS J1527+0652), a bright ($r=20.5$) $z=2.759$ galaxy was first identified as 
part of the SDSS Giant Arcs Survey (SGAS; \citealt{Hennawi2008}) and was spectroscopically 
confirmed by \citet{Koester2010} through detection of Ly$\alpha$ emission and interstellar 
absorption lines.   We have obtained an ESI optical spectrum and multi-color optical imaging.  
Both components of the [CIII], CIII] doublet are detected in the spectrum, with a total S/N=16.1
for the summed doublet flux.   The rest-frame equivalent width (W$_{\rm CIII]}$=1.1~\AA) is 
typical of similarly luminous galaxies at this redshift (e.g., \citealt{Shapley2003, Du2017}).  
We also detect Ly$\alpha$ emission with a 
moderate rest-frame equivalent width (W$_{\rm Ly\alpha}$=$11.8\pm6.2$\AA).  No other nebular 
UV lines are detected in the ESI spectrum.  The upper limit on the OIII]$\lambda\lambda$1661,1666 
components imply equivalent widths below 1~\AA, consistent with the OIII] emission strengths in the 
\citet{Shapley2003} composite of $z\simeq 3$ galaxies with similar Ly$\alpha$ equivalent widths.

 {\it CSWA-38} (SDSS J1226+2152) was confirmed to lie at $z=2.923$ by detection of Ly$\alpha$ 
and metal absorption lines in \citep{Koester2010}.   Like CSWA-39, this source was first identified in 
Sloan Giant Arcs survey.   Here we present deep ESI spectroscopy and imaging.   The optical 
spectrum reveals a 4.1$\sigma$ detection of the CIII]$\lambda$1909 component, but the [CIII]$\lambda$1907 
component is situated on a skyline, precluding a useful limit.   The rest-frame equivalent width of the CIII]$\lambda$1909 component (W$_{\rm CIII]\lambda1909}$=0.4~\AA) is comparable to CSWA-38 and CSWA-19.  
Ly$\alpha$ emission is weak (W$_{\rm Ly\alpha}$=0.4~\AA), and no other nebular UV lines are detected.
 
{\it CSWA-19}  (SDSS J0900+2234) was first confirmed in \citep{Diehl2009} at  $z=2.03$ via detection 
of Ly$\alpha$ emission and several metal absorption features in a spectrum from the dual imaging 
spectrograph (DIS) on the Astrophysical Research Consortium (ARC) 3.5 meter telescope at the Apache 
Point Observatory.  We have since obtained a deep ESI spectrum  and MMT near-infrared imaging of this source.  
The combined optical and near-infrared SED is best-fit by a model with a (magnification-corrected) 
stellar mass of log (M$_{\star}$/M$_{\odot}$) = 10.5, a specific star formation rate of 0.5 Gyr$^{-1}$, and  
V-band attenuation optical depth of $\hat{\tau}_V$=1.09.   The ESI spectrum of the source shows weak detections 
of [CIII], CIII] $\lambda\lambda$1907, 1909 with an integrated S/N=8.6 across both components of the doublet.  
The total rest-frame equivalent width (W$_{\rm CIII]}$=0.7~\AA) is among the lowest in our sample.   
No other nebular rest-UV lines are detected, as Ly$\alpha$ is situated to the blue of the ESI coverage.


 {\it CSWA-128} (SDSS J1958+5950) is a bright (r=20.7) z=2.22 galaxy that was spectroscopically 
 confirmed in \citet{Stark2013b}  through detection of rest-optical emission lines in an LBT/LUCI 
near-infrared spectrum  and interstellar metal absorption lines in an MMT optical spectrum.  
We have since obtained  optical and near-infrared imaging and a deep ESI optical spectrum.  
The magnification-corrected SED (Figure~\ref{fig:sed_cswa}) implies a  stellar mass of log 
(M$_{\star}$/M$_{\odot}$) = $9.9^{+0.3}_{-0.3}$  and an sSFR of 1.0 Gyr$^{-1}$. 
The ESI spectrum reveals a faint 4.8$\sigma$ detection of 
[CIII], CIII] emission at the systemic redshift ($z=2.225$) defined by the 
rest-optical lines, implying a rest-frame equivalent width of W$_{\rm CIII]}$=0.7~\AA\ for the 
doublet.  No other nebular lines are detected in the rest-UV.   

The flux ratios of rest-optical lines (Table~\ref{table:opticallines}) constrain the gas physical conditions, providing a framework to 
understand the weak UV nebular line emission.  We infer the gas-phase oxygen abundance using 
the R23, O32 and O3 calibration presented in \citet{Bian2018}.  
The three indices suggest metallicities of 12+log O/H = 8.2 (O32), 12+log O/H=8.4 (R23) and  
12+log O/H=8.5 (O3).  As discussed in \citet{Sanders2020}, O32 metallicity calibration in \citet{Bian2018} may provide a better estimate of oxygen abundance for high redshift galaxy. 
This suggests that the metallicity of the ionized gas of CSWA-128 is $\sim$0.3-0.4 Z$_\odot$.   The values of O32  
(3.3) and R23 (7.0)  are  a factor of 2.0 and 1.6 smaller than that of the strong CIII] emitter CSWA-141, consistent 
with expectations for nebular gas with a lower ionization parameter and excitation.  
In contrast, the electron density implied by the [SII]$\lambda\lambda$6717,6731 
doublet (200 cm$^{-3}$) is similar to that found in both strong CIII] emitters such as CSWA-141 and 
typical $z\simeq 2-3$ galaxies \citep{Sanders2016}.

{\it CSWA-103} (SDSS J0145-0455) is a $z=1.96$ galaxy that was confirmed in \citet{Stark2013b} through 
detection of metal absorption lines and weak Ly$\alpha$ emission in an MMT blue channel spectrum. 
We have since obtained a moderate resolution Keck/ESI spectrum and optical and near-infrared imaging with LBT and MMT, respectively.  
The broadband SED (Figure~\ref{fig:sed_cswa}) is best fit by a stellar synthesis model with a stellar mass (magnification corrected)  
of log (M$_{\star}$/M$_{\odot}$) = 10.4, sSFR=0.7 Gyr$^{-1}$, and  $\hat{\tau}_V$=0.48. The ESI spectrum constrains rest-frame 
wavelengths 1350-3425~\AA.   Positive emission is detected from both components of the  [CIII], CIII] doublet, 
implying a total rest-frame equivalent width of 0.5~\AA.  No other nebular emission lines are observed in the ESI spectrum.

{\it CSWA-164} (SDSS J0232-0323)  was spectroscopically confirmed in \citet{Stark2013b} 
based on the presence of Ly$\alpha$ emission and interstellar absorption lines in an MMT blue channel spectrum.   
\citet{Stark2013b} also presented detection of [OII], [OIII]$\lambda$5007, and H$\alpha$ in a Magellan/FIRE near-infrared 
spectrum of CSWA-164, revealing a systemic nebular redshift of $z=2.512$.      We have 
since obtained a deep moderate resolution optical spectrum with ESI and optical broadband imaging.  
The ESI spectrum reveals the  Ly$\alpha$ emission (W$_{Ly\alpha}=2.0$~\AA) detected previously together with weak 
emission (S/N=4.1) from the [CIII],CIII]$\lambda\lambda$1907,1909  doublet.   This corresponds to a total 
CIII],CIII] equivalent width of just 0.4~\AA, the smallest measured value in our sample.

The FIRE spectrum provides insight into the ionized gas physical conditions of CSWA-164.  
Unfortunately both H$\beta$ and [NII] are obscured by skylines, precluding a robust 
determination of the oxygen abundance through standard strong line calibrations.  We can however 
estimate oxygen abundance using O32 metallicity calibrations from \citet{Bian2018}. We 
find that O32=0.7 corresponds to the oxygen abundance of 12+log O/H=8.6.  Thus 
the ionized gas appear to be reasonably metal rich in CSWA-164.  
The value of ionization-sensitive line ratios (O32 = 0.7) is among the lowest in our sample. 
In contrast to the metallicity and ionization parameter, the electron density derived from the [OII] doublet (165 cm$^{-3}$) is consistent 
with the range spanned by other galaxies in our sample.  

{\it CSWA-40} (SDSS J0952+3434) is a $z=2.190$ galaxy identified through the 
Sloan Bright Arcs Survey and spectroscopically confirmed by \citet{Kubo2010} using the DIS on the 
ARC 3.5 meter and the RC Spectrograph on the Mayall 4 meter telescope at Kitt Peak National Observatory.  
The  spectra reveal Ly$\alpha$ in absorption along with several other metal absorption lines.  
We have obtained a moderate resolution optical spectrum and imaging with Keck/ESI.  The  
continuum S/N of the CSWA-40 ESI spectrum is lower than other systems.   This is the only ESI 
spectrum that does not reveal detection of the [CIII], CIII] doublet, implying a rest-frame equivalent width below 1.6~\AA\ for the sum of both 
components.  No other nebular emission lines are observed in the spectrum.  

{\it CSWA-163} (SDSS J2158+0257) was spectroscopically confirmed in \citet{Stark2013b} 
based on identification of metal absorption lines and Ly$\alpha$ absorption in an MMT 
blue channel spectrum.  A Magellan/FIRE near-infrared spectrum was also obtained in that paper, 
revealing a systemic nebular redshift of $z=2.079$ based on detection of 
[OII], H$\gamma$, H$\beta$, [OIII], H$\alpha$, and [NII].  
We have obtained new optical and near-infrared imaging of this source, allowing constraints 
to be placed on the broadband SED (Figure~\ref{fig:sed_cswa}). The data are best fit by stellar synthesis 
models with stellar mass log (M$_{\star}/M_\odot$) = $9.9^{+0.1}_{-0.1}$  after magnification 
correction, V-band attenuation optical depth of $\hat{\tau}_V$=0.67, and an sSFR of 1.8 Gyr$^{-1}$, suggesting  
CSWA-163 is a relatively massive galaxy with a fairly evolved stellar population.  Perhaps not surprisingly 
given the low sSFR and lack of Ly$\alpha$ emission, the MMT blue channel spectrum does not show 
any CIII] emission at the systemic redshift.  The 3$\sigma$ flux upper limit  suggests that the rest-frame 
equivalent width of the double must be lower than 3.1~\AA. 

The FIRE spectrum provides useful constraints on the ionized gas of CSWA-163.   The oxygen 
abundance can be derived from  O32 and O3 calibration from \citet{Bian2018}. Both suggest 
moderately enriched gas: 12+log O/H=8.5 (O32) and12+log O/H=8.4 (O3)
implying a nebular oxygen abundance of 0.4-0.5 Z$_\odot$.   The ionization-sensitive line 
ratios (O32=1.1, O3=4.3) are consistent with a relatively low ionization parameter and 
moderate gas-excitation.   The electron density inferred from [OII] is 144 cm$^{-3}$, consistent 
with the other systems studied in this paper.  

{\it CSWA-16} (SDSS J1111+5308) was confirmed to have a redshift of $z=1.95$ based on the presence 
of numerous metal absorption lines in an MMT blue channel spectrum \citep{Stark2013b}.   The discovery spectrum 
shows no emission from the blended [CIII], CIII] doublet.  The 3$\sigma$ upper limit on the total flux 
from the doublet requires the rest-frame equivalent width to be smaller than 2.3~\AA.   We have 
obtained multi-band optical imaging with LBT, allowing more robust constraints to be placed on the 
apparent magnitude (r=21.8) and the magnification-corrected UV absolute magnitude (M$_{\rm{UV}}$=$-20.8$).

{\it CSWA-165} (SDSS J0105+0144) is a $z=2.13$ galaxy  that was first confirmed in \citet{Stark2013} through 
detection of strong metal absorption lines and weak Ly$\alpha$ emission in an MMT 
blue channel spectrum.   We have since obtained a Magellan/FIRE near-infrared spectrum 
and optical and near-infrared imaging from LBT and MMT respectively.   The FIRE 
spectrum reveals detection of [OII], H$\beta$, [OIII]$\lambda$5007, H$\alpha$, and [NII], 
indicating a  systemic redshift of $z=2.128$.  The MMT shows no emission from the [CIII], CIII] 
doublet at the systemic redshift.    The 3$\sigma$ upper limit on the flux of the doublet indicates 
that the total CIII] equivalent width must be lower than 1.3~\AA.  The broadband SED is best-fit by 
a synthesis model with stellar mass of  log(M$_{\star}$/M$_{\odot}$) = 10.0, sSFR of 3.9 Gyr$^{-1}$, 
and $\hat{\tau}_V$=0.62.

The gas-phase metallicity can be inferred from O32, R23, and O3 strong line calibrations. All three indicate 
metal rich ionized gas: 12+log O/H = 8.4 (O32) 8.6 (R23), 8.6 (O3), consistent with a gas-phase metallicity in the range 
0.5-0.8 Z$_\odot$.  The ionization-sensitive ratios (O32=0.6, O3=2.1) suggest an ionization parameter that is lower than 
average among $z\simeq 2-3$ galaxies.  In contrast, the electron density derived from the [OII] doublet flux ratio (220 cm$^{-3}$) is 
similar to that found in other systems at high redshift.

{\it CSWA-11} (SDSS J0800+0812) was confirmed at $z=1.41$ via detection of metal absorption lines and 
[OII] emission in  MMT blue and red channel spectra \citep{Stark2013b}.  
The CIII] doublet is not detected in either the blue or red channel MMT spectra, implying  a rest-frame 
equivalent width below 0.9~\AA\ at 3$\sigma$.   We have obtained  multi-wavelength 
imaging and near-infrared spectroscopy. The magnification-corrected SED  suggests a stellar mass of 
log(M$_{\star}$/M$_{\odot}$) = $10.0^{+0.4}_{-0.3}$, an sSFR of 7.4 Gyr$^{-1}$, and V-band attenuation optical depth of $\hat{\tau}_V$=0.28. 
The FIRE spectrum reveals detections of [OII], [OIII]$\lambda$5007, 
and H$\alpha$, but strong skylines obscure detections of H$\beta$, [OIII]$\lambda$4959 and [NII]$\lambda$6586.
Using the theoretically-expected flux ratio of [OIII]$\lambda$5007/[OIII]$\lambda$4959, we infer that this system has 
O32 = 0.92, relatively low for our sample. Using the O32 calibration, we estimate metallicity of 12+log O/H=8.5.  
The electron density implied by the [OII] doublet flux ratio is 469$^{+416}_{-238}$ cm$^{-3}$, consistent with the 
range spanned by other galaxies in our sample.  

{\it CSWA-116} (SDSS J0143+1607) is a $z=1.50$ galaxy confirmed in \citet{Stark2013b} through 
detection of rest-UV metal absorption lines and [OII] in MMT blue and red channel spectra.  We do 
not detect the CIII] doublet in the MMT spectra, indicating that the rest-frame equivalent width 
is below 0.7~\AA\ at 3$\sigma$.    We have obtained optical and near-infrared imaging of CSWA-116, 
providing constraints on the broadband SED. The data are best fit by a stellar model with     
log(M$_{\star}$/M$_\odot$) = $9.4$, an sSFR of 1.3 Gyr$^{-1}$, and V-band attenuation optical depth of $\hat{\tau}_V$=0.27.

\begin{figure*}
\centering\
\subfloat{\includegraphics[angle=90,scale=0.43]{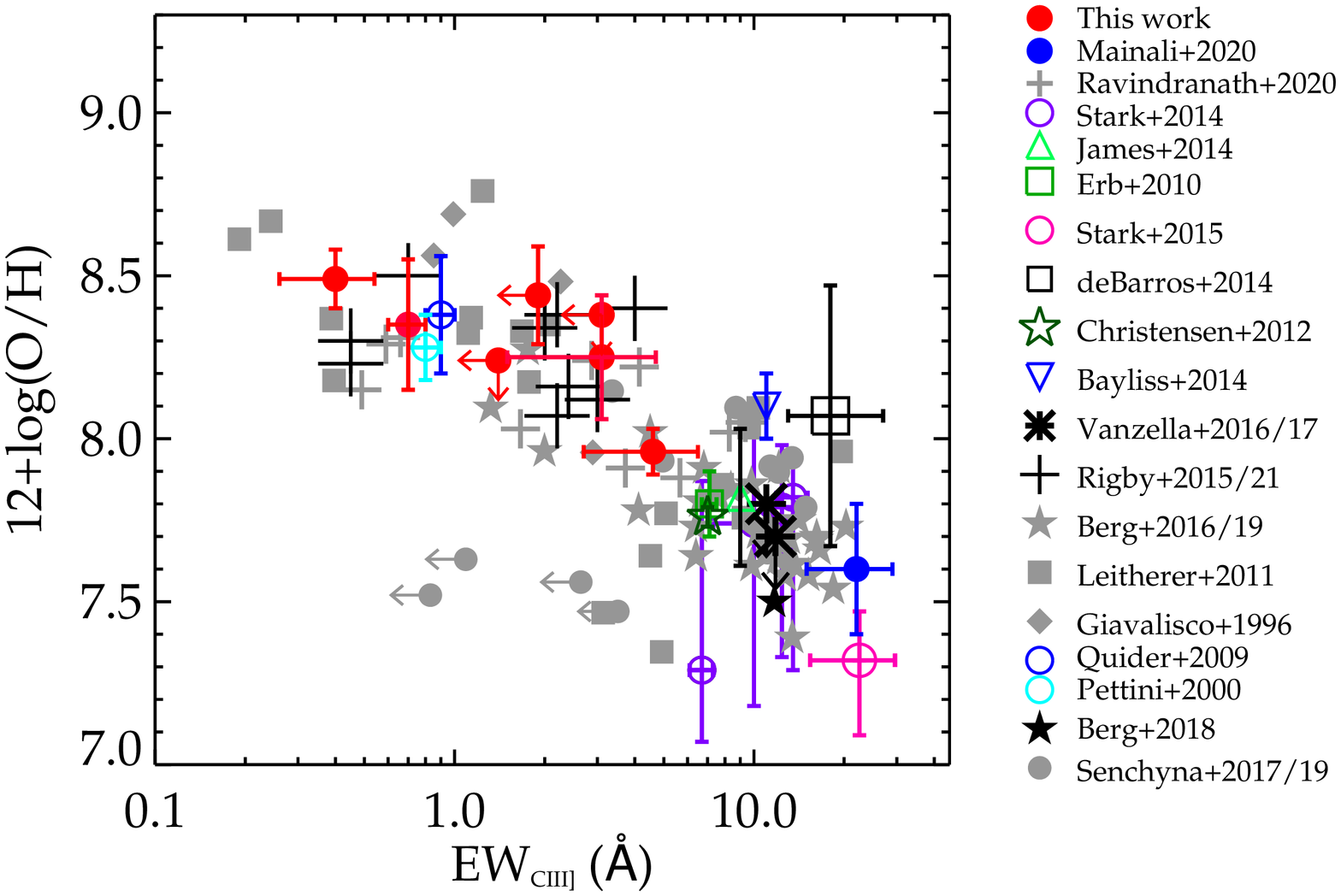}}
\subfloat{\includegraphics[angle=0,scale=0.37]{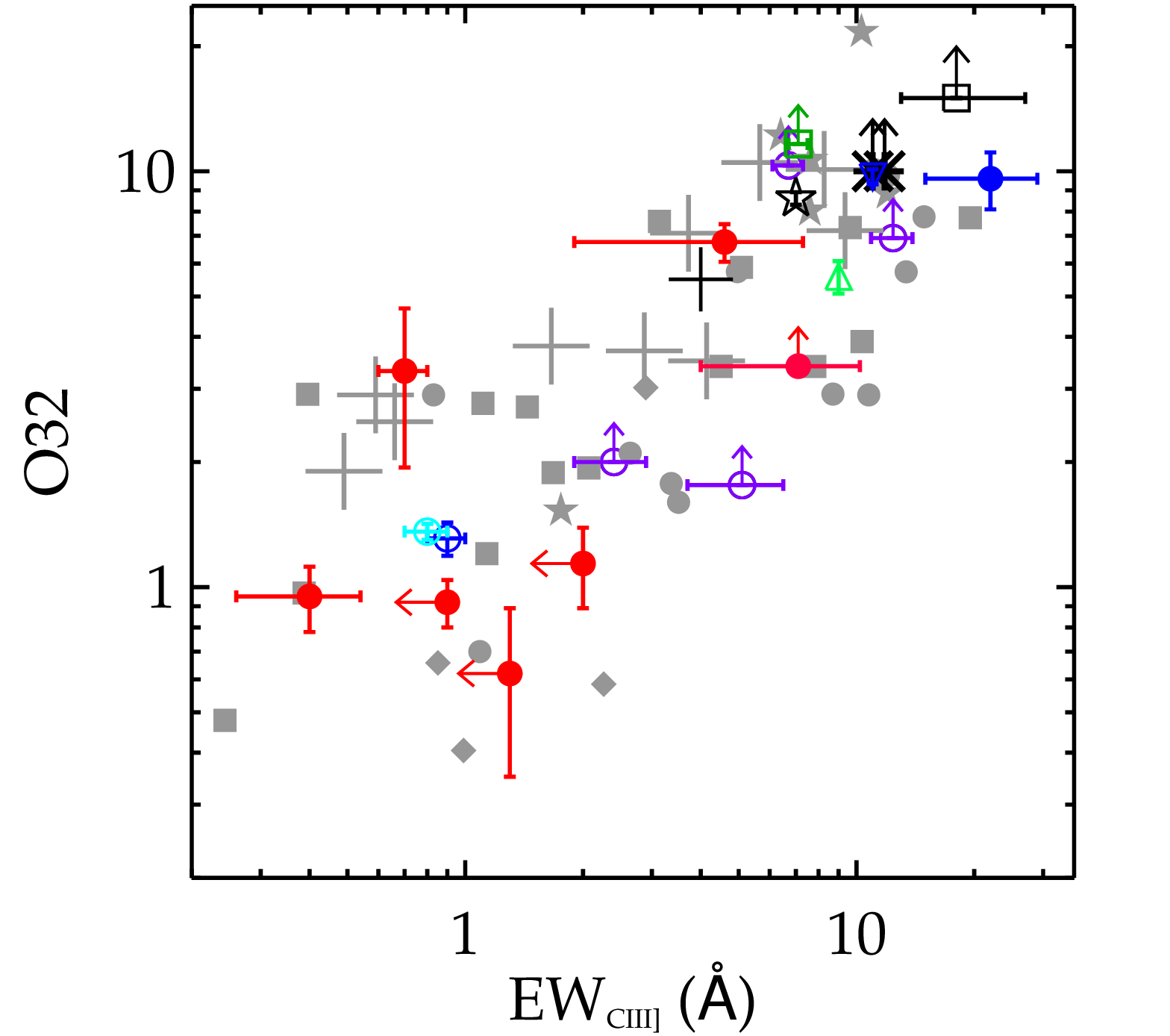}}\\
\caption{(Left:) Empirical relationship between oxygen abundance (12+log(O/H)) and rest frame CIII] equivalent width (EW$\rm_{CIII]}$). (Right:) Empirical relationship between the line ratio of [OIII]$\lambda\lambda$4959,5007 to [OII]$\lambda\lambda$3727,3729 (i.e. O32) and rest-frame CIII] EW (EW$\rm_{CIII]}$). The red symbol represents bright lensed galaxies presented in this paper, while other data points are compilation from the literature.}
\label{fig:ciiivsmetallicity}
\end{figure*}

\subsection{Physical Properties of CASSOWARY galaxy sample}

The data obtained for this paper provide new constraints on the ionized gas 
properties and the stellar populations of lensed galaxies at $z\simeq 1-3$ identified by the CASSOWARY selection in 
SDSS.  Here we briefly describe what these data reveal about the average properties in this sample, providing a 
concise summary of the source-by-source description presented in \S3.1.  
Following correction for magnification, the rest-frame UV absolute magnitudes are found to range between 
M$_{\rm{UV}}$=$-$20.2 and M$_{\rm{UV}}$=$-$23.0, with a median value (M$_{\rm{UV}}$=$-$21.9) that is roughly 
three times the value of L$^\star_{\rm{UV}}$ at $z\simeq 2$ (e.g., \citealt{Reddy2009}).  
Both optical and near-infrared imaging exist for 11 of the 16 galaxies considered in this paper, allowing 
the stellar content to be characterized through SED fitting.  The median stellar mass and sSFR of 
this subset is 1.3$\times$10$^{10}$ M$_\odot$ and 2.1 Gyr$^{-1}$, respectively.  The latter is similar to the average sSFR of $z\simeq 2-3$ galaxies (e.g., \citealt{Reddy2009}).

Rest-optical line measurements exist for seven of the CASSOWARY galaxies 
shown in Figure~\ref{fig:ciiimosaic}.  The median gas-phase oxygen abundance derived from 
the rest-optical line ratios is 12+log O/H = 8.33, i.e., ionized gas metallicity of 0.4 Z$_\odot$. 
The ionization-sensitive ratio O32 ranges between 0.7 and 6.7 with a median value of O32=1.1.  
While this is slightly lower than the median O32 in the KBSS and MOSDEF surveys, it is 
well within the range spanned by galaxies in these samples (e.g., \citealt{Sanders2016,Strom2016,Steidel2016}).   The Balmer decrements range between H$\alpha$/H$\beta$=3.47 and 4.97.
The median electron density derived from  the [OII] or [SII] doublet flux ratios (198 cm$^{-3}$) is close to 
the average electron density of $z\simeq 2.3$ galaxies from the MOSDEF survey \citep{Sanders2016}.

All sixteen galaxies in our sample have constraints on rest-UV line emission.   Most sources 
show very weak [CIII], CIII] emission.  The median equivalent width for the summed doublet 
is 1~\AA, similar to that seen in composite spectra of $z\simeq 3$ LBGs \citep{Shapley2003, Llerena2021}.  
The two strongest [CIII], CIII] emitters (CSWA 141, CSWA -13) have rest-frame equivalent widths 
of 4.6~\AA\ and 4.4~\AA, respectively.   The rest-UV spectrum of CSWA-141 also shows prominent nebular 
emission from OIII], Si III] and Mg II.  While these lines are absent in CSWA-13, its spectrum does 
reveal broad He II emission, indicative of a significant Wolf Rayet 
population.  
No high ionization nebular lines are detected in our sample.  

Our sample allows to compare and contrast ISM conditions expected in strong and weak  CIII] emitters.  A typical CIII] equivalent widths of star-forming galaxies at $z\sim2-3$ is $\sim$1.7~\AA\ \citep{Shapley2003, Du2017}. From here on, we refer to those galaxies with CIII] equivalent widths $>2\times$ 
typically seen at $z\sim2-3$ ($\gtrsim3.5$~\AA) as strong CIII] emitters, whereas objects with CIII] equivalent widths below this threshold are described as weak CIII] emitters. 
Together with galaxies presented in 
this paper, we compile sources from the literature having constraints from both CIII] equivalent widths and optical spectra. The majority of the literature objects are lower redshifts galaxies \citep{Giavalisco1996,Leitherer2011,Berg2016,Berg2019,Senchyna2017,Senchyna2019,Ravindranath2020}, while majority of sample at $z>1$ are comprised of gravitationally lensed systems \citep{Mainali2020,deBarros2016,Stark2014,Vanzella2017,Berg2018,Bayliss2014,Vanzella2016,James2014,Erb2010,Christensen2012,Rigby2021,Quider2009,Hainline2009,Pettini2000,Teplitz2000,Jones2013}. We present a compilation of $z>1$ sources in Table~\ref{table:compilation}.

\begin{figure*}
\centering
\subfloat{\includegraphics[scale=0.3]{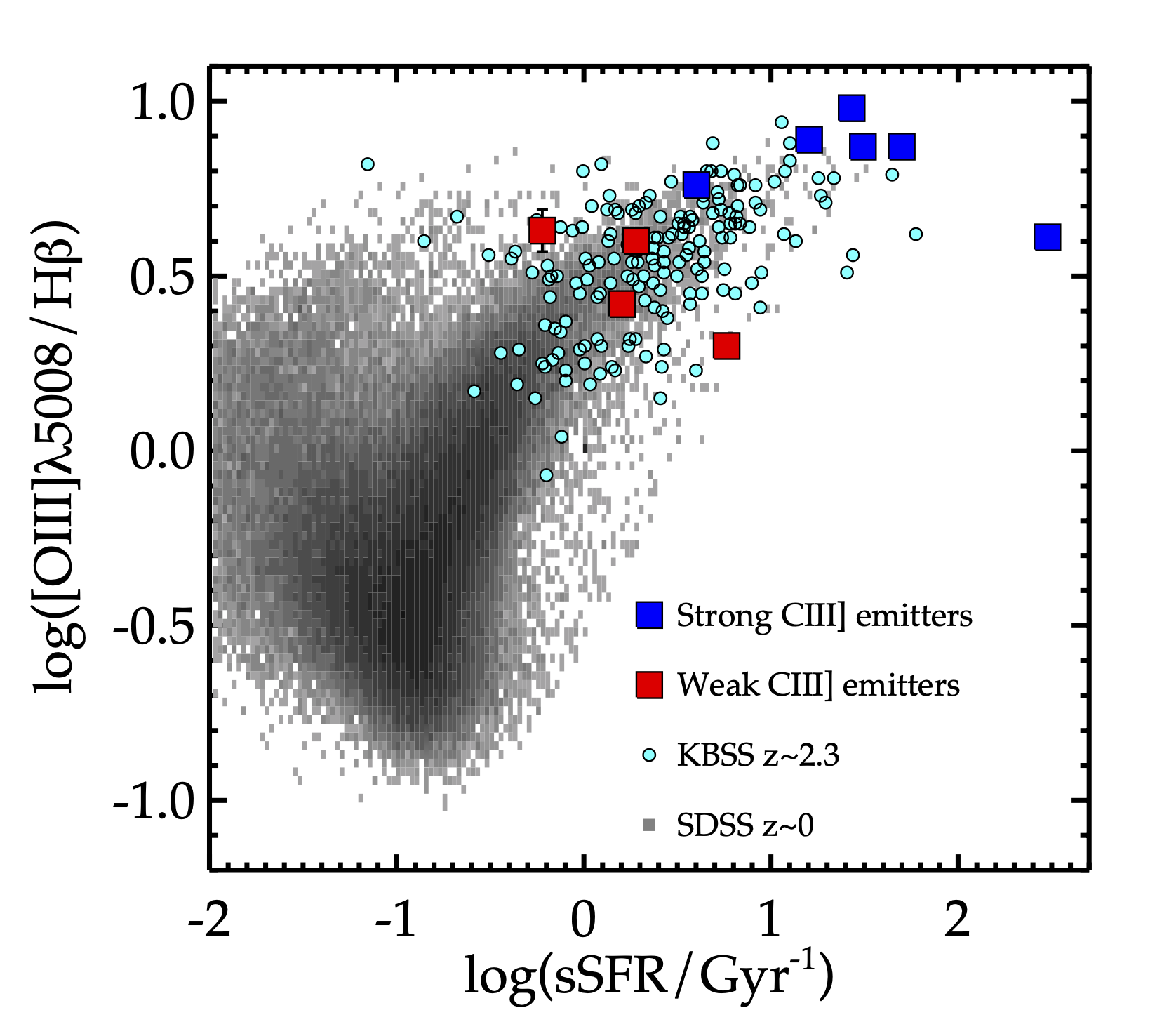}}
\subfloat{\includegraphics[scale=0.265]{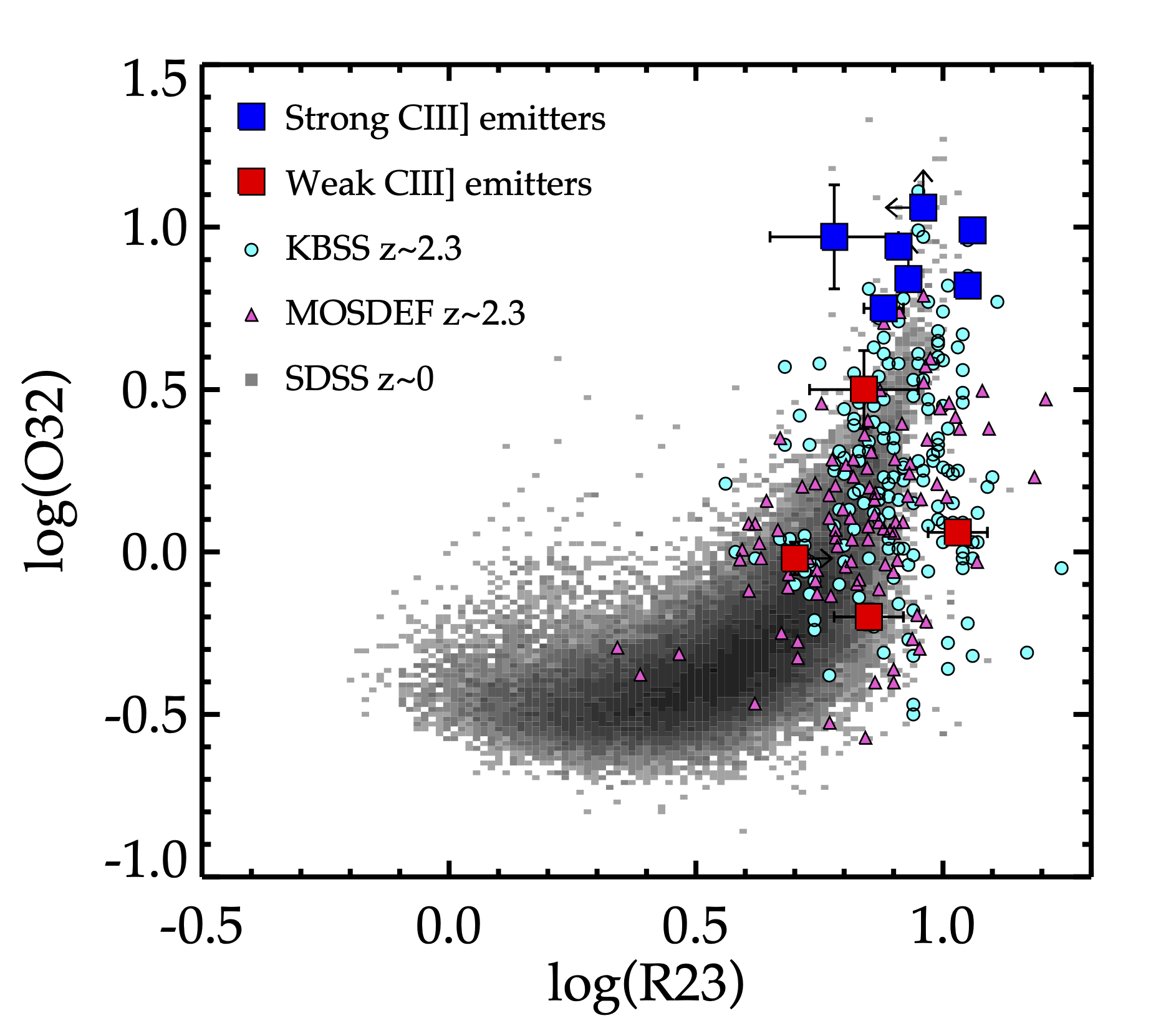}}\\
\caption{The strong (blue) and weak (red) CIII] emitters (see \S3.2) in O3 vs sSFR (left) and O32 vs R23 (right) plots. Star forming galaxies at $z\sim2.3$ from KBSS survey \citep{Strom2016} are shown in cyan circle whereas MOSDEF survey \citep{Sanders2016} are shown in violet triangle. Local galaxies from SDSS survey are shown in grey. Galaxies with properties similar to those at z$>$6 tend to have strongest C~III] emission.}
\label{fig:r23vso32}
\end{figure*}

In Figure~\ref{fig:ciiivsmetallicity}, we plot empirical relationships between CIII] equivalent width and oxygen abundance (left panel), and CIII] equivalent width and O32 (right panel). The oxygen abundance of the strong CIII] emitters are mostly measured using direct metallicity measurements, while strong optical line calibrations are used for the weaker CIII] emitters (see Table~\ref{table:compilation}). As can be seen in the Figure~\ref{fig:ciiivsmetallicity}, the gas-phase metallicity of the strong CIII] emitters is consistently below 12+log(O/H)=8.0, suggesting metallicities below 20\% Z$_{\odot}$. In contrast, most of the weaker CIII] emitters imply higher metallicities than this value. This result is consistent with previous investigations in the literature (e.g., \citealt{Rigby2015,Jaskot2016,Maseda2017,Nakajima2018,Schaerer2018,Senchyna2019,Ravindranath2020,Du2020,Tang2021a}). The right panel of  Figure~\ref{fig:ciiivsmetallicity} shows that the O32 values of the stronger CIII] emitters ($\gtrsim3.5$~\AA) are on an average six times larger than the those of the weaker CIII] emitters. The median O32 value of the weaker CIII] emitters are similar to typical galaxies at $z\sim2-3$ (O32=1.3,\citealt{Sanders2016}). 
A variety of factors can influence the O32 value of a galaxy. The elevated values seen in the strongest of CIII] emitters tend to be found in galaxies dominated by the 
light of a very young stellar population \citep{Tang2021a, Sanders2020}, as expected in galaxies undergoing a burst of star formation. Overall these empirical relationships support a physical picture where galaxies with metal poor gas and 
young stellar populations are able to power strong CIII] emission \citep[e.g.][]{Stark2014,Rigby2015,Senchyna2017}. The scatter seen in the CIII] EW at a given metallicity or O32 value may stem from differences in ISM conditions (relative carbon abundances (C/O), ionization parameters) and stellar age and metallicity.  

The compilation of strong CIII] emitters further provides information on the global spectral  properties expected from typical reionization era systems. In Figure~\ref{fig:r23vso32}, we show strong CIII] emitters (denoted by red square) and weak CIII] emitters (denoted by blue square) in log([OIII]/H$\beta$) vs sSFR (left panel) and 
log(O32) vs log(R23)  (right panel) plots. We compare them to typical line ratios expected at $z\sim2$ using dataset from KBSS survey \citep{Steidel2014,Strom2016} and MOSDEF survey \citep{Sanders2016}, as well as at $z\sim0$ (local SDSS galaxies). The strong CIII] emitting galaxies appears to be distinct from local SDSS galaxies as well as typical galaxies at $z\sim2$ in both the plots. 
However, the position of weaker CIII] emitters on both the diagrams is similar to typical galaxies at $z\sim2$. Taken together, this further demonstrates that strong CIII] emitters show highly ionized gas conditions from a large sSFR systems. Assuming these strong CIII] emitters representative of a typical reionization era systems, we might expect a similar ISM conditions in galaxies at $z>6$.

\section{Discussion}
 \label{sec:discussion}


In this paper, we present spectra of some of the brightest-known 
gravitationally-lensed galaxies at $z\simeq 2-3$, discovered over the 
footprint of SDSS. Included in this sample are CSWA-13 and CSWA-141, two exceptionally bright systems with sSFRs ($>20$ Gyr$^{-1}$) that are similar to those of the  reionization-era. Their spectra reveal rest-frame CIII] equivalent widths more than twice what is typical at $z\simeq 2-3$ (e.g., \citealt{Shapley2003,Du2018,Llerena2021}).
In this section, we explore the ionized gas conditions and the properties of the 
outflowing gas, taking advantage of emission and absorption 
lines that are often too faint to be detected in individual high redshift 
galaxies with similarly intense emission lines. Our analysis will primarily focus on CSWA-141, as the redshift of CSWA-13 places the strong rest-optical lines 
in regions of low atmospheric transmission.  For CSWA 141, the rest-optical lines 
are similar in equivalent width to those found in the reionization era 
(e.g., \citealt{deBarros2017,Endsley2020a,Boyett2022}), classifying this galaxy 
as an EELG and revealing a stellar population dominated by a recent burst 
or upturn in star formation. 

The gas properties in such intense line emitting galaxies have been the 
subject of a number of spectroscopic investigations in recent years. These studies demonstrate that the nebular gas is generally under extreme ionization 
conditions in EELGs \citep{Tang2018}, with the ionization parameter reaching its 
largest values in the galaxies powered by the youngest 
stellar populations (or  the largest equivalent width 
rest-optical nebular lines). The gas in CSWA-141 is consistent with this 
picture, showing a large O32 ratio (6.7) as expected given its elevated sSFR (31.2 Gyr$^{-1}$) and [OIII]+H$\beta$ EW (730~\AA). Its large ionization parameter implied 
by the photoionisaiton models (log U = -2.1) suggests a large photon density is impingent 
on the ionised gas. This can be driven be a variety 
of factors, including efficient ionizing photon production (e.g., \citealt{Chevallard2018, Tang2021a}) and a compact configuration of ionized gas around the sources of ionizing radiation. The latter is the norm for galaxies dominated by very young stellar clusters 
(e.g., \citealt{Whitmore2011,Hannon2019,Chen2023}), 
and thus may be expected in EELGs like CSWA-141.

\begin{figure}
\centering
\subfloat{\includegraphics[scale=0.36]{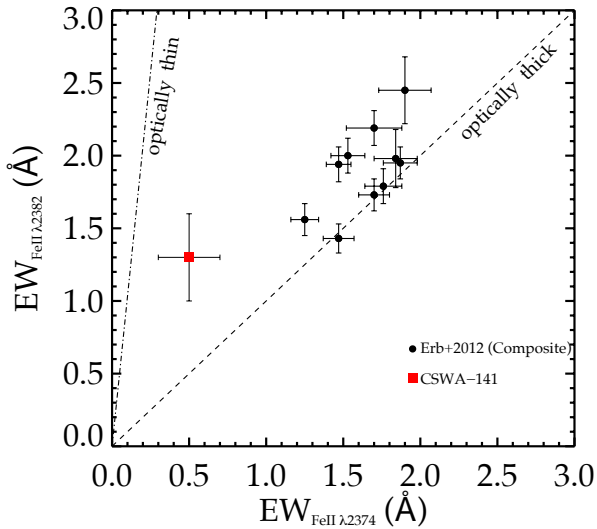}}
\caption{A comparison between equivalent widths of Fe~II $\lambda$2374 and Fe~II $\lambda$2382. The red square represents CSWA-141 data point whereas black circles are composite data points presented in \citealt{Erb2012}. The two dashed black lines represent optically thin and optically thick cases. This reflects that Fe~II is not optically thick in CSWA-141 and therefore has a lower column density c.f. black points}
\label{fig:feii_plot}
\end{figure}

Further insight into the gas conditions of EELGs is made possible by detection of three density-sensitive emission lines in CSWA-141.  As we discussed in \S3.1, the [CIII], CIII] doublet implies 
very high gas densities ($16500^{+12100}_{-7800}$cm$^{-3}$), perhaps again reflecting a very compact or concentrated gas geometry surrounding the extremely young star clusters that power EELGs. Although the uncertainty in CIII] density is currently large,  such high densities are routinely seen in $z\simeq 2-4$ galaxies with spectrally-resolved CIII] measurements (e.g., \citealt{Christensen2012,James2014,James2018,Bayliss2014,Acharyya2019}) and are also starting to be seen in the first handful of $z\gsim 7$ galaxies with CIII] doublet measurements \citep[][]{Stark2017,Jiang2021}. As such, the high C~III] density is less likely due to a statistical fluctuation, although a larger sample should confirm this. In most of the cases, the CIII] density is also significantly in excess of that inferred from lower-ionization 
species. The same trend is apparent in CSWA-141, with the CIII] density being nearly two orders of magnitude higher than those derived from [SII] and [OII]. This is consistent with a picture (e.g., \citealt{Kewley2019,Acharyya2019,Berg2021}) dictated by the ionization structure of the nebulae, with the lower ionization rest-optical lines (i.e., [OII] and [SII]) probing primarily the outer layers which are preferentially dominated by lower density gas. The higher ionization lines (like CIII]) probe the inner regions of the nebula, where densities are expected to be higher. Variations in the critical density of the ions will additionally contribute to CIII] probing higher density gas than [OII] (e.g., \citealt{James2018}). A key question is whether very young EELGs like CSWA-141 (and many of the $z>7$ galaxies) might have a large fraction of their ionized gas in very dense clumps, as is expected at the earliest evolutionary phases following the formation of star clusters (e.g., \citealt{Kim2018, Rigby2004}). Currently samples with CIII] densities tend to be those that have at least moderately large CIII] 
EW, generally indicative of a relatively young stellar population. 
Larger CIII] density samples are required to test whether the presence of very large densities is at all sensitive to the luminosity-weighted stellar population age. 

\begin{figure}
\centering
\subfloat{\includegraphics[scale=0.6]{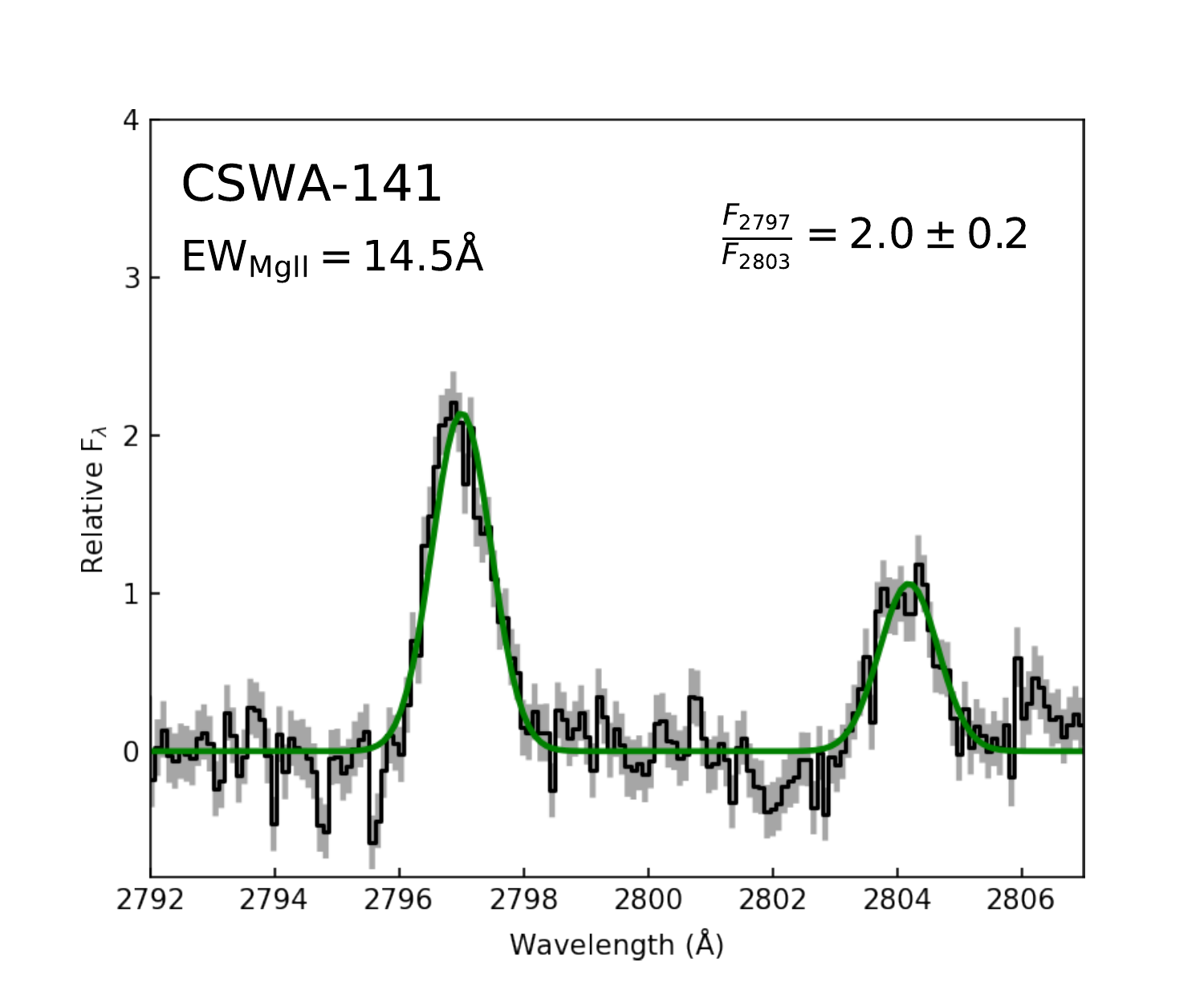}}
\caption{Mg II 2797,2803 line profile in CSWA-141. The black curve represents continuum subtracted flux level while the green solid line is gaussian fit to the observed data. Both the Mg II doublet are well fitted by a single component gaussian. Mg II 2797 is clearly stronger than Mg II 2803 line, suggesting a Mg II doublet intensity ratio ($F_{2797}$/$F_{2803}$) of 2.0$\pm$0.2. The doublet ratio is consistent with the intrinsic recombination value (2.0) indicating minimal scattering by low-ionization ISM along the line of sight. Based on the observed Mg II doublet ratio, CSWA-141 has implied escape fraction of $f\rm_{esc}$(LyC) = 27$\pm$4\%.}
\label{fig:mgii_plot}
\end{figure}

CSWA 141 also provides a unique window on the kinematics 
and covering fraction of the outflowing gas in EELGs, a population 
which becomes commonplace at $z>6$. As it is this  outflowing 
gas which regulates the escape of ionising radiation, systems like CSWA 141 
offer potential for understanding the likely contribution of EELGs 
to reionisation. At low redshift, \citep{Jaskot2017, Chisholm2017} find that the most 
extreme [OIII] emitting galaxies (the Green Peas) tend to 
have low outflow velocities and suggest that this is consistent with models of suppressed superwinds, where catastrophic cooling prevents the development of large scale outflows (e.g., \citealt{Silich2007,Silich2017,Gray2019}). 
The low ionization absorption lines in the CSWA 141 
spectra are all blueshifted with respect to the systemic 
redshift, indicating the presence of outflows. In the ESI spectrum, we detect Fe II and Mg II absorption lines, while the MODS spectrum probes slightly bluer wavelengths, allowing detection of Al II$\lambda$1670. The average velocity of low ionization outflowing gas in CSWA 141 is 76 km s$^{-1}$.
This is slightly lower than outflow velocities of 
100-300 km s$^{-1}$ that are commonly observed in galaxies at $z\sim1-3$ (e.g., \citealt{Shapley2003,Weiner2009,Steidel2010,Jones2012}). However given the low stellar mass of CSWA-141, the measurement is consistent with expectations from known trends between galaxy mass and outflow velocity (e.g., \citealt{Martin2005,Weiner2009,Erb2012, Chisholm2016}).
 
 The strength of the absorption lines in CSWA 141 provide insight into 
 the opacity of neutral gas along the line of sight to the young 
 star clusters which dominate the light. The Fe~II $\lambda$2374 and Fe~II $\lambda$2382 absorption lines are very weak with equivalent widths of 0.5~\AA\ and 1.5~\AA, respectively. \citet{Jones2018} measured a column density N(Fe II) = 1.5$^{+0.7}_{-0.2} \times 10^{14}$ cm$^{-2}$, which is among the lowest of the sample. In Figure~\ref{fig:feii_plot}, we compare these equivalent widths with those measured from composite spectra of more typical star forming galaxies at $1<z<2$ \citep{Erb2012}. As is clear in the figure, most star forming  galaxies at these redshifts show similar Fe~II $\lambda$2374 and Fe~II $\lambda$2382 equivalent widths, as expected 
for optically thick neutral gas. In contrast, the Fe~II $\lambda$2374 
absorption in CSWA-141 is significantly weaker than  Fe~II $\lambda$2382. 
With an Fe~II $\lambda$2374 EW that is roughly three times weaker than 
that found in the $z\simeq 1-2$ composites, the lines are much closer to expectations for optically thin neutral gas. We additionally note that 
the Fe~II $\lambda$2382 line is susceptible to emission filling (e.g., \citealt{Erb2012}) which tends to weaken the observed Fe~II $\lambda$2382 equivalent widths. This is particularly likely to be the case for 
EELGs like CSWA-141. In such a scenario, the line ratio would move 
even closer to that expected for optically thin conditions, implying  a low covering fraction and potentially low column density of neutral gas in the outflow. \\
 
 Further indications that CSWA 141 has a low column density of neutral gas comes from the
 resonant Mg II$\lambda\lambda$ 2797,2803 emission line. Both components of the  doublet are
 confidently detected (S/N$>$10) in emission with total equivalent widths of 14.5 \AA\
 (Figure~\ref{fig:mgii_plot}). This is not only one of the highest measured Mg II EWs at
 z$>$1, but it is also one of the brightest Mg II emission lines, making detailed (and 
 resolved) study of the line uniquely possible. Recent studies have pointed out that the
 flux ratio of the doublet is strongly sensitive to the neutral gas column density in the
 galaxy. As such it is thought to correlate closely with the ionizing photon escape fraction 
 \citep{Henry2018,Chisholm2020, Xu2022, Seive2022,Xu2023}, perhaps providing one of the best indirect indicators of
 photon leakage.  At low HI column densities ($<$10$^{17.2}$ cm$^{-2}$), galaxies become
 optically thin to the resonant Mg II line photons \citep{Chisholm2020}.  This leads to strong
 nebular Mg II emission, and it drives the doublet 
 ratio (R=F$_{2797}$/F$_{2803}$) to its intrinsic value of $R=2$. If the line photons are
 resonantly scattered by Mg$^+$ ions in the neutral gas along the line of sight, the doublet ratio will decrease, asymptoting to a value of R=1 if the gas is optically thick to Mg II photons  (see \citealt{Chisholm2020} for a detailed discussion). 
 
The measured Mg II doublet ratio in CSWA-141 is R=2.0$\pm$0.2, consistent with the intrinsic value 
produced in the HII regions. This suggests that optically thin channels along the line of sight to the young star clusters are powering the nebular emission. 
Both components of the doublet are  well-fit by single Gaussians, suggesting that the line 
profile is not significantly impacted by resonant scattering. The very large EW of the line 
also points to minimal attenuation of line photons. We note that the Mg II profile does 
show very weak blue-shifted absorption (see Figure~\ref{fig:mgii_plot}), suggesting the 
presence of some neutral gas along the line of sight, but this gas must be either 
optically thin or clumpy with low density channels to result in the observed line profile.

Given the derived oxygen abundance of CSWA-141 (see \S3.1) and nominal assumptions on the Mg/O ratio \citep{Chisholm2020}, this can be converted to an estimate of the hydrogen 
column density (See equation 14 of \citealt{Chisholm2020}). Given the measured 
doublet ratio is consistent with the intrinsic value, CSWA-141 is formally consistent 
with a negligible hydrogen column density. Within the measurement 
errors of the flux ratio, we find a 1$\sigma$ upper limit on the HI column density
of 3.8$\times$10$^{16}$ cm$^{-2}$. This value is well below the HI column density at 
which galaxies become optically thin to LyC radiation ($<$10$^{17.2}$ cm$^{-2}$), suggesting 
that CSWA-141 may be a likely candidate for LyC leakage. While more realistic geometries  (i.e., clumpy gas) would alter the derived column densities, the observed line profile 
requires there to be low density channels along the line of sight where resonant line 
photons (and potentially LyC emission) are transmitted \citep{Gazagnes2020, Saldana-Lopez2022}. Because of the extreme brightness 
of CSWA-141, it presents a unique opportunity to spatially resolve the absorbing gas in an 
extreme line emitter that is very similar in its properties to those systems at $z>6$.
An upcoming {\it HST} UVIS grism observations (GO-16710, PI: Mainali) will
provide more direct constraints on the escape of ionizing radiation from the galaxy.

 \begin{table*}
\begin{tabular}{lcccccccc}
\hline Object & EW$\rm_{CIII]}$ & R23 & O32 & O3  &Electron Density&   12 + log (O/H)   & Reference \\
&($\rm \AA$)& &&& (cm$^{-3}$) &   && \\
\hline \hline

CSWA-141 &4.6$\pm$1.9 & 11.4$\pm$1.0 & 6.6$\pm$1.0 & 7.4$\pm$0.2 & 160$^{+76}_{-74}$$^{a}$, 350$^{+294}_{-206}$$^{b}$, 16500$^{+12100}_{-7800}$$^{c}$ & 7.95$\pm$0.08$^{d}$  & This work  \\ 
CSWA-2 & 3.1$\pm$1.6& \ldots &  \ldots & 5.2$\pm$2.1 & \ldots & 8.4$\pm$0.2$^{g}$  & This work, 16  \\ 
CSWA-128 & 0.7$\pm$0.1 & 7.0$\pm$1.2 & 3.3$\pm$0.8 & 4.0$\pm$1.1 & 198$^{+85}_{-76}$$^{b}$ & 8.2$\pm$0.2$^{e}$, 8.4$\pm$0.2$^{f}$, 8.5$\pm$0.2$^{g}$  & This work \\ 
CSWA-164 & 0.4$\pm$0.1 & \ldots & 0.7$\pm$0.2 & \ldots & 165$^{+65}_{-46}$$^{a}$ &  8.6$\pm$0.2$^{e}$  & This work  \\ 
CSWA-163 & $<3.1$ & 10.9$\pm$1.1 & 1.1$\pm$0.1 & 4.3$\pm$0.6 & 144$^{+45}_{-33}$$^{a}$ &  8.5$\pm$0.2$^{e}$, 8.4$\pm$0.2$^{g}$ & This work \\
CSWA-165 & $<1.9$ & 7.1$\pm$1.3 & 0.6$\pm$0.1 & 2.1$\pm$0.4 & 221$^{+75}_{-64}$$^{a}$ &  8.4$\pm$0.2$^{e}$, 8.6$\pm$0.2$^{f}$, 8.6$\pm$0.2$^{g}$  & This work  \\ 
CSWA-11& $<1.4$ & \ldots & 0.8$\pm$0.2 & \ldots & 469$^{+120}_{-156}$$^{a}$ & 8.5$\pm$0.2$^{e}$  & This work  \\ \hline\hline
RXCJ0232-588 &21.7 & 6.0 & 9.4 & 4.1 & 80$^{a}$  & 7.61 $^{d}$ & 1 \\ 
$\rm{Ion2}$ &18& \ldots & $>15$&14.7 & \ldots & 8.07$^{h}$  & 2 \\ 
860$\_$359 & 12.4 & $<9.2$ & $>6.91$ &6.0 & \ldots & $<8.1^{f}$ & 3  \\
ID14 & 11.8 & $<11.7$ &$>10$ &7.6 & \ldots & $<7.8^{d}$  & 4 \\
SL2S0217 & 11.7 & \ldots & \ldots & 3.1 & 300$^{c}$ & $7.5^{d}$  & 5  \\
SGAS J1050+0017 & 11 &11.4 & 9.7 & 7.9 & 2-3$\times100^{a}$ & $>8.1^{d}$  &6 \\
ID11 & 11 & $<11.8$ &$>10$ & 8.4 & \ldots & $7.7^{d}$ & 7 \\
CSWA-20 & 9.1 & 7.7 & 5.6 & 4.9 & 276$^{a}$, 17100$^{b}$ & 7.82$^{d}$ & 8\\
BX418 &7.1& $<9.2$ & $>11.6$ & 6.4 & \ldots & 7.8$^{d}$  & 9  \\ 
A1689 31.1 & 7& 7.8 & 8.2 & 4.7 & 330$^{a}$, 2900$^{c}$& 7.76$^{d}$ & 10 \\ 
MACS 0451-1.1&6.7& $<5.9$  & $>8.4$ & 3.9 & \ldots  & $<8.0^{f}$   & 3  \\ 
SDSS J1723+3411 & 4.0 & 8.6 & 5.5 & 5.4 & 47$^{a}$, 1950$^{c}$ & 8.4$^{f}$ & 11 \\
MACS 0451-3.1 &2.4& \ldots & $>2.0$ & \ldots & \ldots & \ldots  & 3 \\
Cosmic Horseshoe & 0.9 & 5.5 & 1.31 & 2.4 & 840-6900$^{b}$ & 8.65$^{f}$ & 12,13 \\ 
MS 1512-cB58 & 0.8 & 7.9 & 1.36 &3.6& \ldots & 8.47$^{f}$ & 14,15  \\ \hline 
\end{tabular}
\\
$^{a}$Derived from [OII] doublet, $^{b}$Derived from [SII] doublet, $^{c}$Derived from CIII] doublet.\\
 $^{d}$Direct (T$_{e}$) method, $^{e}$O32 method, $^{f}$R23 method, $^{g}$O3 method,\\
 $^{h}$ Using HII-CHI-mistry code (Perez-Montero 2014)\\
$^{1}$\citet{Mainali2020},$^{2}$\citet{deBarros2016},$^{3}$ \citet{Stark2014},$^{4}$\citet{Vanzella2017},$^{5}$\citet{Berg2018},$^{6}$\citet{Bayliss2014},$^{7}$\citet{Vanzella2016},$^{8}$\citet{James2014}, $^{9}$\citet{Erb2010},$^{10}$\citet{Christensen2012}, $^{11}$\citet{Rigby2021}, $^{12}$\citet{Quider2009}, $^{13}$\citet{Hainline2009}, $^{14}$\citet{Pettini2000}, $^{15}$\citet{Teplitz2000}, $^{16}$\citet{Jones2013}
 
\caption{Rest-optical emission line properties of galaxies at high redshift with CIII] emission constraints. The upper half shows data from this paper whereas lower half shows measurements from the literature.}
\label{table:compilation}
\end{table*}

\section{Summary}

We present new spectroscopic and photometric  observations of sixteen bright 
gravitationally lensed galaxies originally identified in SDSS via the CASSOWARY program.
Observations were conducted using LBT, Keck, MMT and Magellan. Included in this 
sample is the $z=1.42$ galaxy CSWA-141, one of the brightest known EELGs at high 
redshift, with an [OIII]+H$\beta$ EW (730~\AA) nearly identical to the average value seen at $z\simeq 7-8$. In this paper, we focus on the rest-UV spectral properties of the sample, leveraging high quality Keck/ESI data. Owing to the brightness of our targets ($g\simeq 19$-21), we are able to detect rest-UV metal line emission in the Keck spectra down to very low EW values. While most systems have  weak line emission (median CIII] EW =1.7~\AA), CSWA 141 shows relatively strong emission  (CIII] EW 4.6~\AA) together with detections of a variety of UV lines (OIII], Si III], Fe II$^\star$, Mg II). 

We  compare the properties of the strong ($\gsim 3.5$~\AA) and weak ($\lsim 3.5$~\AA) CIII] emitters in our sample and in the literature. We find that the stronger 
CIII] emitters have larger sSFR and lower 
gas-phase oxygen abundances. We 
find that the strong CIII] emitters can 
be easily separated by their rest-optical line ratios, with larger values of O32 at 
roughly fixed R23.  Overall, these results 
suggest that CIII] tends to be strong 
in galaxies dominated by young stellar populations with low metallicity and 
extreme ionization conditions. This is 
consistent with trends found in observations at low and high redshift (e.g., \citealt{Rigby2015,Senchyna2017,Senchyna2019,Maseda2017,Nakajima2018,Schaerer2018,Du2020,Ravindranath2020,Tang2021a}) and in 
photoionization models (e.g., \citealt{Jaskot2016,Nakajima2018}). 

The brightness of CSWA-141 enables a detailed investigation of an EELG with properties similar to that which become common at $z>6$. This galaxy is characterized by low stellar mass (4.0$\times$10$^{8}$ M$_\odot$), large sSFR (31.2 Gyr$^{-1}$), low gas-phase metallicity (12+log O/H=7.95) 
and relatively highly ionized gas (O32=6.7) and has likely undergone a recent upturn or burst of star formation. We find that the electron density traced by the CIII] doublet (1.65$\times$10$^{4}$ cm$^{-3}$) is higher than that traced by [OII] and [SII] doublet (160 and 350 cm$^{-3}$, respectively), a discrepancy that is also found in other systems (e.g., \citealt{James2014,Acharyya2019}).  While this is likely to reflect the ionization structure of the HII regions powering the lines \citep[][]{Kewley2019}, it may also indicate that CSWA-141 contains a significant fraction of its ionized gas in very dense 
clumps, as is expected in the earliest stages following the formation of star clusters \citep[e.g.][]{Kim2018}.

The spectra of CSWA-141 provide several probes of the 
neutral gas opacity in the galaxy, including both 
low ionization absorption lines and the resolved Mg II doublet. The Fe II$\lambda\lambda$2374, 2382 absorption lines indicate the presence of outflowing gas with average 
velocity of 76 km s$^{-1}$. The lines are much weaker 
than in typical star forming galaxies at $z\simeq 1-2$, implying a low covering fraction and potentially low column density of neutral gas in the outflow. The resonant Mg II$\lambda\lambda$ 2797,2803 emission line supports this 
picture.  The Mg II doublet ratio in CSWA-141 ( R= F$_{2797}$/F$_{2803}$) is  2.0$\pm$0.2, consistent with the intrinsic value produced in the HII regions. When 
combined with the very large EW of Mg II and the near-Gaussian profiles of the doublet components, this suggests minimal resonant scattering, consistent with 
a very low column density of neutral hydrogen. These 
indirect indicators suggest CSWA-141 may be a likely 
candidate for LyC leakage.

\section*{Acknowledgments}
We would like to thank Ryan Endsley for helping with MMT observations of some of the sources presented in this paper. We also thank Stephane Charlot and Jacopo Charlot for making the BEAGLE population synthesis  tool available to us 
for this paper.

DPS acknowledges support from the National Science Foundation
through the grant AST-2109066.  TJ acknowledges support from the National Science Foundation through the grant AST-2108515. RSE acknowledges funding from the European Research Council (ERC) under the European Union’s Horizon 2020 research and innovation 0programme (grant agreement No 669253). Y.H. acknowledges support from the National Sciences and Engineering Council of Canada grant RGPIN-2020-05102, the Fonds de recherche du Québec grant 2022-NC-301305, and the Canada Research Chairs Program.

Observations presented in this 
paper were obtained from the Keck Observatory, which was made possible by the generous financial support of the W. M. Keck Foundation. The material is based upon work supported by NASA under award number 80GSFC21M0002. The authors acknowledge the very significant cultural role that the summit of Mauna Kea has always had within the indigenous Hawaiian community.
We are most fortunate to have the opportunity to conduct observations from this mountain.  Some of the observations reported here were
obtained at the MMT Observatory, a joint facility of the University
of Arizona and the Smithsonian Institution. This paper includes data gathered with the 6.5 meter Magellan Telescopes located at Las Campanas Observatory, Chile. Some of the data presented in this paper were obtained using the Large Binocular Telescope (LBT). The LBT is an international collaboration among institutions in the United States, Italy and Germany. The LBT Corporation partners are: The University of Arizona on behalf of the Arizona university system; Istituto Nazionale di Astrofisica, Italy;  LBT Beteiligungsgesellschaft, Germany, representing the Max Planck Society, the Astrophysical Institute Potsdam, and Heidelberg University; The Ohio State University; The Research Corporation, on behalf of The University of Notre Dame, University of Minnesota and University of Virginia.


\section*{DATA AVAILABILITY}
The data underlying this article will be shared on reasonable request
to the corresponding author.



\bibliographystyle{mnras}
\bibliography{references} 




\appendix


\bsp	
\label{lastpage}
\end{document}